\RequirePackage{silence}
\RequirePackage{rotating}
\documentclass{jfm}

\let\proof\relax

\usepackage{newtxtext}
\usepackage{newtxmath}
\usepackage{bbding}
\usepackage{natbib}
\usepackage{alphalph}
\usepackage{xfrac}
\usepackage{graphicx}
\usepackage{comment}
\usepackage{float}
\usepackage{subcaption}
\usepackage{placeins}
\usepackage{array}
\usepackage[colorlinks=true,urlcolor=blue,citecolor=blue,linkcolor=blue,filecolor=blue,pdfcreator={},pdfproducer={}]{hyperref}
\usepackage[nameinlink,noabbrev]{cleveref}
\usepackage[normalem]{ulem}
\usepackage{xcolor, soul}

\crefname{section}{§}{§§}
\Crefname{section}{§}{§§}
\crefformat{section}{§#2#1#3}
\crefformat{subfigure}{#2figure~#1#3}
\Crefformat{subfigure}{#2Figure~#1#3}

\newcommand{\RomanNumeralCaps}[1]
\linenumbers

\definecolor{red-dark}{rgb}{0.6350 0.0780 0.1840}
\definecolor{orange-dark}{rgb}{0.8500 0.3250 0.0980}
\definecolor{yellow-dark}{rgb}{0.9290 0.6940 0.1250}
\definecolor{blue-dark}{rgb}{0 0.4470 0.7410}
\definecolor{blue-light}{rgb}{0.3010 0.7450 0.9330}

\title{Turbulent heat transfer in open-channel flows with a thermally conductive porous wall}

\author{Seyed Morteza Habibi Khorasani\aff{1,2}\corresp{\email{smhk2@mech.kth.se}}, Geert Brethouwer\aff{1} \and Shervin Bagheri\aff{1}}

\shortauthor{S. M. Habibi Khorasani, G. Brethouwer and S. Bagheri}

\affiliation{\aff{1}FLOW, Department of Engineering Mechanics, KTH Royal Institute of Technology, SE-100 44 Stockholm, Sweden
\aff{2}Group Trucks Technology, Volvo Group, SE-405 08 Gothenburg, Sweden
}

\begin{document}
\maketitle

\begin{abstract}
 Results of direct numerical simulations (DNS) of porous-wall turbulent flows in open channels with conjugate heat transfer are reported in this work. For the conductive porous walls considered here, the change in heat transfer is not monotonic. The heat flux initially decreases when going from a conductive smooth wall to slightly porous walls. In this initial porous-wall turbulence regime, the near-wall flow remains smooth-wall like and the heat transfer is dominated by molecular diffusion. As such, a reduction of the more favorably conducting solid material diminishes the overall heat transfer performance. Beyond a certain level of permeability however, the near-wall flow transitions to the K-H-like regime \citep{gomez2019,Khorasani_Luhar_Bagheri_2024} marked by the presence of cross-stream rollers, and the heat flux undergoes an increasing trend until it eventually surpasses that of smooth-wall turbulence. Neglecting the thermal behavior of the solid material can therefore result in overestimations of any gains in heat transfer.
 Additionally, thermal performance is assessed in terms of the Reynolds analogy breakdown, which is the disparity between the fractional increases in the Stanton number, $St$, and the fractional increases in the skin-friction coefficient, $C_f$, relative to smooth-wall flow \citep{Bunker_2013,rouhi_endrikat_modesti_sandberg_oda_tanimoto_hutchins_chung_2022}. Similar to rough walls, the breakdown is unfavorable for porous walls. The unfavorable breakdown in Reynolds analogy is due to growing dissimilarities between the transfer of momentum and heat in the vicinity of the porous wall as it becomes more permeable. Turbulent sweep and ejection type events contribute more significantly to momentum transfer across the permeable surface than they do to heat transfer. However, unlike for rough walls, a saturation limit for heat transfer is not observed for the porous walls considered here. How much of a maximum increase in heat transfer can be achieved is something that remains to be determined.
\end{abstract}


% {\bf MSC Codes }  {\it(Optional)} Please enter your MSC Codes here

\section{Introduction}\label{sec:intro}

\subsection{Passive modification of momentum and heat transfer using surfaces}

 Passive means of affecting turbulent flows for flow control purposes is an area of intense research. What makes the passive approach attractive is the lack of any additional power expenditure which is needed when using active means. Different surfaces such as riblets, roughness, and porous walls can alter the dynamics of flow turbulence \citep{endrikat_2021,CHUNG2021,gomez2019,Khorasani_Luhar_Bagheri_2024}. This, in turn, has motivated studies on how turbulent heat transfer over such surfaces may differ from its smooth-wall counterpart. \cite{leonardi_orlandi_djenidi_antonia_2015} investigated turbulent heat transfer over surfaces with transverse bar\nobreakdash-type roughness using direct numerical simulations (DNS). They reported that the heat flux over the rough surfaces was higher compared to a smooth-wall, although it was always accompanied by an even greater increase in drag. This difference in turbulent momentum and heat transfer was due to the flow separation occurring in the vicinity of the roughness elements, leading to increased contributions from pressure drag which has no heat transfer analogue.
 
 Dissimilarity in the transport mechanisms of heat and momentum transfer represents a breakdown of the Reynolds analogy \citep{VonKarman_1939}. This analogy is based on the phenomenological understanding that the turbulent transport of both heat and momentum over smooth walls are similar and thus parity between the heat and momentum fluxes should exist. The analogy commonly takes the form of a scaling factor which is the ratio between the wall heat flux and wall shear stress, with the former expressed in terms of the Stanton number, $St$, and the latter the skin-friction coefficient, $C_f$. Therefore, for practical engineering purposes knowledge of the Reynolds analogy factor allows for estimating the heat transfer of a system when equipped only with knowledge of the skin friction. The analogy also serves as a measure of heat\nobreakdash-transfer efficiency, since it measures the fractional increase in heat transfer relative to the fractional increase in skin friction. Deviations from the analogy factor of a target surface represent a breakdown in the analogy, which can be favorable (greater rate of increase in $St$ relative to the rate of increase in $C_f$) or unfavorable (lower rate of increase in $St$ relative to the rate of increase in $C_f$). As such, for the bar\nobreakdash-type roughness investigated by \cite{leonardi_orlandi_djenidi_antonia_2015}, the breakdown in the analogy relative to a smooth-wall turbulent flow is unfavorable.
 
 Further investigations of rough-wall turbulent heat transfer were carried out by \cite{PEETERS_2019}, who extended the conclusions of \cite{leonardi_orlandi_djenidi_antonia_2015} to the more general category of irregularly-rough surfaces. Their results also indicated that heat transfer does not increase indefinitely as the size of the viscous-scaled roughness elements increases. An observation that was also made in the study of \cite{macdonald_hutchins_chung_2018}, who conducted minimal-channel DNS of turbulent heat transfer over walls with sinusoidal roughness of different heights. They observed that for a given physical roughness size, the Stanton number achieves a maximum in the transitionally rough regime and subsequently decays monotonically in the form-drag dominated fully rough regime.
 
 In a recent study, \cite{rouhi_endrikat_modesti_sandberg_oda_tanimoto_hutchins_chung_2022} gathered and reported the heat-transfer efficiencies reported throughout the literature for both active cooling methods and different surface types (see figure 1 of their manuscript). This highlighted that active methods such as impingement jets and wall blowing/suction have demonstrated the greatest favorable performance. Rough surfaces, due to the reasons described above, consistently demonstrate unfavorable performance. \cite{rouhi_endrikat_modesti_sandberg_oda_tanimoto_hutchins_chung_2022} also investigated turbulent heat transfer over riblets. Overall, the majority of the investigated riblets led to an unfavorable breakdown of the Reynolds analogy, with only certain triangular-shaped riblets achieving a somewhat favorable breakdown.

 The purpose of this study is to investigate porous-wall turbulent heat transfer using DNS that also take into account the heat transfer in the solid phase. The objective being to analyze the heat transfer enhancement with respect to the effects of porous walls on flow turbulence. As such, a brief overview of porous-wall turbulence is provided next. For the interested reader, more extensive details may be found in the provided references.   

\subsection{Porous-wall turbulence}\label{sec:porous_wall_flow}

 % \begin{figure}
 %  \centerline{\includegraphics[width=0.9\linewidth]{Figures/Regimes}}
 %  \caption{Conceptual sketch of porous-wall turbulence regimes. Reproduced from \cite{khorasani_2023}. \todo[inline]{Seyed: Placeholder image. Final one will be different.}}
 %  \label{fig:regime_sketch}
 % \end{figure}

 In their investigation of porous-wall turbulence, \cite{gomez2019} used analytical solutions derived from Brinkman's equation \citep{brinkman_1949}, $\nabla{p}=-~\nu\,{\mathsfbi{K}}^{-1}\,\boldsymbol{u}~+~\tilde{\nu}\,\nabla^2{\boldsymbol{u}}$, as boundary conditions to introduce the effect of anisotropic porous walls on a turbulent channel flow. Assuming a permeability tensor, $\mathsfbi{K}$, with only the principle components, $K_x$, $K_y$, and $K_z$ being non-zero (zero off-diagonal components), and streamwise\nobreakdash-preferential configurations (${K_x}\hspace{-1pt}>\hspace{-1pt}{K_y}$), they demonstrated the existence of a limited regime where turbulence remains smooth-wall-like for small values of the viscous\nobreakdash-scaled effective wall\nobreakdash-normal permeability ($\sqrt{K_y^+}\hspace{-1pt}<\hspace{-1pt}0.6$). Beyond this regime, instabilities in the flow become triggered, giving rise to cross\nobreakdash-stream rollers and the flow transitioning to the so-called K\nobreakdash-H\nobreakdash-like regime. The existence of these cross\nobreakdash-stream rollers had earlier been demonstrated by \cite{jimenez_uhlmann_pinelli_kawahara_2001}. Through linear stability analysis, they attributed their emergence to the shear waves of the inviscid mean velocity profile which grow unstable in the presence of walls with small to moderate degrees of permeability. For very large permeabilities, the relevant instability becomes the Kelvin\nobreakdash-Helmholtz instability of shear layers. This distinction between high and low permeability flow characteristics was also made in the study of \cite{MANES2011} where turbulent boundary layers over isotropic porous foams were experimentally investigated. They observed that only for the foam with the largest permeability ($\sqrt{K^+}\hspace{-3pt}=\hspace{-3pt}17.2$) did the flow exhibit the characteristic frequency associated with the Kelvin\nobreakdash-Helmholtz instability of turbulent mixing layers, such as those over plant canopies \citep{FINNIGAN2000}. Therefore, even though the Kelvin\nobreakdash-Helmholtz instability does not give rise to the cross\nobreakdash-stream rollers observed over low- to intermediately\nobreakdash-permeable walls, owing to the similarity of the rollers to those of the well-known K\nobreakdash-H inviscid instability, this flow regime is commonly referred to as the K\nobreakdash-H\nobreakdash-like regime.

 In a preceding study, \cite{Khorasani_Luhar_Bagheri_2024} conducted scale-resolving simulations of turbulent open-channel flows over porous lattices with low to intermediate permeabilities. The low-permeability cases retained the near-wall flow characteristics of smooth-wall turbulence while the intermediate-permeability cases fell into the K-H-like regime.
 % An illustration of these regimes is sketched in \cref{fig:regime_sketch}.
 The approximate permeability threshold across which this regime change occurred was also in reasonable agreement with the linear stability predictions of \cite{sharma2017,gomez2018} using their Brinkman-derived permeable-wall boundary conditions.

  \begin{figure}
   \centerline{\includegraphics[width=0.9\linewidth]{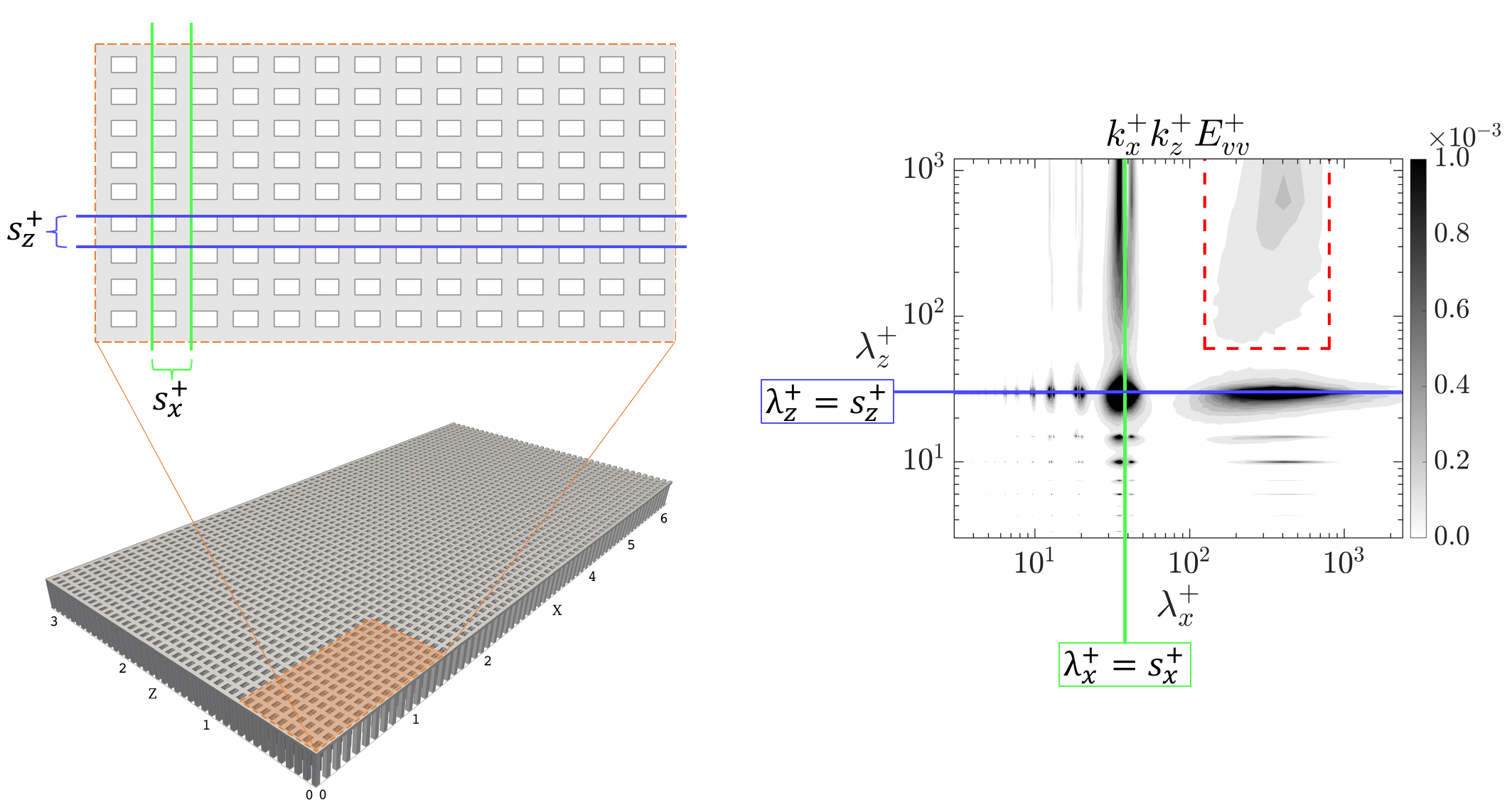}}
   \vspace*{-46ex}
   \begin{center}
   \hspace*{-85ex}(\emph{a})
   \end{center}
   \vspace*{46ex}
   \vspace*{-43ex}
   \begin{center}
   \hspace*{5ex}(\emph{b})
   \end{center}
   \vspace*{43ex}
   \vspace*{-15pt}
   \captionsetup{width=0.98\textwidth,justification=justified}
   \caption{Illustration of how the texture-coherent flow scales are related to the physical pore dimensions: The region of the porous wall magnified in (\emph{a}) has its streamwise and spanwise pitches, $s_x^+$ and $s_z^+$, marked using the \textcolor{green}{green} and \textcolor{blue}{blue} lines, respectively. The wall-normal premultiplied spectral energy is shown in (\emph{b}), where the area enclosed by the \textcolor{red}{red} dashed-line represents the ambient turbulence scales. The contours indicated by the \textcolor{blue}{blue} and \textcolor{green}{green} lines in (\emph{b}) are the principle energetic scales of the texture-coherent flow component that occur at wavelengths equal to the porous wall pitches. The similar-looking but less intense contours at lower wavelengths are the harmonics of the texture-coherent flow.}
  \label{fig:pore_coherent_flow_illustration}
 \end{figure}

 Aside from the permeability-triggered flow regime change, there is also the flow component induced by the presence of the porous wall. Any solid structure introduces spatial inhomogeneities into the flow-field since the flow is forced to navigate around the solid obstructions. The resulting flow component has scales which conform to the spatial wavelengths of the porous wall. This flow component is commonly referred to as the texture-coherent flow for a given surface texture or structure \citep{Fairhall_Abderrahaman-Elena_García-Mayoral_2019}. Its presence is demonstrated in \cref{fig:pore_coherent_flow_illustration}, which shows the premultiplied spectral energy of the wall-normal velocity, $k_x^+k_z^+E_{vv}^+$, (\hyperref[fig:pore_coherent_flow_illustration]{figure 1b}) for the turbulent flow over the depicted porous wall (\hyperref[fig:pore_coherent_flow_illustration]{figure 1a}). The energy spectrum reveals that in addition to the turbulent scales of the flow (the area enclosed by the dashed red lines), energetic scales exist at wavelengths equal to the wall-parallel pore spacings $s_x^+$ and $s_z^+$, which are the signature of the texture-coherent flow. The texture-coherent flow undergoes modulation by the surrounding turbulence \citep{Abderrahaman2019} and the two can also directly interact with one another should their scales overlap. Such interactions make additional non-linear contributions to the near-wall dynamics. 

\subsection{Purpose of this study}\label{sec:study_purpose}

 In this study, turbulent heat transfer over conductive porous lattices comprised of cuboid pores covering a range of permeabilities, from small ($\sqrt{K^+}\approx1$) to large ($\sqrt{K^+}\approx19$), are numerically investigated. In terms of flow, these porous walls span the range of regimes from smooth-wall-like to K-H-like, and also include configurations where scale-overlap between the ambient turbulence and texture-coherent flow occurs. This way, we can examine how heat transfer evolves over porous walls as we increase the permeability and pass through different stages of porous-wall turbulence. It also allows for comparisons to be made with heat transfer over rough walls across their transitional and fully rough regimes, including whether or not heat transfer saturates over porous walls as it does over rough walls. Additionally, inclusion of conjugate heat transfer permits examining whether increasing permeability always benefits heat transfer or can have a detrimental effect on it.     
 As mentioned previously, heat-transfer efficiency is assessed using the Reynolds analogy, following \cite{Bunker_2013} and \cite{rouhi_endrikat_modesti_sandberg_oda_tanimoto_hutchins_chung_2022}. A preliminary effort is also made to examine the dissimilarity in momentum and and heat transfer (Reynolds analogy breakdown) over the porous walls in terms of near-wall flow events that contribute to the turbulent transport of heat and momentum.

\section{DNS setup}\label{sec:numerics_and_cases}

 \begin{figure}
  \begin{center}
   \hspace*{5mm}\includegraphics[width=1\linewidth]{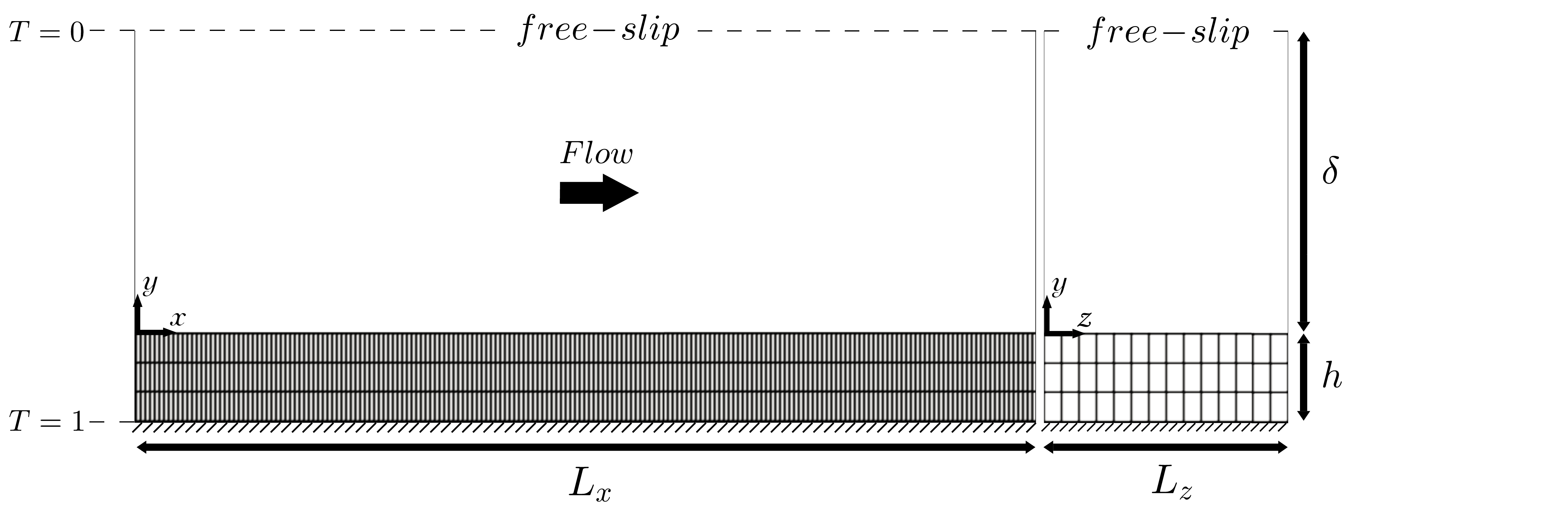}
   \vspace*{-3mm}
   % \captionsetup{justification=justified}
   \caption{Sketch of the computational domain, similar to that of \cite{Khorasani_Luhar_Bagheri_2024}.}
   \label{fig:domain}
  \end{center}
 \end{figure}

 The numerical setup used here is similar to that of \cite{Khorasani_Luhar_Bagheri_2024} and the interested reader can refer to that work for details. However, for the sake of clarity, the most pertinent aspects as well as any differences are described in this section.
 
 DNS of open channel flow were conducted in the numerical domain depicted in \cref{fig:domain}. The dimensions of the domain are $(L_x,L_y,L_z)=(2\pi\delta,\delta + h,\pi\delta)$ where $x, y ,z$ refer to the streamwise, wall-normal and spanwise directions. It has a channel height of $\delta$ and a porous wall thickness of $h=0.3\delta$. The flow is governed by the incompressible Navier\nobreakdash-Stokes equations,
 \begin{subequations}\label{eq:NS}
 \begin{gather}
 \nabla\cdot{\boldsymbol{u}} = 0,\\
 \cfrac{\partial \boldsymbol{u}}{ \partial t} + \boldsymbol{u}\cdot\nabla{\boldsymbol{u}} = -\cfrac{1}{\rho}\nabla{p} + \nu{\nabla}^2{\boldsymbol{u}}+\Pi{\boldsymbol{e_x}},
 \end{gather}
 \end{subequations}
 where ${\boldsymbol{u}(x, t)}=(u,v,w)$ is the velocity vector, ${\boldsymbol{p}(x, t)}$ is the pressure, $\rho$ is the density, and $\nu$ is the kinematic viscosity. The vector $\boldsymbol{e_x}$ is a unit vector in the streamwise direction and $\Pi$ is the spatially uniform but temporally varying forcing term, adjusted such that a constant flow\nobreakdash-rate is maintained along the $x$ direction in the channel region. The domain is periodic along $x$ and $z$ directions. No-slip and impermeability conditions are imposed at $y=-h$ while at $y=\delta$ free-slip and impermeability conditions are imposed. Heat transfer is incorporated by solving a convection-diffusion equation for a scalar temperature field,
 \begin{gather}\label{eq:Heat}
 \cfrac{\partial T}{\partial t} + (\boldsymbol{u}\cdot\nabla){T} = \alpha{\nabla}^2{T},
 \end{gather}
 where $T$, $\alpha={\rho}/(c_p\,\kappa)$, $c_p$, and $\kappa$ are the temperature, thermal diffusivity, specific heat capacity, and thermal conductivity, respectively. Fixed-temperature conditions $T\rvert_{y=-h} = 1$ and $T\rvert_{y=\delta} = 0$ are imposed at the wall-normal domain boundaries so that heat is transferred from the bottom of the porous wall to the top of the channel region with the flow acting to cool the porous wall.
 
 Equations \eqref{eq:NS}-\eqref{eq:Heat} are solved using the open-source solver CaNS \citep{COSTA_2018,COSTA_2021}, which discretizes them using second-order central finite differences on structured Cartesian grids using a staggered arrangement. The grids used are uniform along $x$ and $z$ but non-uniform along $y$, such that it is stretched in the channel region of the domain using a hyperbolic tangent function but is uniform across the porous region. The porous walls are numerically represented using the immersed-boundary method (IBM) of \cite{breugem_boersma_2005} for grid-conforming Cartesian geometries. It involves modifying the discretized advective and diffusive fluxes of cells identified as solids such that the no-slip condition becomes exactly imposed, giving second-order spatial accuracy. A detailed description of the IBM can be found in \cite{Paravento}. The grid resolution used is the same as that in \cite{Khorasani_Luhar_Bagheri_2024}, with the number of grid points being $(N_x,N_y,N_z)=(1620, 324, 810)$. For further descriptions and assessments of resolution the reader is referred to that work. CaNS integrates the equations in time using a fractional-step pressure\nobreakdash-correction algorithm. To remove the diffusive constraint on the timestep used for temporal integration, the diffusive terms can be treated implicitly by solving a Helmholtz equation for each of the velocity components. However, to benefit from this in our conjugate heat transfer simulations that have a spatially varying thermal diffusivity, we use the method of \cite{Dodd_2014} to split the diffusive term into a variable term and a constant term, with the former treated implicitly and the latter explicitly. This prevents the greater thermal diffusivity in the solid phase from imposing a strict restriction on the timestep.    

 The relevant non-dimensional quantities for characterising the flow are the bulk Reynolds number, $\Rey_b = U_b\delta/\nu$, and the Prandtl number, $Pr = \nu/\alpha$. The simulations were conducted at a targeted $\Rey_{\tau} = u_{\tau}\delta/\nu = 360$, defined based on the channel region of the domain ($0 \le y \le \delta$) and achieved by adjusting $\Rey_b$. Here, $u_{\tau}=\sqrt{{\tau_{w}}/{\rho}}$ is the friction velocity at $y=0$ (the porous wall surface).
 For the thermal property of the working fluid, $Pr = 0.71$ was used, which corresponds to air. A thermal diffusivity of $\alpha_s = 4.4\alpha_f$ was set for the solid phase to make it analogous to aluminum.
 Variables are decomposed using the Reynolds decomposition \citep{Reynolds_1895}
 \begin{gather}\label{Reynolds_decomposition}
 {\phi}(x,y,z,t) = {\Phi}(y) + \phi^{\prime}(x,y,z,t),
 \end{gather}
 where $\phi$ is the instantaneous value, $\Phi$ is the time\nobreakdash- and plane\nobreakdash-averaged mean value, and ${\phi^{\prime}}$ is the full time-and-space varying fluctuation. The latter can be further decomposed into \citep{reynolds_hussain_1972}
 \begin{gather}\label{Triple_decomposition}
 \phi^{\prime}(x,y,z,t) = \Tilde{\phi}(x,y,z) + \phi^{\prime\prime}(x,y,z,t),
 \end{gather}
 where $\Tilde{\phi}(x,y,z)$ is the time-invariant but spatially\nobreakdash-varying dispersive component induced by the porous wall and $\phi^{\prime\prime}(x,y,z,t)$ the remaining time\nobreakdash-and-space varying fluctuation due to turbulence.
 
 Plane-averaged flow statistics are reported as fluid-averaged and solid-averaged quantities,
 \begin{gather}\label{eq:phase_averaging}
 {\Phi}(y)_i = \cfrac{1}{A_i}\iint_{A_i}{\overline{{\phi}(x,y,z)}}\,dx\,dz,
 \end{gather}
 where ${\overline{{\phi}(x,y,z)}}$ is the time-averaged variable and $i = f, s$ indicates over which phase the plane averaging has been done ($f$ for fluid and $s$ for solid). Since the flow velocities are zero in the solid-occupied regions, all velocity-related statistics are fluid-phase averages (also called `intrinsic' averages as in the work of \citealt{nikora2007double}), and so the subscript $f$ is dropped for them. The temperature field however is present throughout both phases and the distinction between them is made clear by using the appropriate subscript. The `$+$' superscript indicates viscous\nobreakdash-scaled quantities using $u_{\tau}$ and $\nu$.
 The temperature field, $T$ , is normalized by the difference between the lower and upper boundary temperatures, giving the non-dimensional temperature
 \begin{gather}
 \theta = \cfrac{T}{(T\rvert_{y=-h} - T\rvert_{y=\delta})}
 \end{gather}
 % Since the heat flux at the surface of the wall interfacing the bulk flow changes across the different cases, thermal quantities are not normalized using this heat flux as is commonly done in smooth-channel simulations (using the so-called friction temperature $\T_{\tau} = {q_w}/({\rho\,c_p\,u_{\tau}})$). In this way, the differences in thermal behavior between the different cases can be interpreted more clearly.

 \begin{table}
   \centering
   \begin{tabular}[t]{ m{1.2cm}m{0.9cm}m{1.0cm}m{0.9cm}m{0.8cm}m{0.8cm}m{0.8cm}m{0.8cm}m{0.8cm}m{0.8cm}m{0.2cm} }
    \hspace{-0.25mm}Case & \hspace{-3.2mm}Symbol & \hspace{-0.1mm}$Re_{\tau}$ & \;\;$\varphi$ & ${s_{x}}^+$ & ${s_{y}}^+$ & ${s_{z}}^+$ & \hspace{-2mm}$\sqrt{K_{x\phantom{y}}^+}$ & \hspace{-1mm}$\sqrt{K_{y\phantom{y}}^+}$ & \hspace{-1mm}$\sqrt{K_{z\phantom{y}}^+}$ & \hspace*{-1mm}$\sqrt{\cfrac{K_{x\phantom{y}}^+}{K_{y\phantom{y}}^+}}$\\
    &&&&&&&&&&\\
    ${KP1}$                & \Large$\textcolor{red-dark}{\bullet}$                            & $364$ & $0.75$ & $36.0$ & $108.0$& $28.0$ & $3.68$ & $3.96$ & $5.45$ & $0.93$\\
    ${KP1}^{\prime}$       & \hspace{0.1mm}\normalsize$\textcolor{red-dark}{\blacksquare}$    & $361$ & $0.86$ & $56.0$ & $108.0$& $42.0$ & $7.96$ & $8.20$ & $10.09$ & $0.97$\\
    ${KP1}^{\prime\prime}$ & \hspace{-0.25mm}\Large$\textcolor{red-dark}{\star}$              & $364$ & $0.95$ & $108.0$ & $108.0$& $84.0$ & $18.97$ & $18.93$ & $19.84$ & $1.00$\\
    ${KP2}$                & \Large$\textcolor{orange-dark}{\bullet}$                         & $361$ & $0.68$ & $36.0$ & $54.0$ & $28.0$ & $3.31$ & $3.17$ & $4.64$ & $1.04$\\
    ${KP2}^{\prime}$       & \hspace{0.1mm}\normalsize$\textcolor{orange-dark}{\blacksquare}$ & $362$ & $0.81$ & $56.0$ & $54.0$ & $42.0$ & $6.43$ & $6.46$ & $7.43$ & $1.00$\\
    ${KP2}^{\prime\prime}$ & \hspace{-0.25mm}\Large$\textcolor{orange-dark}{\star}$           & $363$ & $0.92$ & $108.0$ & $54.0$ & $84.0$ & $12.66$ & $15.32$ & $12.46$ & $0.83$\\
    ${KP3}$                & \Large$\textcolor{yellow-dark}{\bullet}$                         & $364$ & $0.61$ & $36.0$ & $36.0$ & $28.0$ & $2.79$ & $2.79$ & $3.57$ & $1.00$\\
    ${KP3}^{\prime}$       & \hspace{0.1mm}\normalsize$\textcolor{yellow-dark}{\blacksquare}$ & $365$ & $0.76$ & $56.0$ & $36.0$ & $42.0$ & $4.80$ & $5.77$ & $5.09$ & $0.83$\\
    ${KP3}^{\prime\prime}$ & \hspace{-0.25mm}\Large$\textcolor{yellow-dark}{\star}$           & $362$ & $0.86$ & $108.0$ & $36.0$ & $84.0$ & $7.71$ & $12.69$ & $6.99$ & $0.61$\\
    {\emph{TP\phantom{1}}} & \Large$\textcolor{blue-light}{\bullet}$                          & $357$ & $0.50$ & $28.0$ & $27.0$ & $28.0$ & $2.11$ & $2.16$ & $2.11$ & $0.98$\\
    {\emph{SP\phantom{1}}} & \Large$\textcolor{blue-dark}{\bullet}$                           & $359$ & $0.43$ & $21.0$ & $54.0$ & $21.0$ & $1.59$ & $1.11$ & $1.59$ & $1.43$\\
    {\emph{ZP\phantom{1}}} & \Large$\textcolor{black}{\bullet}$                               & $363$ & $0.0$ & $0.0$ & $0.0$ & $0.0$ & $0.0$ & $0.0$ & $0.0$ & $-$\\
  \end{tabular}
  \vspace*{-1mm}
  \captionsetup{width=0.98\textwidth,justification=justified}
  \captionof{table}{Parameters of the DNS cases. The porosity for each wall is given by $\varphi$. The pore pitch-lengths are ${s_{x}}^+$, ${s_{y}}^+$ and ${s_{z}}^+$, while $\sqrt{K_{x\phantom{y}}^+}$, $\sqrt{K_{y\phantom{y}}^+}$ and $\sqrt{K_{z\phantom{y}}^+}$ are the effective permeabilities which are equivalent to the permeability Reynolds number, $\Rey_{K}$, used throughout the literature. The ratio of streamwise to wall-normal permeability serves as the measure of a wall's anisotropy.}
  \label{tab:DNS}
 \end{table}

 A summery of all the DNS cases and their relevant parameters are given in \cref{tab:DNS}. The \emph{zero-permeability} case, $ZP$, is the baseline smooth-wall case where the wall region is occupied by a solid impenetrable block. The \emph{smooth-wall-like permeable} case, $SP$, has low permeability and retains the smooth-wall-like structure of near-wall turbulence. The \emph{transitionally permeable} case, $TP$, shows signs of departing the smooth-wall-like regime. The \emph{Kelvin-Helmholtz permeable} cases, $KP\langle\rangle$, fall into the K-H-like regime of wall turbulence. 
 The walls of cases $SP$, $TP$, $KP3$, $KP2$, and $KP1$ respectively correspond to cases $LP1$, $MP$, $HP3$, $HP2$, and $HP1$ of \cite{khorasani_2023}, where it was established to which turbulence regime each case belonged to. The diagonal permeabilities, $\sqrt{K_{i}^+}$, were obtained using Darcy's law, $\nabla{p}=-\nu\,{\mathsfbi{K}}^{-1}\cdot\boldsymbol{u}$, by conducting Stokes simulations in representative-elementary-volumes of each porous wall. The $KP\langle\rangle^{\prime}$ and $KP\langle\rangle^{\prime\prime}$ cases retain the same wall-normal pitches, $s_y$, as the $KP\langle\rangle$ case but have successively larger wall-parallel pitches, $s_x$ and $s_z$. The purpose of this successive increase in $s_x$ and $s_z$, in addition to increasing permeability, was to cause the texture-coherent scales to overlap with those of the ambient turbulence as discussed in \cref{sec:porous_wall_flow}.

 \begin{figure}
    \begin{center}
    \begin{subfigure}{.665\textwidth}
        {\captionsetup{labelfont=it,textfont=normalfont,singlelinecheck=false,justification=raggedright,labelformat=parens}\caption{}\label{fig:bulk_temperature_history}}%
        \includegraphics[width=1\linewidth]{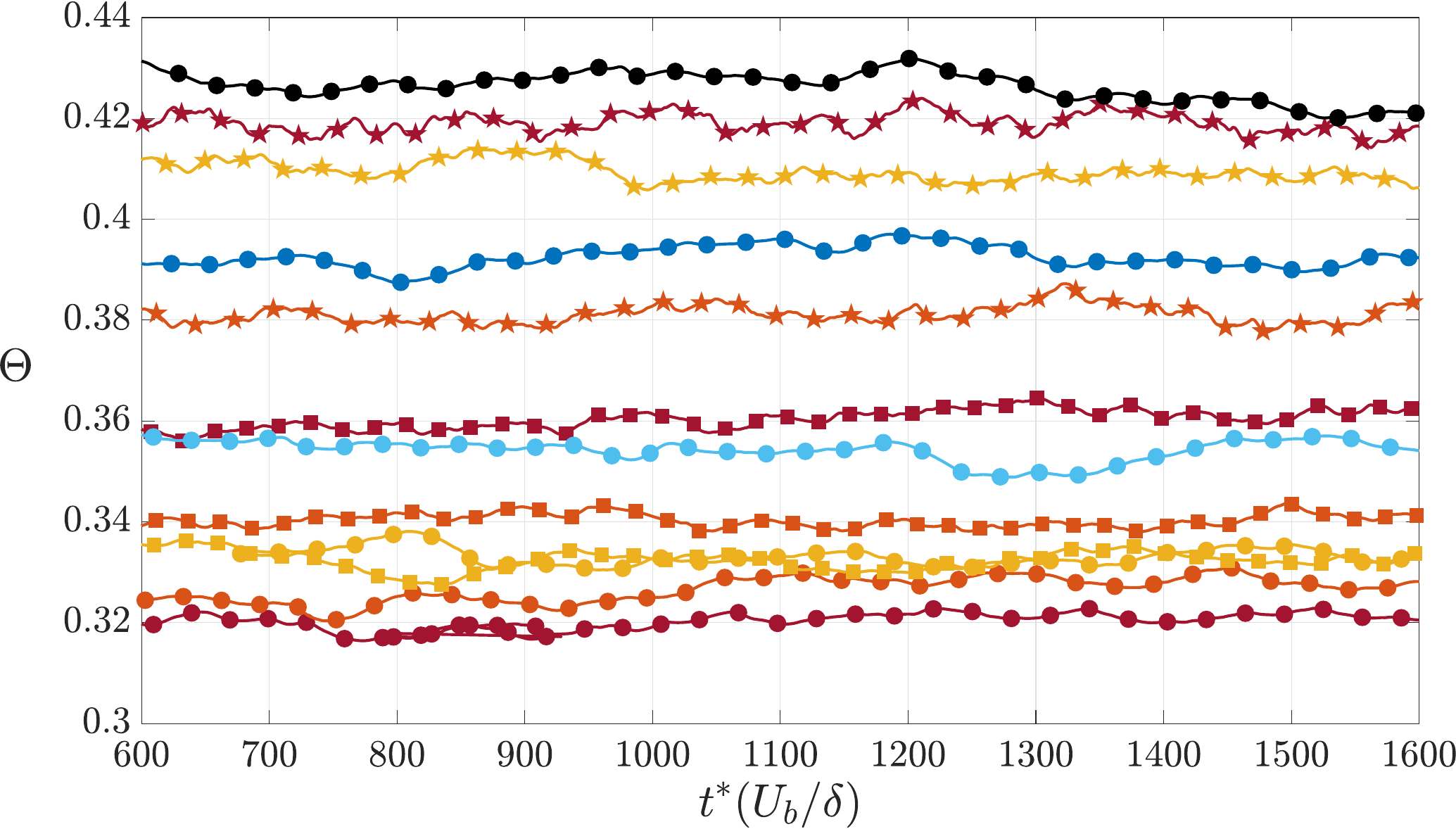}
    \end{subfigure}\\
    \begin{subfigure}{.68\textwidth}
        {\captionsetup{labelfont=it,textfont=normalfont,singlelinecheck=false,justification=raggedright,labelformat=parens}\caption{}\label{fig:heat_flux_history}}%
        \includegraphics[width=1\linewidth]{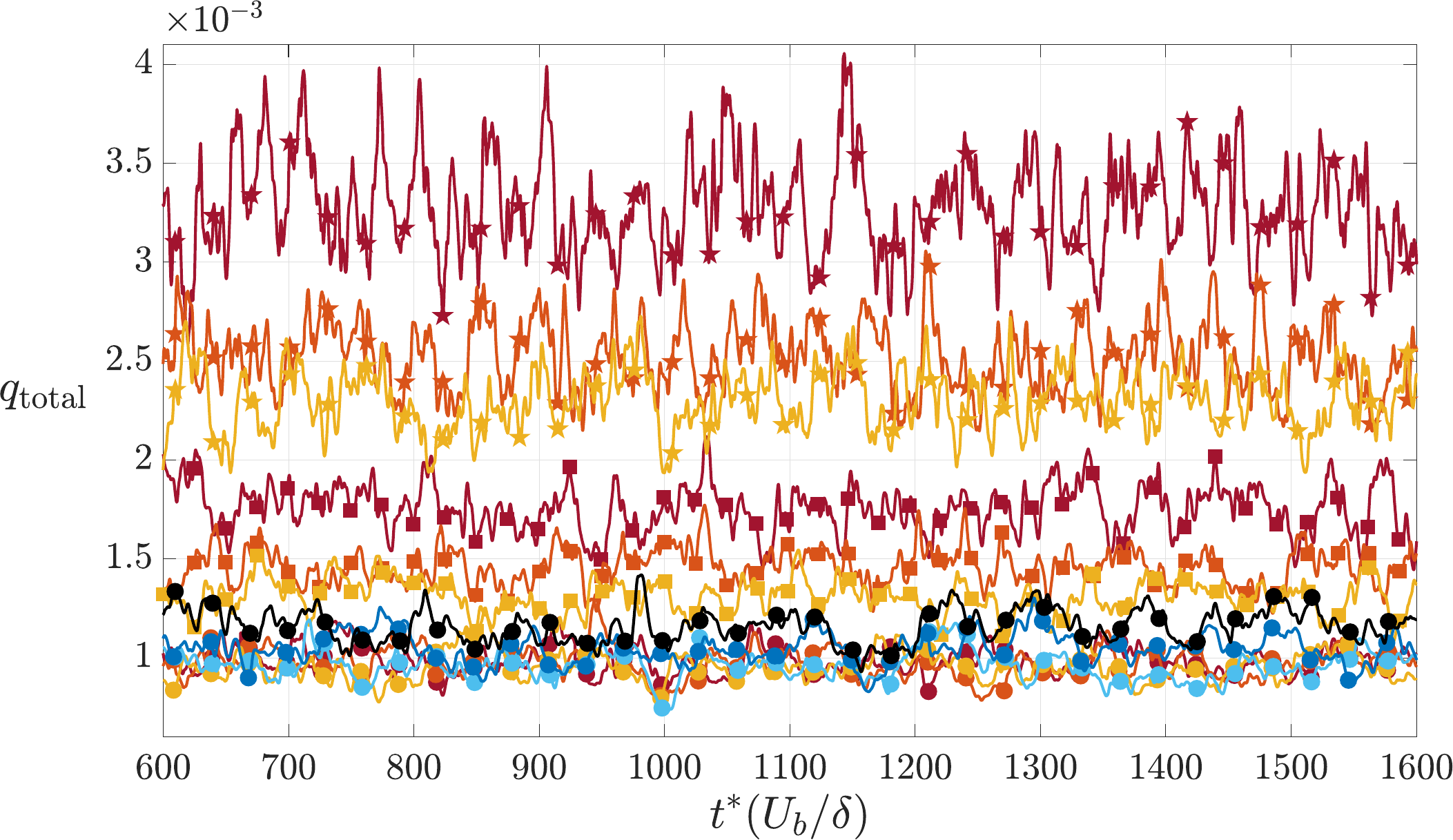}
    \end{subfigure}
    \captionsetup{width=1\textwidth,justification=justified}
    \vspace{-3pt}
    \caption{Time history of (\emph{a}) the bulk temperature and (\emph{b}) the total heat flux at the top boundary.}
    \label{fig:temperature_history}
    \end{center}
    \vspace{-3pt}
 \end{figure}

 For each case of \cref{tab:DNS}, a precursor simulation was first ran for $2000 \delta/U_b$ so the flow and temperature fields became fully developed and achieved statistical steadiness. This then served as the initial condition for the production run where data was acquired, with flow statistics collected over an additional $2000 \delta/U_b$. Statistical steadiness is evidenced in \cref{fig:temperature_history}, which shows a portion of the time history of the bulk temperature and total heat flux. 

\section{Results}\label{results}

\subsection{Flow differences over the different porous walls}\label{results:velocity}

 \begin{figure}
    \begin{center}
    \begin{subfigure}{.47\textwidth}
     {\captionsetup{labelfont=it,textfont=normalfont,singlelinecheck=false,justification=raggedright,labelformat=parens}\caption{}\label{fig:mean_velocity_channel}}%
     \includegraphics[width=1\linewidth]{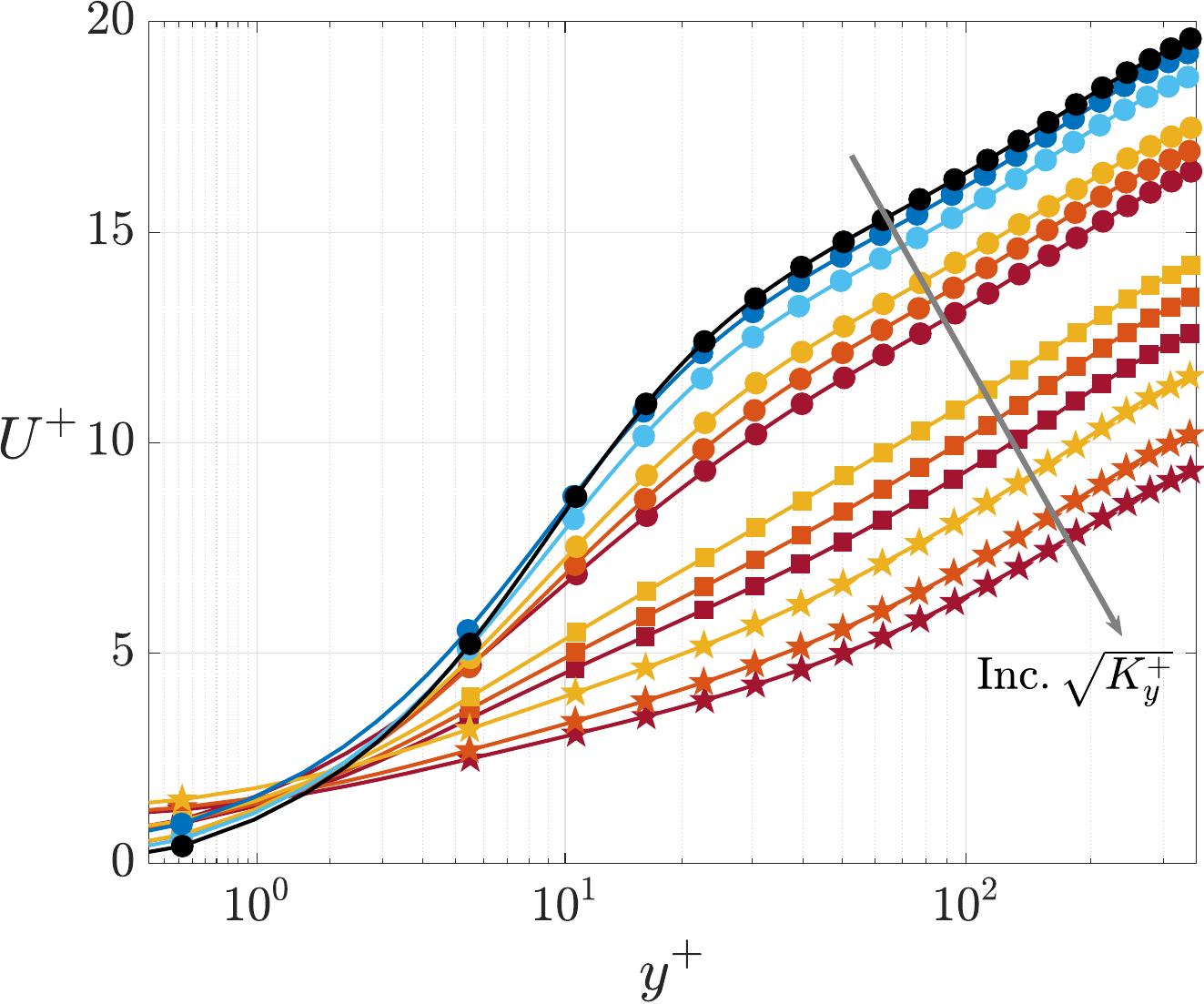}
    \end{subfigure}
    \hspace{1pt}
    \begin{subfigure}{.475\textwidth}
     {\captionsetup{labelfont=it,textfont=normalfont,singlelinecheck=false,justification=raggedright,labelformat=parens}\caption{}\label{fig:Reynolds_stress_channel}}%
     \includegraphics[width=1\linewidth]{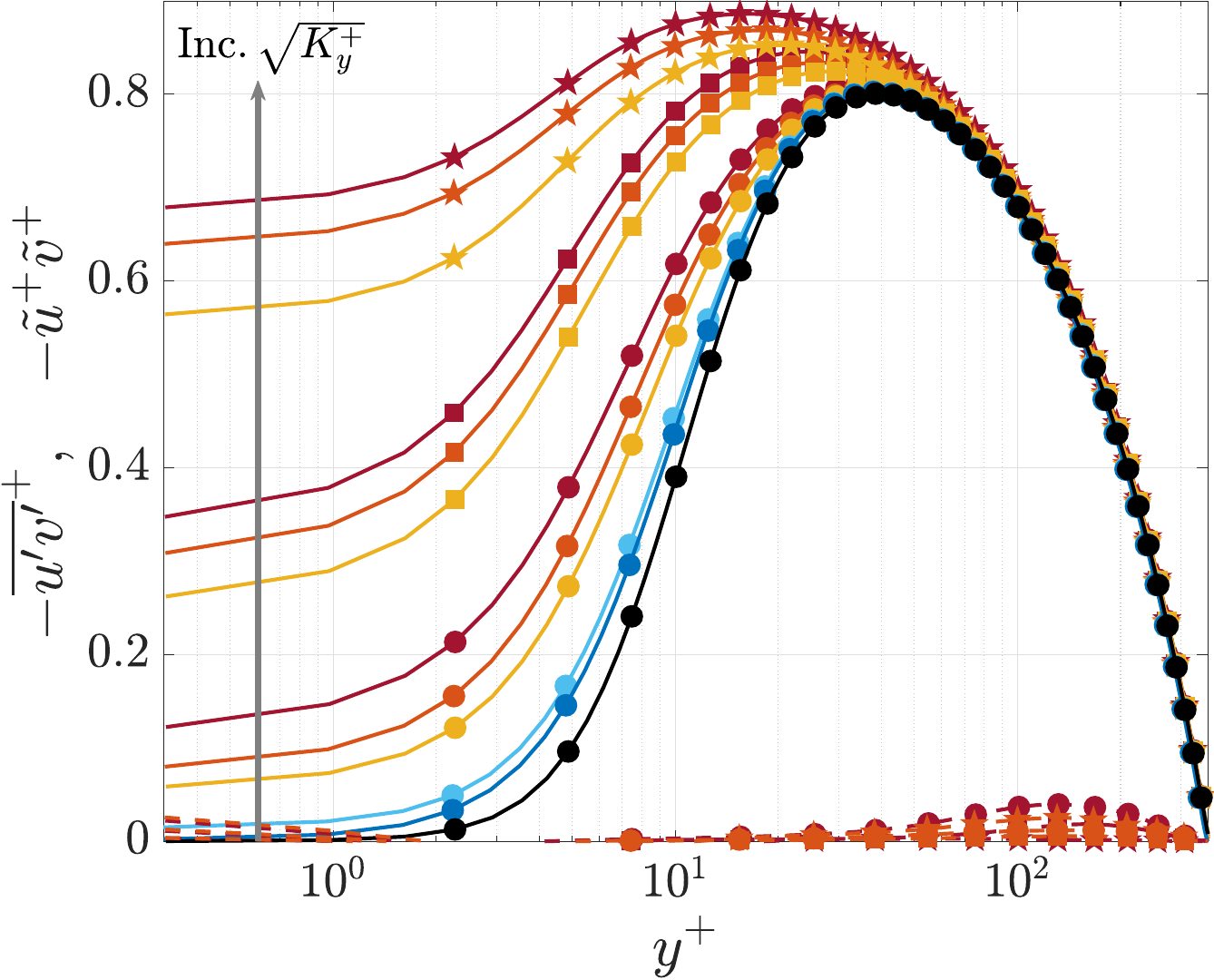}
    \end{subfigure}\\
    \begin{subfigure}{.475\textwidth}
     {\captionsetup{labelfont=it,textfont=normalfont,singlelinecheck=false,justification=raggedright,labelformat=parens}\caption{}
     \label{fig:rms_channel}}%
     \includegraphics[width=1\linewidth]{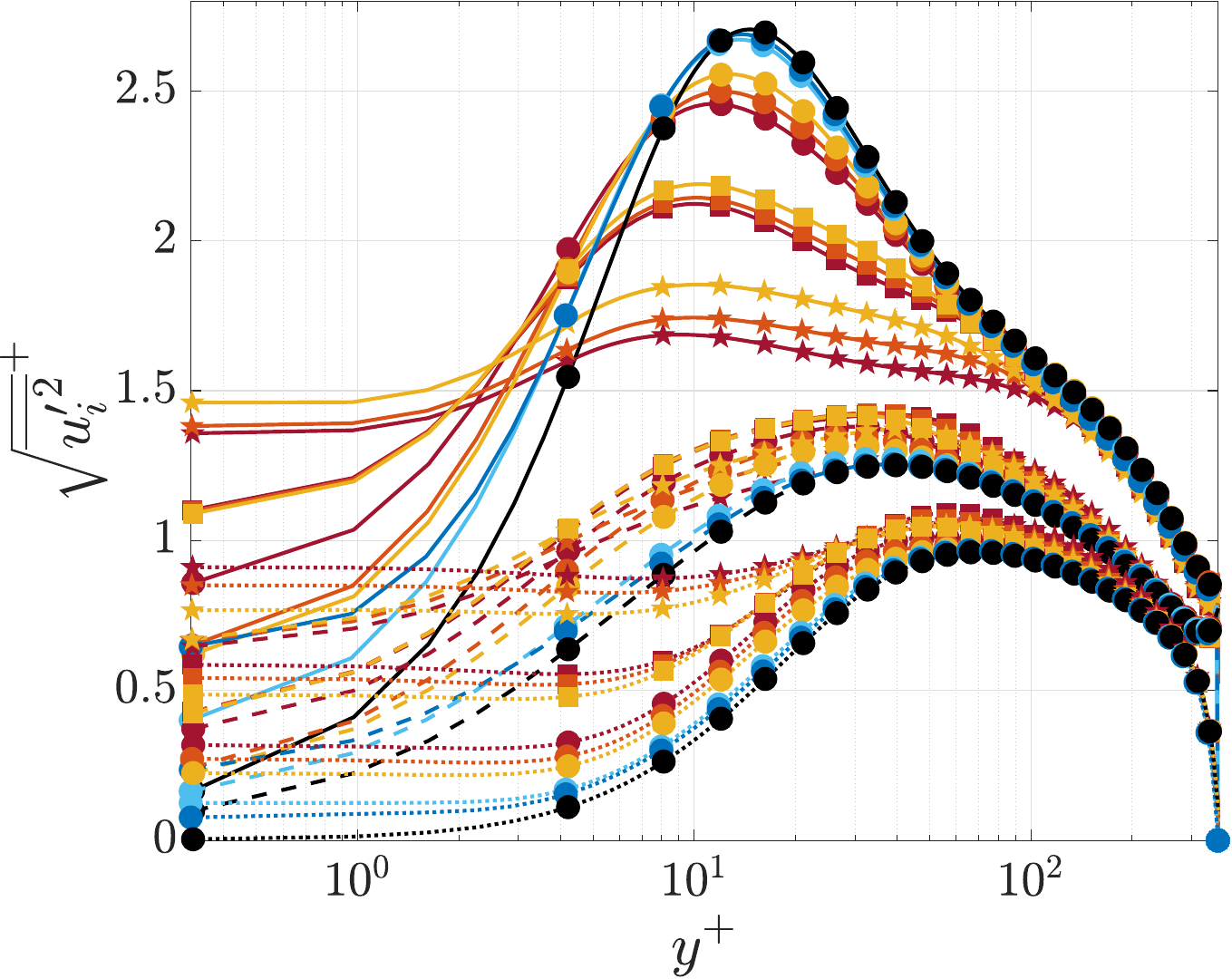}
    \end{subfigure}
    % \hspace{1pt}
    \captionsetup{width=1\textwidth,justification=justified}
    \vspace{-3pt}
    \caption{Profiles of the (\emph{a}) mean velocity, (\emph{b}) Reynolds shear stress, (\emph{c}) root-mean-square velocity fluctuations. The direction of the gray arrow in (\emph{a}) and (\emph{b}) indicates increasing wall\nobreakdash-normal permeability. In (\emph{b}); \textemdash, total Reynolds stress; $--$, dispersive Reynolds stress. In (\emph{c}); \textemdash, total streamwise fluctuations; $--$, total spanwise fluctuations; $\cdot\!\cdot\!\cdot$, total wall-normal fluctuations.}
    \label{fig:velocity_statistics_channel}
    \end{center}
    \vspace{-3pt}
 \end{figure}

 Before analyzing the heat transfer, the velocity fields for the cases of \cref{tab:DNS} in the channel region ($y>0$) are briefly examined to establish their features. Since heat transfer is passive, structural changes in the velocity field will largely determine any differences that emerge in the temperature fields for the different walls.

 \Cref{fig:mean_velocity_channel} shows the plane- and time-averaged mean velocity profiles. It is clear that as the permeability grows larger, the mean velocity deficit relative to the smooth-wall case, $ZP$, also becomes larger. The low-permeability case of $SP$ does not deviate too much from $ZP$, whereas the large permeability cases of $KP\langle\rangle^{\prime\prime}$ differ considerably in terms of their overall mean momentum. The differences observed in \cref{fig:mean_velocity_channel} for the mean velocities are reflected in \cref{fig:Reynolds_stress_channel} for the total Reynolds shear stress profiles, $-\overline{{u^{\prime}}{v^{\prime}}}^+$. What immediately stands out is how the $KP\langle\rangle$, $KP\langle\rangle^{\prime}$, and $KP\langle\rangle^{\prime\prime}$ profiles fall into separate groups in near-wall region. The strength of the Reynolds shear stress progressively grows in this region as the wall becomes more permeable. The profile shape close to the surface also becomes flatter, with the stress not undergoing too much dampening relative to its peak value farther away from the surface. These enhanced levels of stress producing turbulence activity at the surface cause the reduction of the mean momentum depicted in \cref{fig:mean_velocity_channel}. Differences in the shapes of the $-\overline{{u^{\prime}}{v^{\prime}}}^+$ profiles are indicative of structural changes in turbulence, which can also be inferred from the root-mean-square (r.m.s.) velocity fluctuations in \cref{fig:rms_channel}. There, what is immediately apparent is how the near-wall peak for the $u$ velocity becomes diminished as the wall becomes more permeable. The fluctuations for $w$ and $v$ however grow more intense, such that the overall intensity for the three velocity components draw close to one another. At the surface ($y=0$), more permeability permits stronger turbulent activity to penetrate into the wall, which registers as a rise in the r.m.s. values for all three velocity components. While not shown here, it can readily be deduced from figures \hyperref[fig:Reynolds_stress_channel]{\ref*{fig:Reynolds_stress_channel}} and \hyperref[fig:rms_channel]{\ref*{fig:rms_channel}} that in the K-H-like $KP$ cases, the turbulent stresses penetrate into the wall. This penetration is quite deep for the highly permeable $KP\langle\rangle^{\prime\prime}$ cases, as will become clear when later examining the turbulent heat flux.
 
 Figure \hyperref[fig:Reynolds_stress_channel]{\ref*{fig:Reynolds_stress_channel}} also shows the dispersive component of the Reynolds shear stress for the highly permeable cases of $KP1$, $KP1\langle\rangle^{\prime}$, $KP1\langle\rangle^{\prime\prime}$, $KP2$, $KP2\langle\rangle^{\prime}$, and $KP2\langle\rangle^{\prime\prime}$. It is evident that even for the most permeable walls, the dispersive contribution to the total Reynolds shear stress throughout the bulk flow region is negligible. Although, the dispersive stress does appear to grow around $y+\approx150$ with growing permeability (and hence stronger K-H-like rollers). Additionally, the dispersive Reynolds shear stress is negative close to the porous wall, causing it to have a diminishing effect on the total Reynolds shear stress and not an enhancing one. This negative shear contribution is attributable to the K-H-like rollers, which can induce negative shear \citep{endrikat_2021,rouhi_endrikat_modesti_sandberg_oda_tanimoto_hutchins_chung_2022}. Also notable is that, unlike in the case of the canopy structures studied by \citet{Chen_2023}, strong levels of dispersive stress do not occur at the very surface of the porous walls.

\begin{figure}
\vspace*{-20mm}
 \begin{center}
    \hspace*{-13mm}
    \begin{subfigure}[tbp]{.40\textwidth}
        {\captionsetup{position=bottom, labelfont=it, textfont=it, singlelinecheck=false, justification=raggedright, labelformat=parens}
        \caption{KP1}\label{fig:u_velocity_contour_y_0:KP1}}
        \vspace*{-0.8mm}
        \includegraphics[width=1\linewidth]{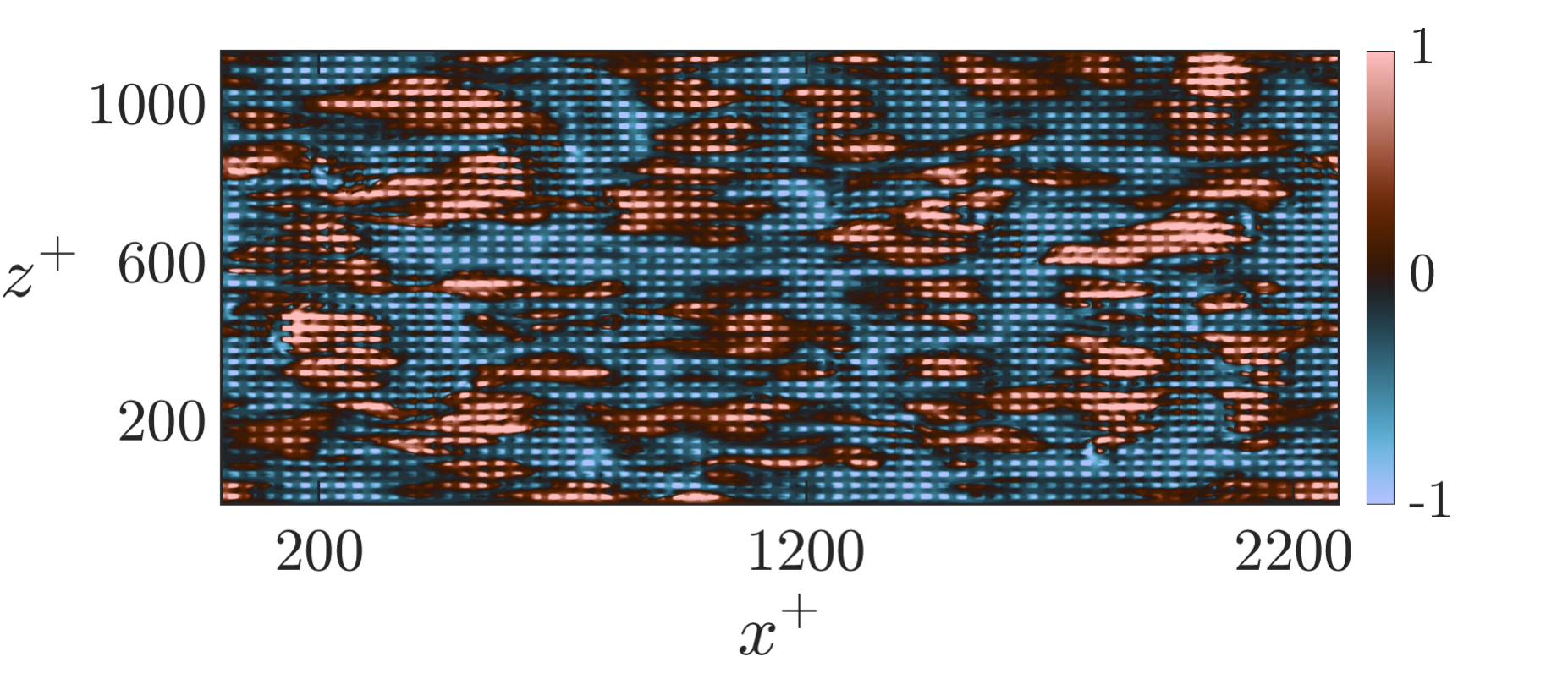}
    \end{subfigure}%
    \hspace*{-2mm}
    \begin{subfigure}[tbp]{.40\textwidth}
        {\captionsetup{position=bottom, labelfont=it,textfont=it,singlelinecheck=false,justification=raggedright,labelformat=parens}
        \caption{$KP1^\prime$}\label{fig:u_velocity_contour_y_0:KP1-1}}
        \vspace*{-0.8mm}
        \includegraphics[width=1\linewidth]{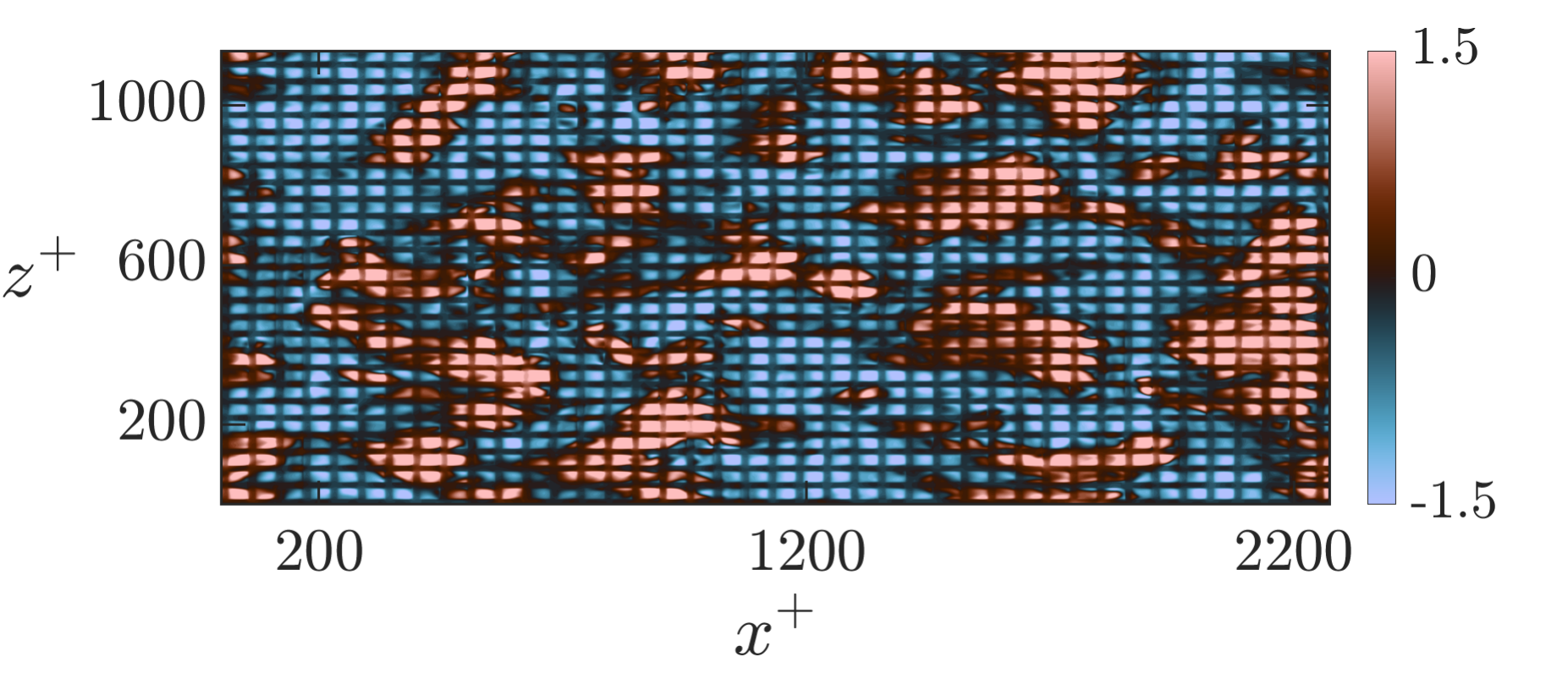}
    \end{subfigure}%
    \hspace*{-2mm}
    \begin{subfigure}[tbp]{.40\textwidth}
        {\captionsetup{position=bottom, labelfont=it,textfont=it,singlelinecheck=false,justification=raggedright,labelformat=parens}
        \caption{$KP1^{\prime\prime}$}\label{fig:u_velocity_contour_y_0:KP1-2}}
        \vspace*{-0.8mm}
        \includegraphics[width=1\linewidth]{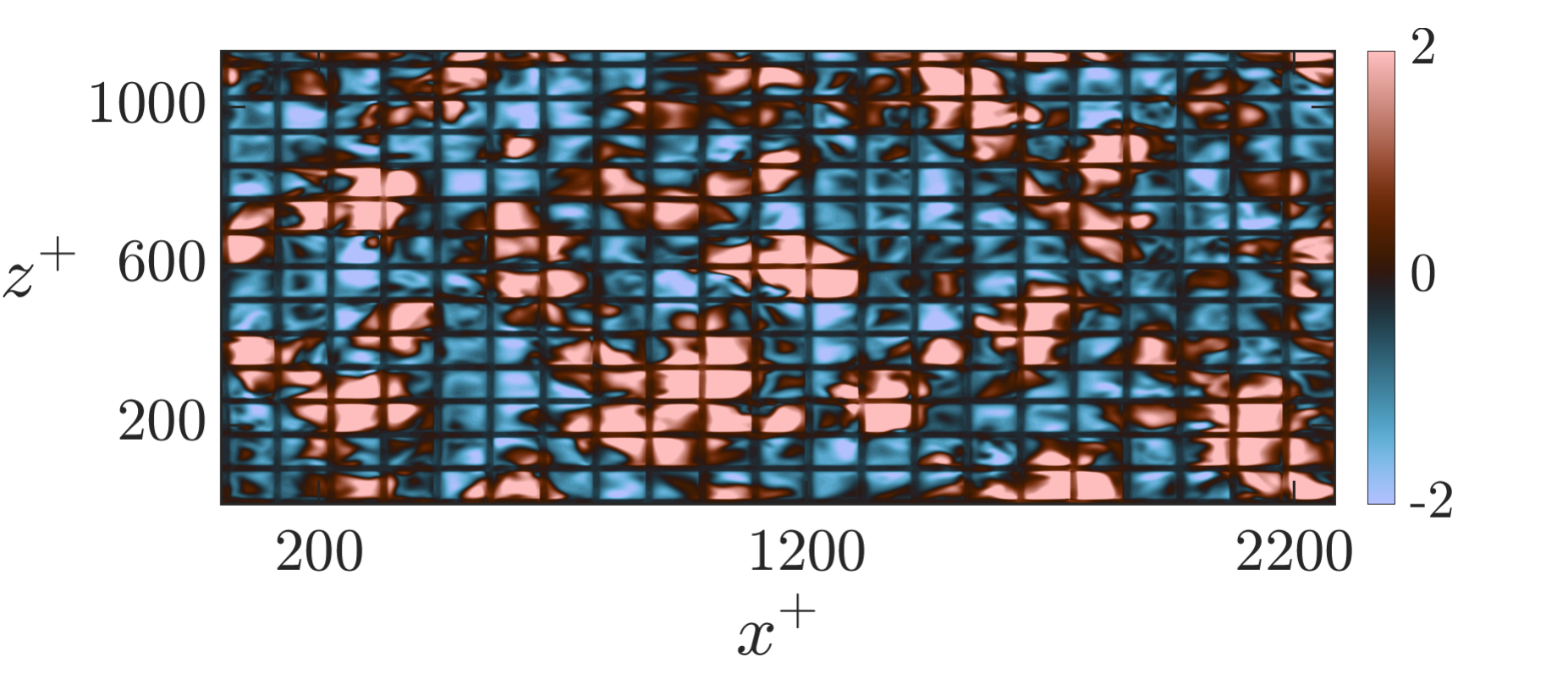}
    \end{subfigure}
    \hspace*{-13mm}
    \begin{subfigure}[tbp]{.40\textwidth}
        {\captionsetup{position=bottom, labelfont=it,textfont=it,singlelinecheck=false,justification=raggedright,labelformat=parens}
        \caption{KP2}\label{fig:u_velocity_contour_y_0:KP2}}
        \vspace*{-0.8mm}
        \includegraphics[width=1\linewidth]{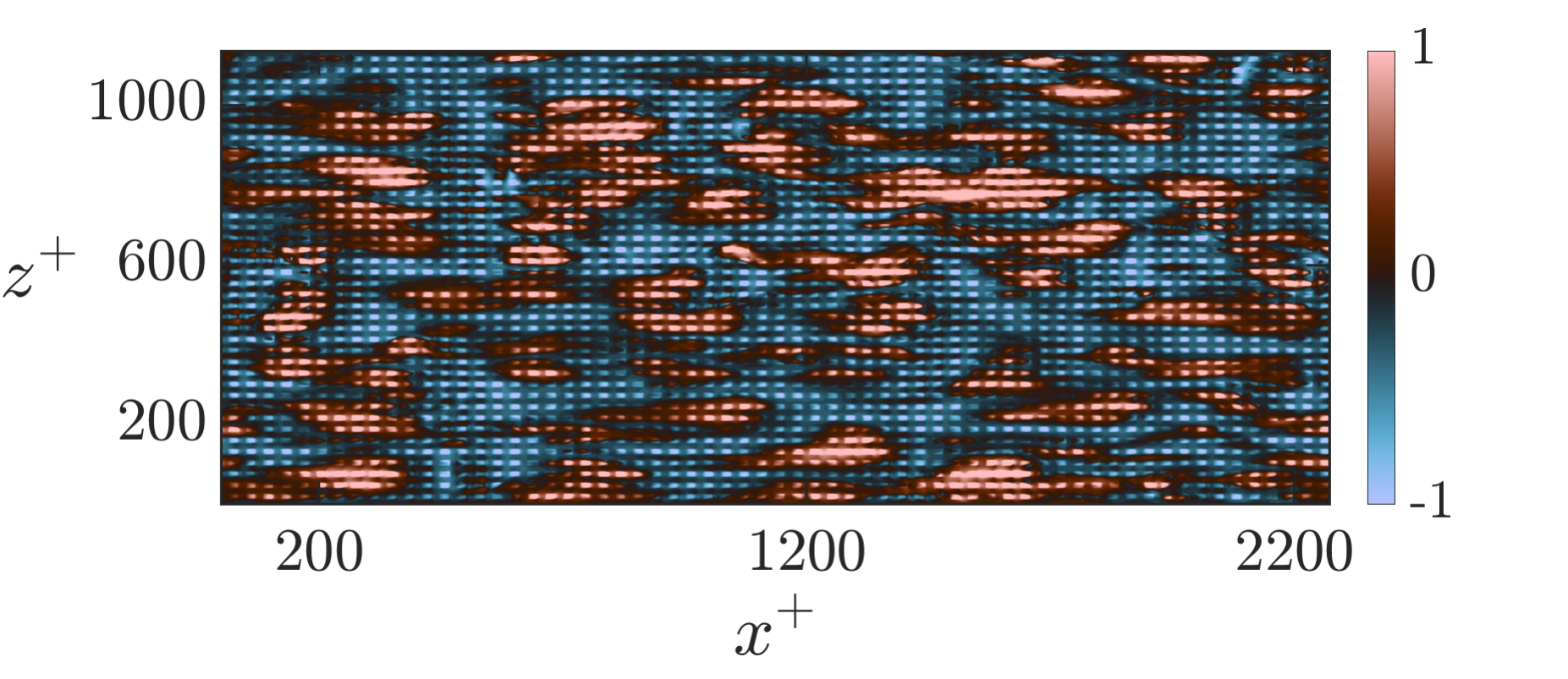}
    \end{subfigure}%
    \hspace*{-2mm}
    \begin{subfigure}[tbp]{.40\textwidth}
        {\captionsetup{position=bottom, labelfont=it,textfont=it,singlelinecheck=false,justification=raggedright,labelformat=parens}
        \caption{$KP2^\prime$}\label{fig:u_velocity_contour_y_0:KP2-1}}
        \vspace*{-0.8mm}
        \includegraphics[width=1\linewidth]{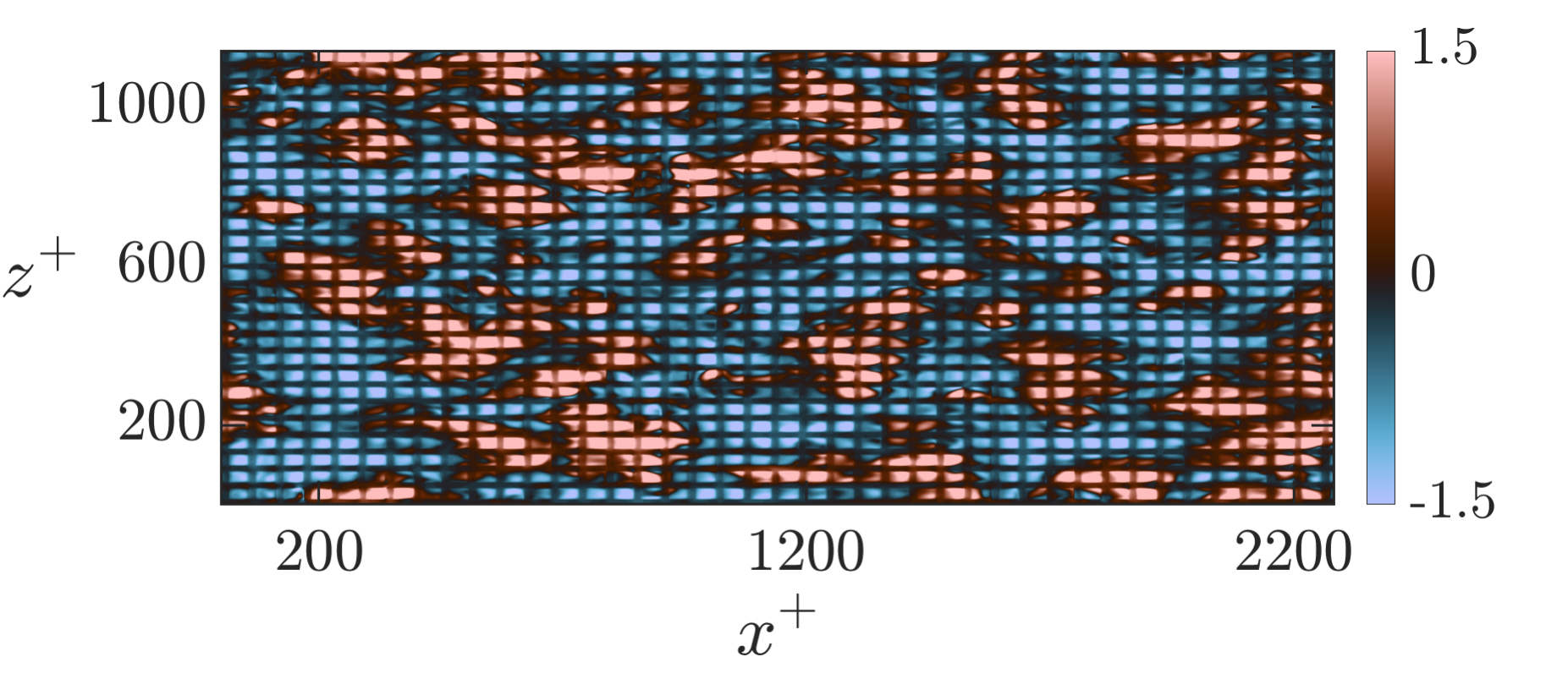}
    \end{subfigure}%
    \hspace*{-2mm}
    \begin{subfigure}[tbp]{.40\textwidth}
        {\captionsetup{position=bottom, labelfont=it,textfont=it,singlelinecheck=false,justification=raggedright,labelformat=parens}
        \caption{$KP2^{\prime\prime}$}\label{fig:u_velocity_contour_y_0:KP2-2}}
        \vspace*{-0.8mm}
        \includegraphics[width=1\linewidth]{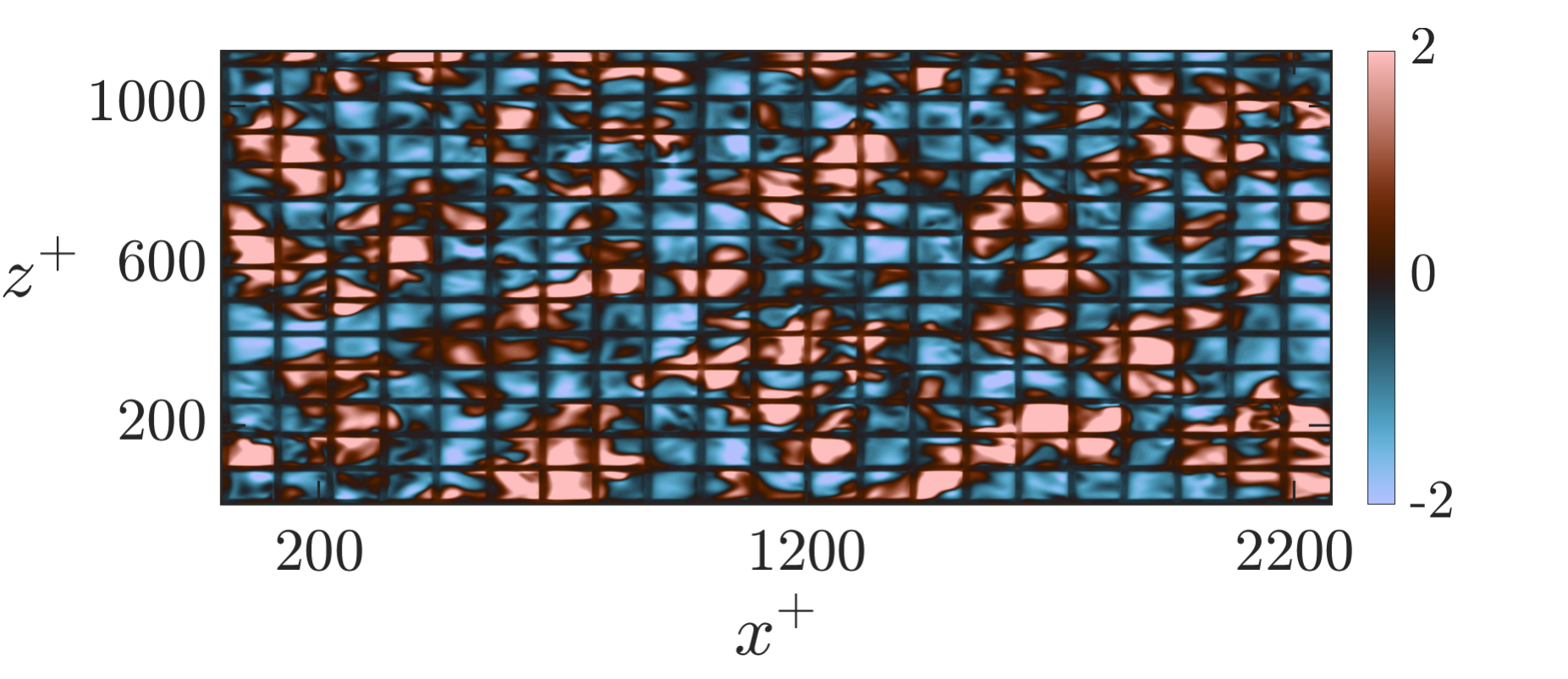}
    \end{subfigure}
    \hspace*{-13mm}
    \begin{subfigure}[tbp]{.40\textwidth}
        {\captionsetup{position=bottom, labelfont=it,textfont=it,singlelinecheck=false,justification=raggedright,labelformat=parens}
        \caption{KP3}\label{fig:u_velocity_contour_y_0:KP3}}
        \vspace*{-0.8mm}
        \includegraphics[width=1\linewidth]{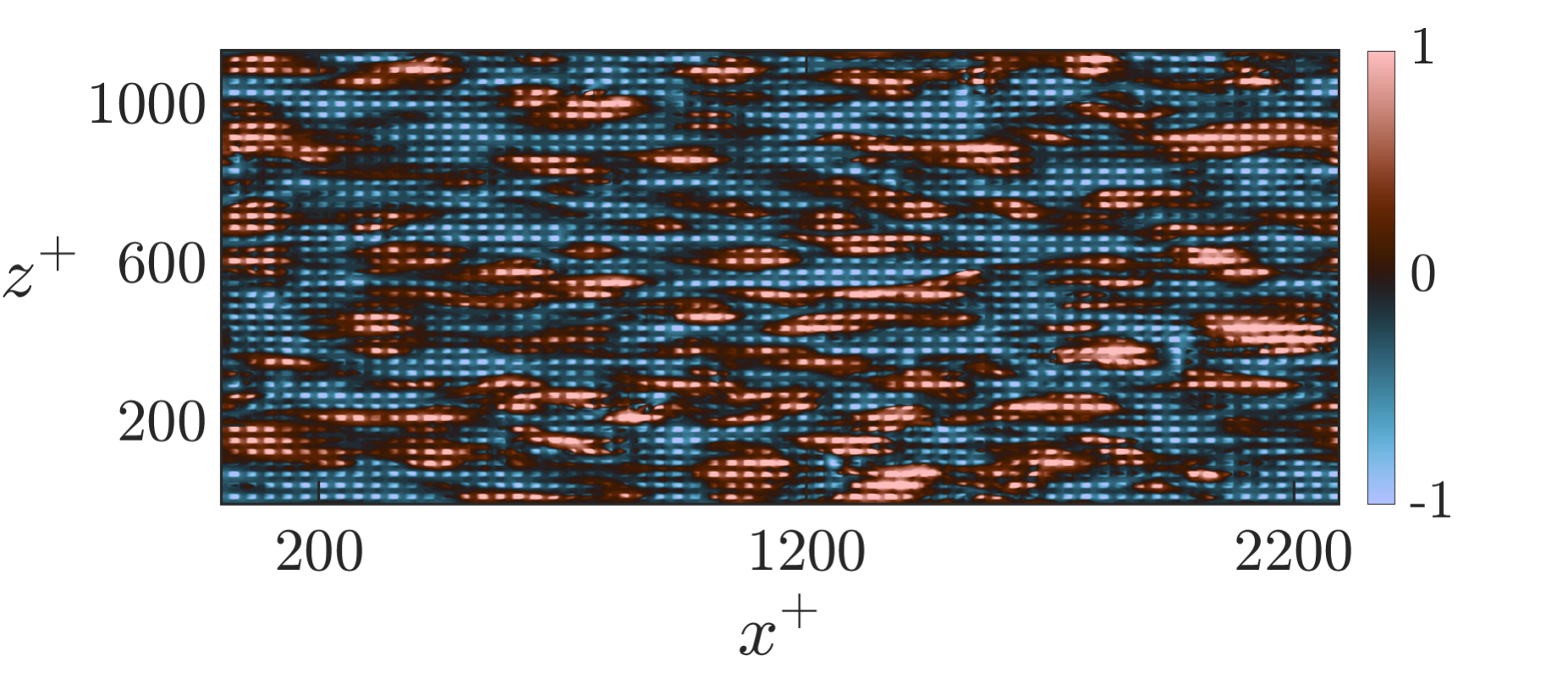}
    \end{subfigure}%
    \hspace*{-2mm}
    \begin{subfigure}[tbp]{.40\textwidth}
        {\captionsetup{position=bottom, labelfont=it,textfont=it,singlelinecheck=false,justification=raggedright,labelformat=parens}
        \caption{$KP3^\prime$}\label{fig:u_velocity_contour_y_0:KP3-1}}
        \vspace*{-0.8mm}
        \includegraphics[width=1\linewidth]{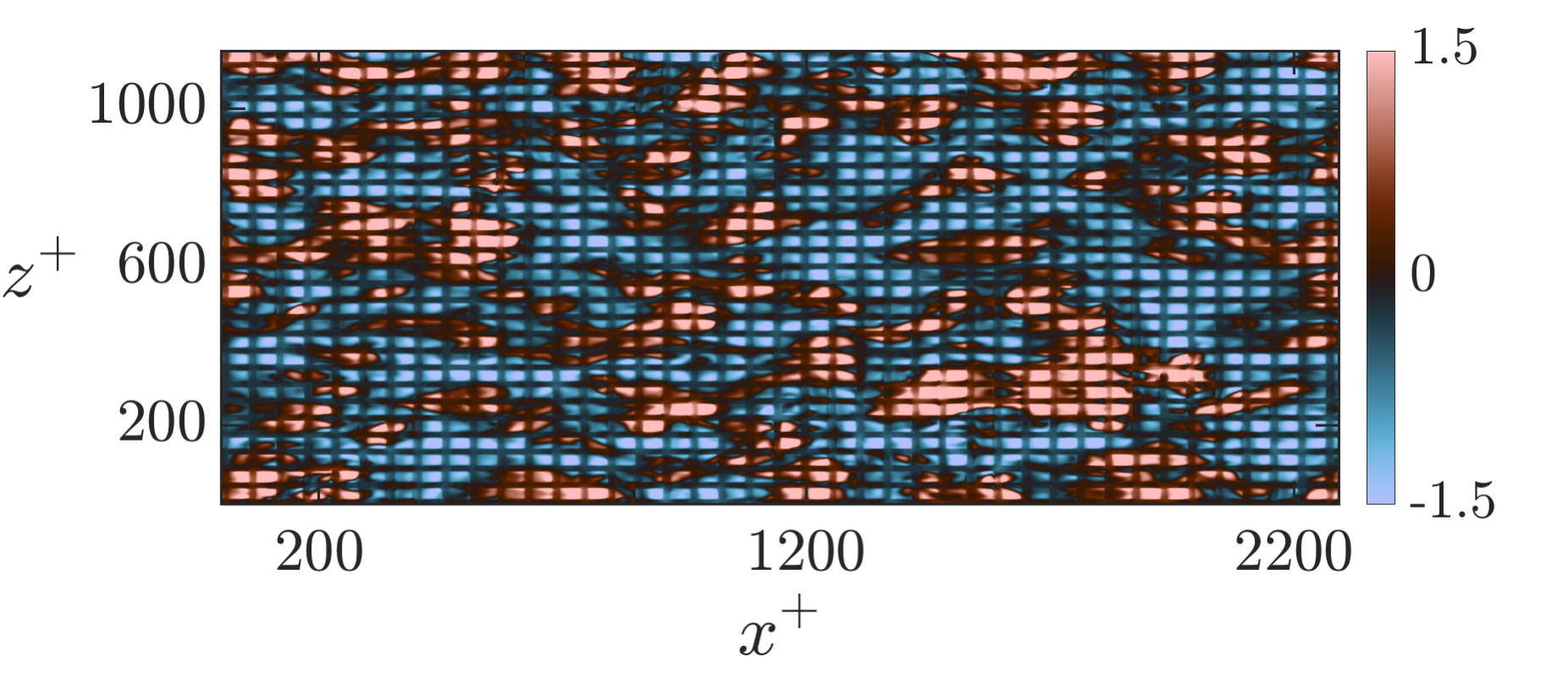}
    \end{subfigure}%
    \hspace*{-2mm}
    \begin{subfigure}[tbp]{.40\textwidth}
        {\captionsetup{position=bottom, labelfont=it,textfont=it,singlelinecheck=false,justification=raggedright,labelformat=parens}
        \caption{$KP3^{\prime\prime}$}\label{fig:u_velocity_contour_y_0:KP3-2}}
        \vspace*{-0.8mm}
        \includegraphics[width=1\linewidth]{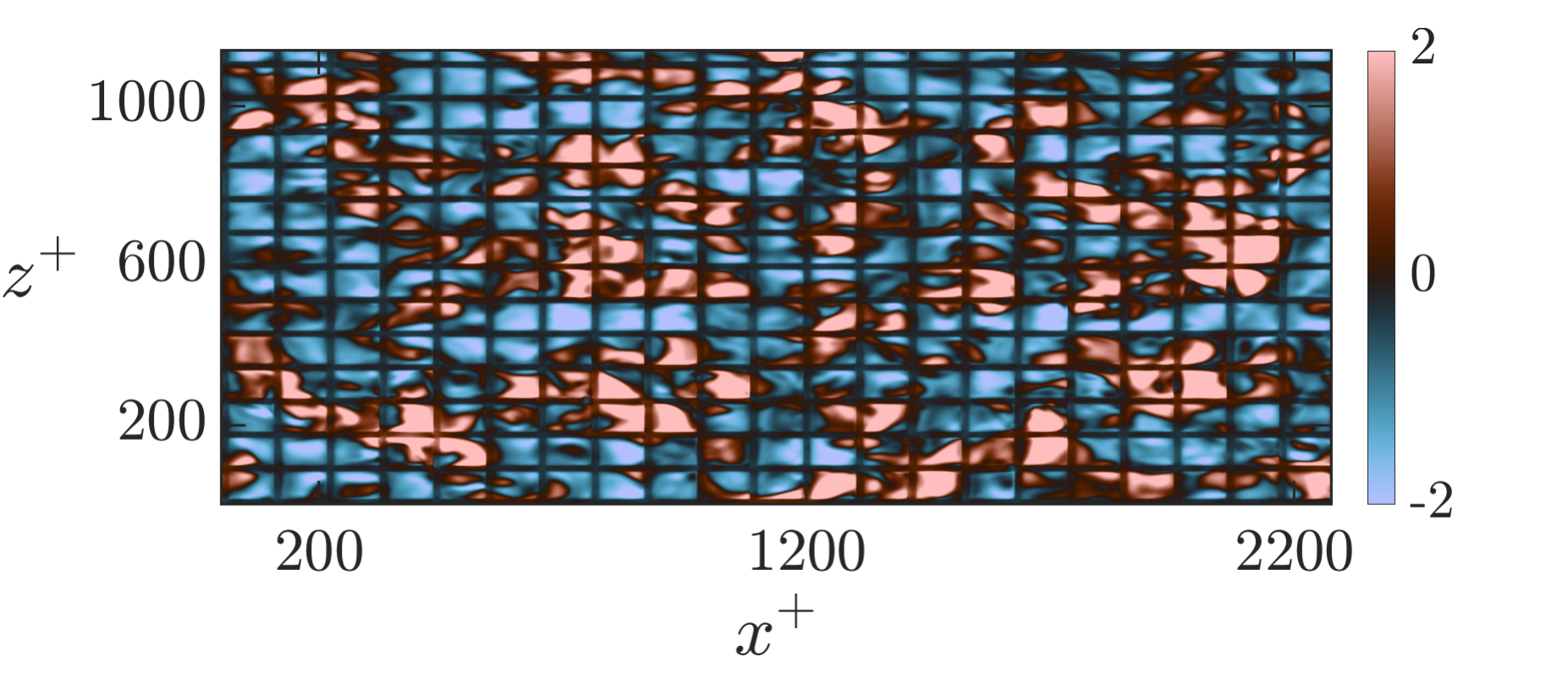}
    \end{subfigure}
    \hspace*{0mm}
    \begin{subfigure}[tbp]{.40\textwidth}
        {\captionsetup{position=bottom, labelfont=it,textfont=it,singlelinecheck=false,justification=raggedright,labelformat=parens}
        \caption{SP}\label{fig:u_velocity_contour_y_0:SP}}
        \vspace*{-0.8mm}
        \includegraphics[width=1\linewidth]{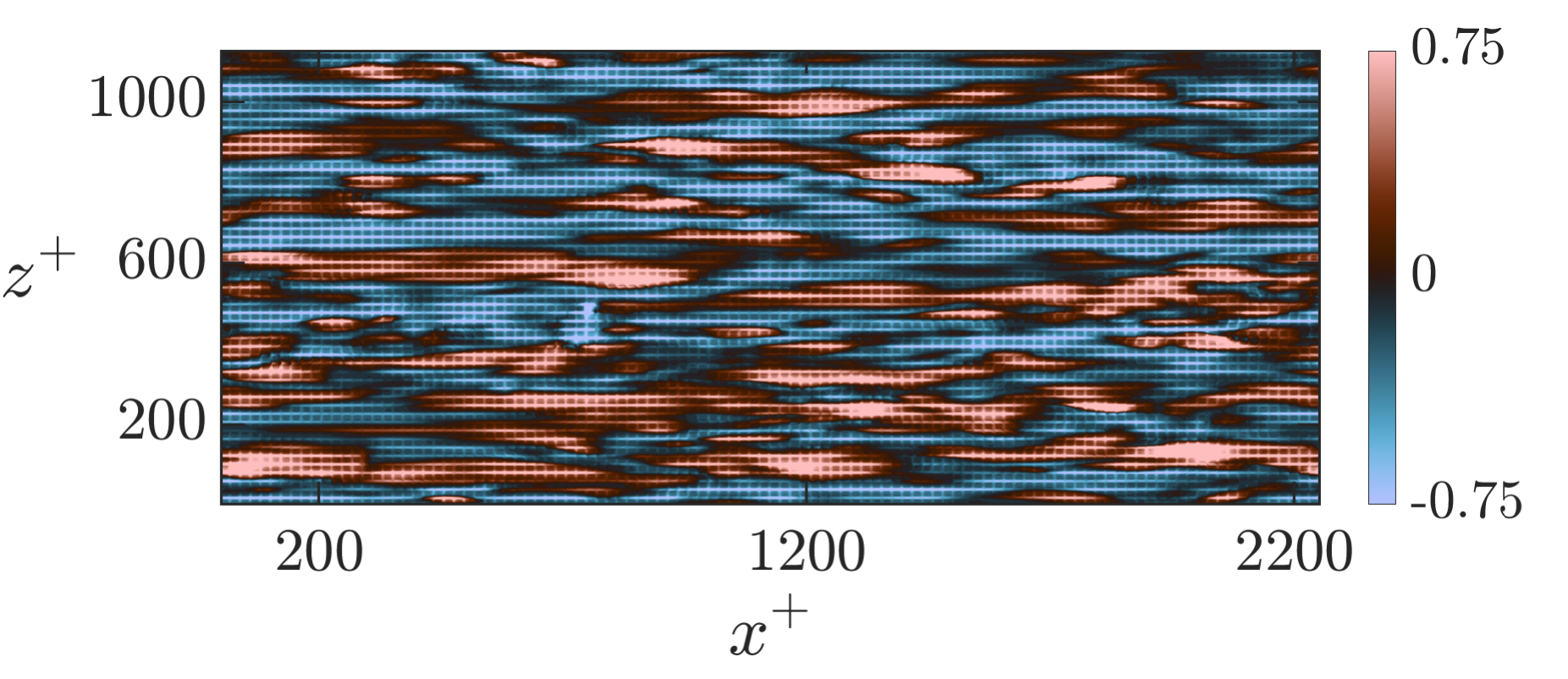}
    \end{subfigure}%
    \hspace*{-1mm}
    \begin{subfigure}[tbp]{.40\textwidth}
        {\captionsetup{position=bottom, labelfont=it,textfont=it,singlelinecheck=false,justification=raggedright,labelformat=parens}
        \caption{TP}\label{fig:u_velocity_contour_y_0:TP}}
        \vspace*{-0.8mm}
        \includegraphics[width=1\linewidth]{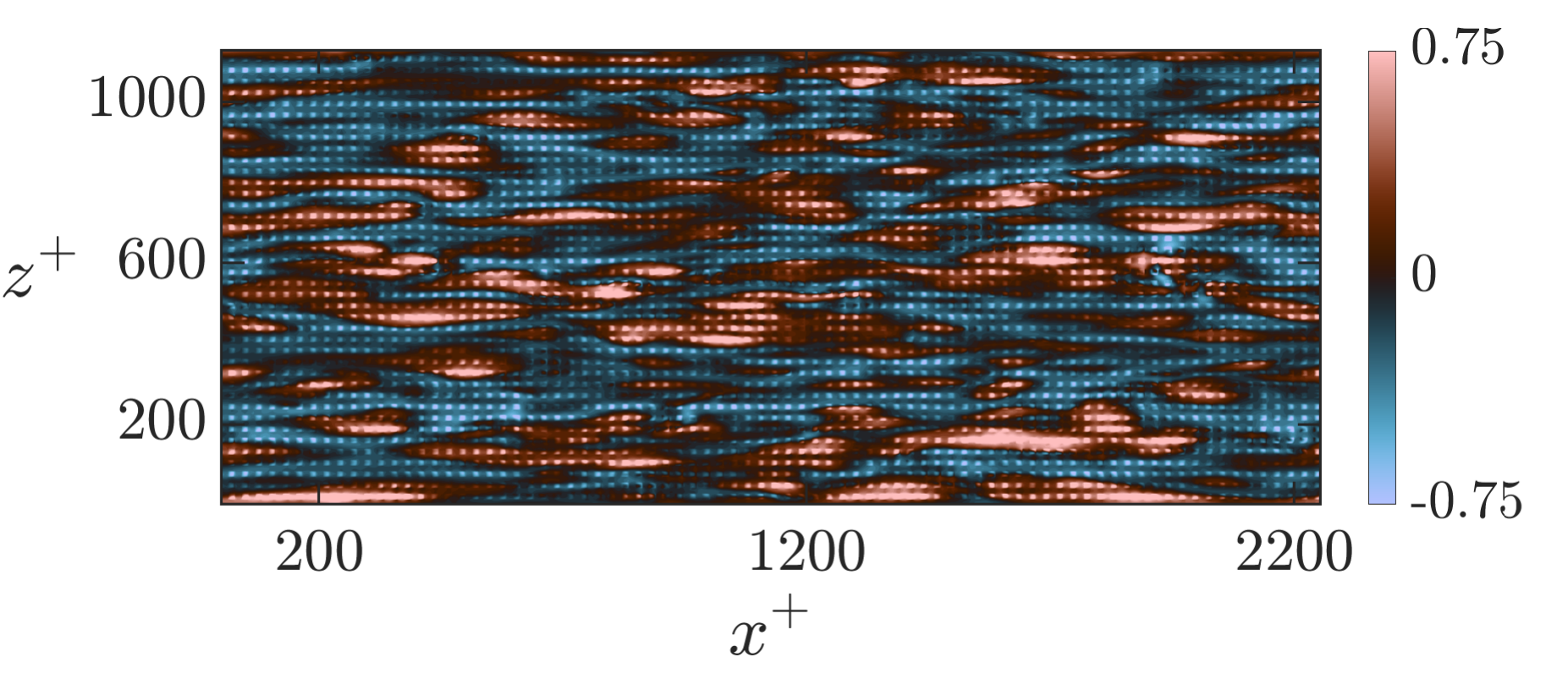}
    \end{subfigure}
    \vspace*{-2mm}
    \captionsetup{width=1\textwidth, justification=justified}
    \caption{Instantaneous snapshots of streamwise velocity fluctuations at $y=0$ normalized by $u_{\tau}$.}
    \label{fig:u_velocity_contours_y_0}
 \end{center}
 \begin{center}
    \hspace*{-13mm}
    \begin{subfigure}[tbp]{.40\textwidth}
        {\captionsetup{position=bottom, labelfont=it,textfont=it,singlelinecheck=false,justification=raggedright,labelformat=parens}
        \caption{KP1}\label{fig:v_velocity_contour_y_0:KP1}}
        \vspace*{-0.8mm}
        \includegraphics[width=1\linewidth]{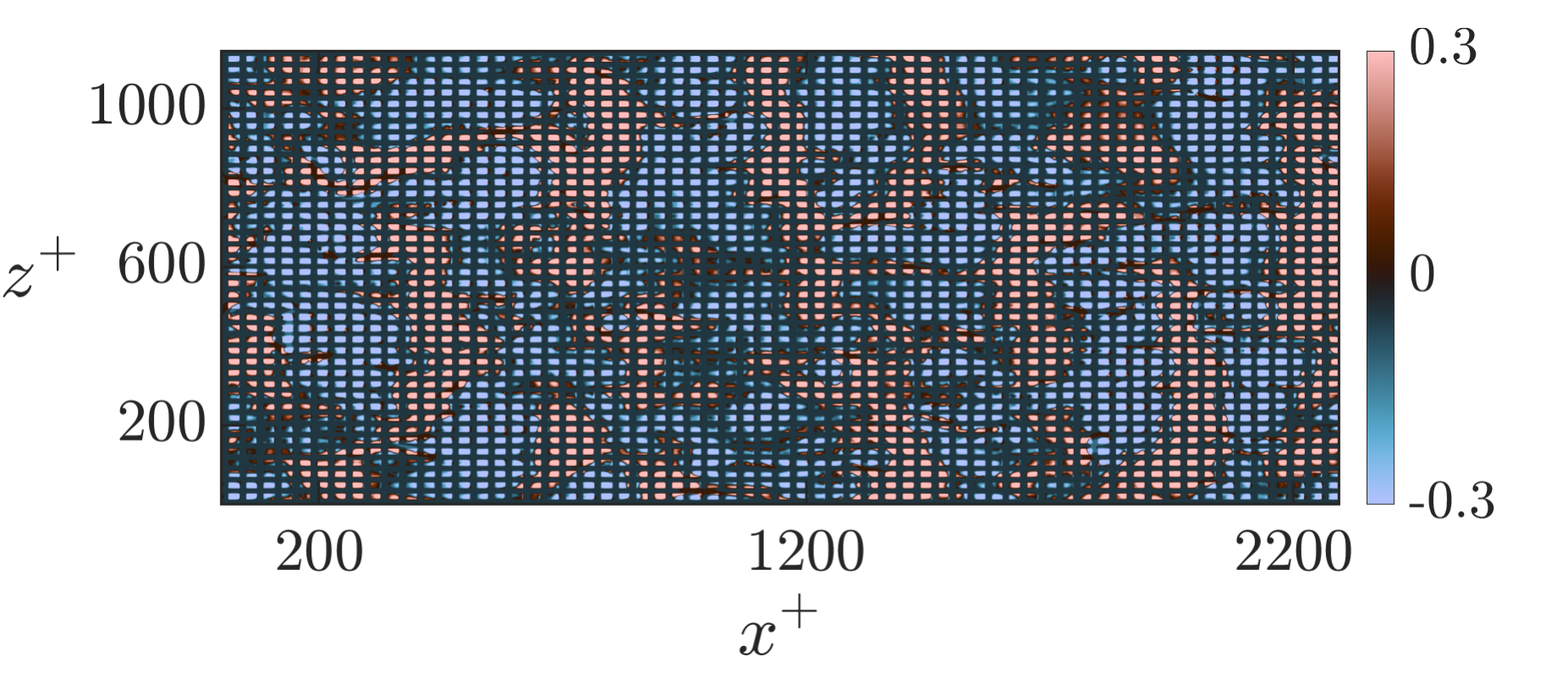}
    \end{subfigure}%
    \hspace*{-2mm}
    \begin{subfigure}[tbp]{.40\textwidth}
        {\captionsetup{position=bottom, labelfont=it,textfont=it,singlelinecheck=false,justification=raggedright,labelformat=parens}
        \caption{$KP1^{\prime}$}\label{fig:v_velocity_contour_y_0:KP1-1}}
        \vspace*{-0.8mm}
        \includegraphics[width=1\linewidth]{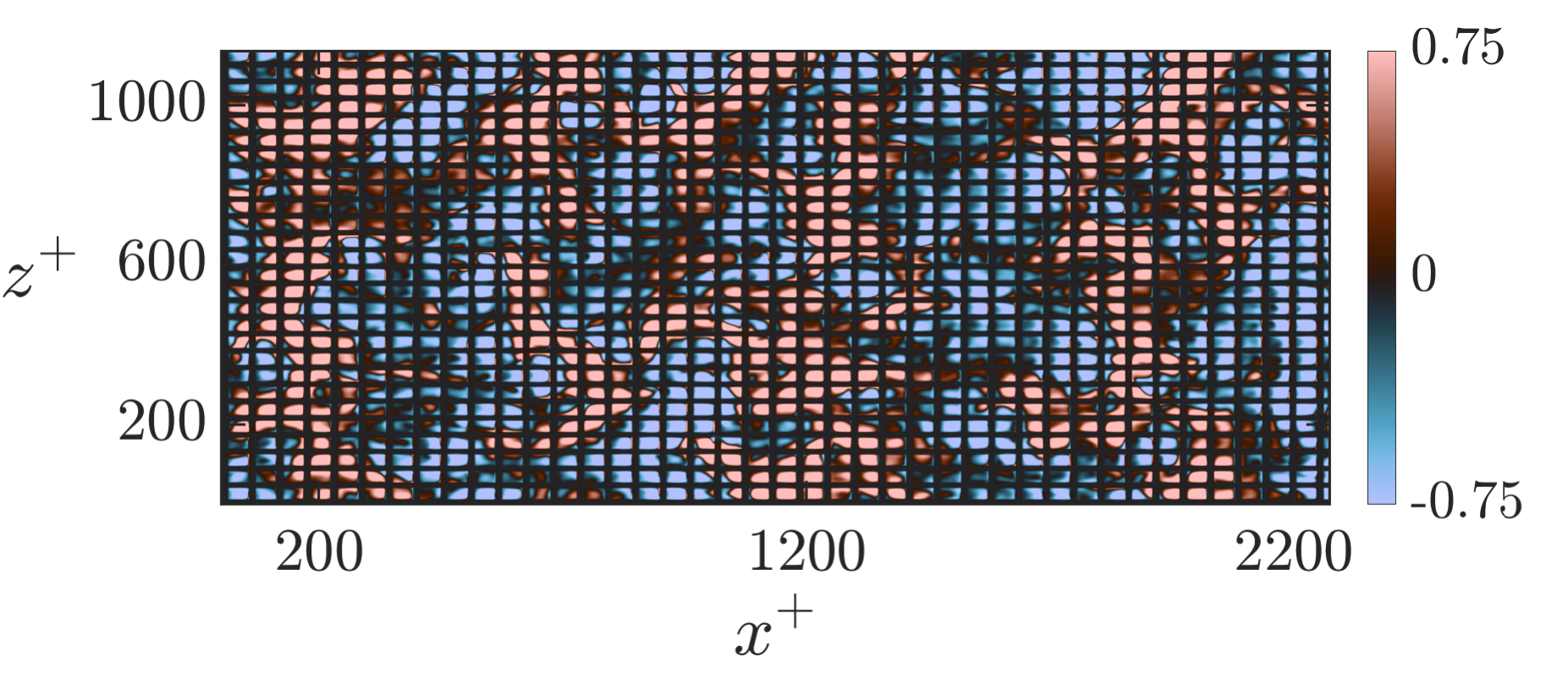}
    \end{subfigure}%
    \hspace*{-2mm}
    \begin{subfigure}[tbp]{.40\textwidth}
        {\captionsetup{position=bottom, labelfont=it,textfont=it,singlelinecheck=false,justification=raggedright,labelformat=parens}
        \caption{$KP1^{\prime\prime}$}\label{fig:v_velocity_contour_y_0:KP1-2}}
        \vspace*{-0.8mm}
        \includegraphics[width=1\linewidth]{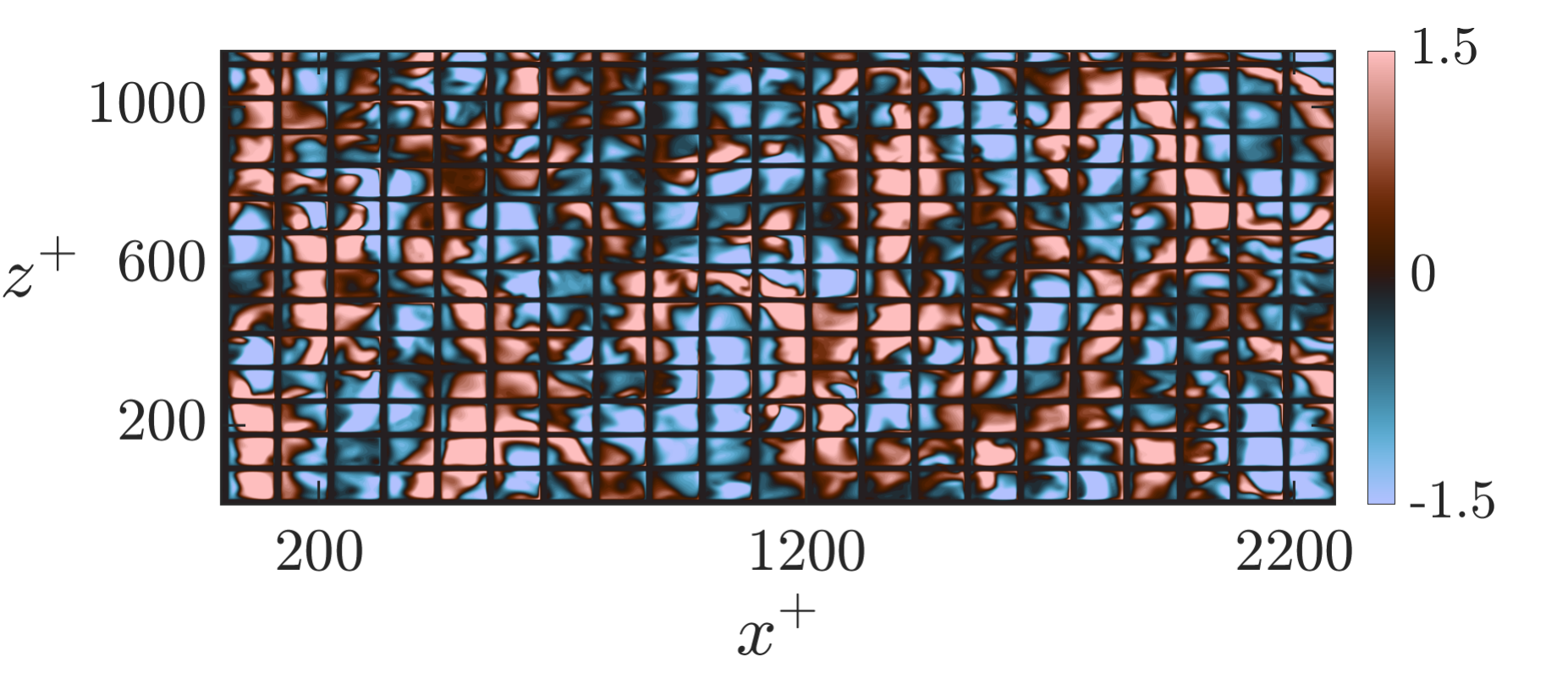}
    \end{subfigure}
    \hspace*{-13mm}
    \begin{subfigure}[tbp]{.40\textwidth}
        {\captionsetup{position=bottom, labelfont=it,textfont=it,singlelinecheck=false,justification=raggedright,labelformat=parens}
        \caption{KP2}\label{fig:v_velocity_contour_y_0:KP2}}
        \vspace*{-0.8mm}
        \includegraphics[width=1\linewidth]{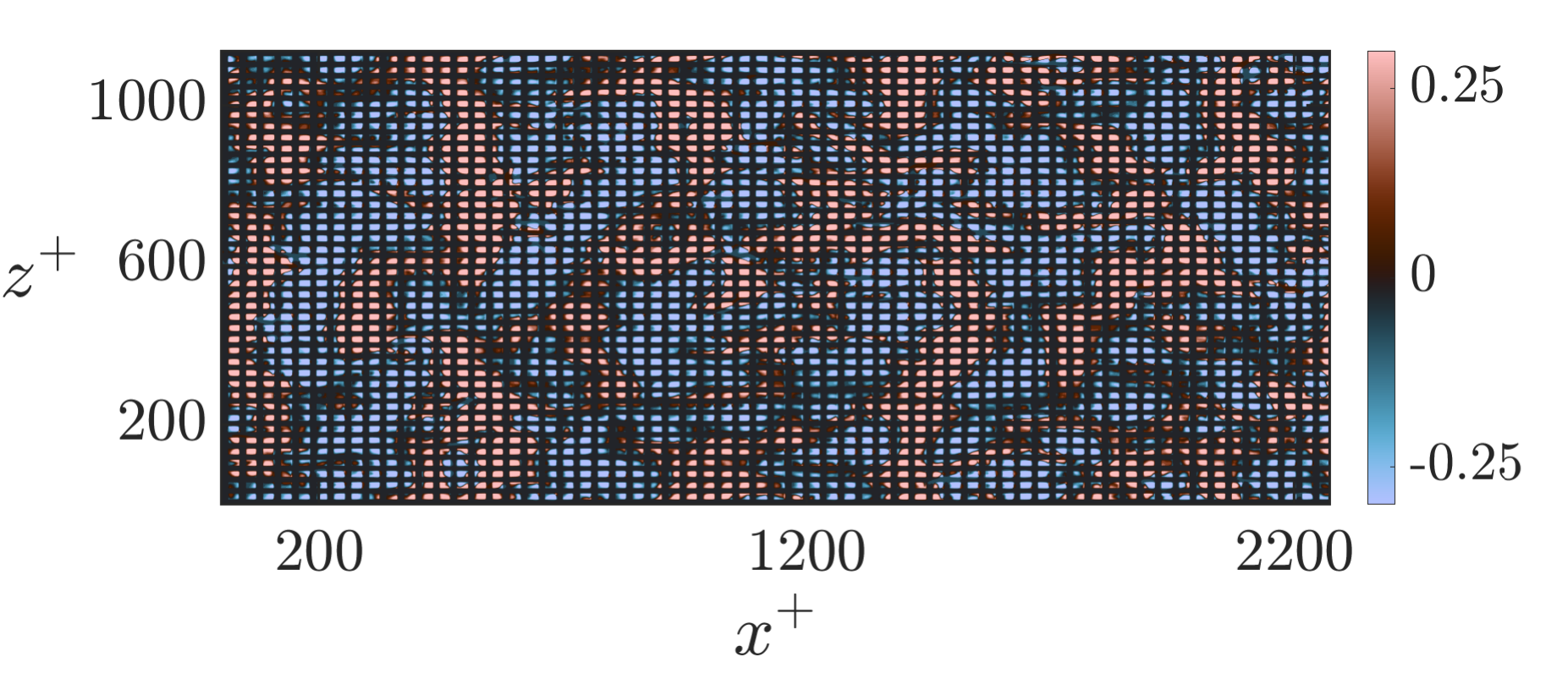}
    \end{subfigure}%
    \hspace*{-2mm}
    \begin{subfigure}[tbp]{.40\textwidth}
        {\captionsetup{position=bottom, labelfont=it,textfont=it,singlelinecheck=false,justification=raggedright,labelformat=parens}
        \caption{$KP2^{\prime}$}\label{fig:v_velocity_contour_y_0:KP2-1}}
        \vspace*{-0.8mm}
        \includegraphics[width=1\linewidth]{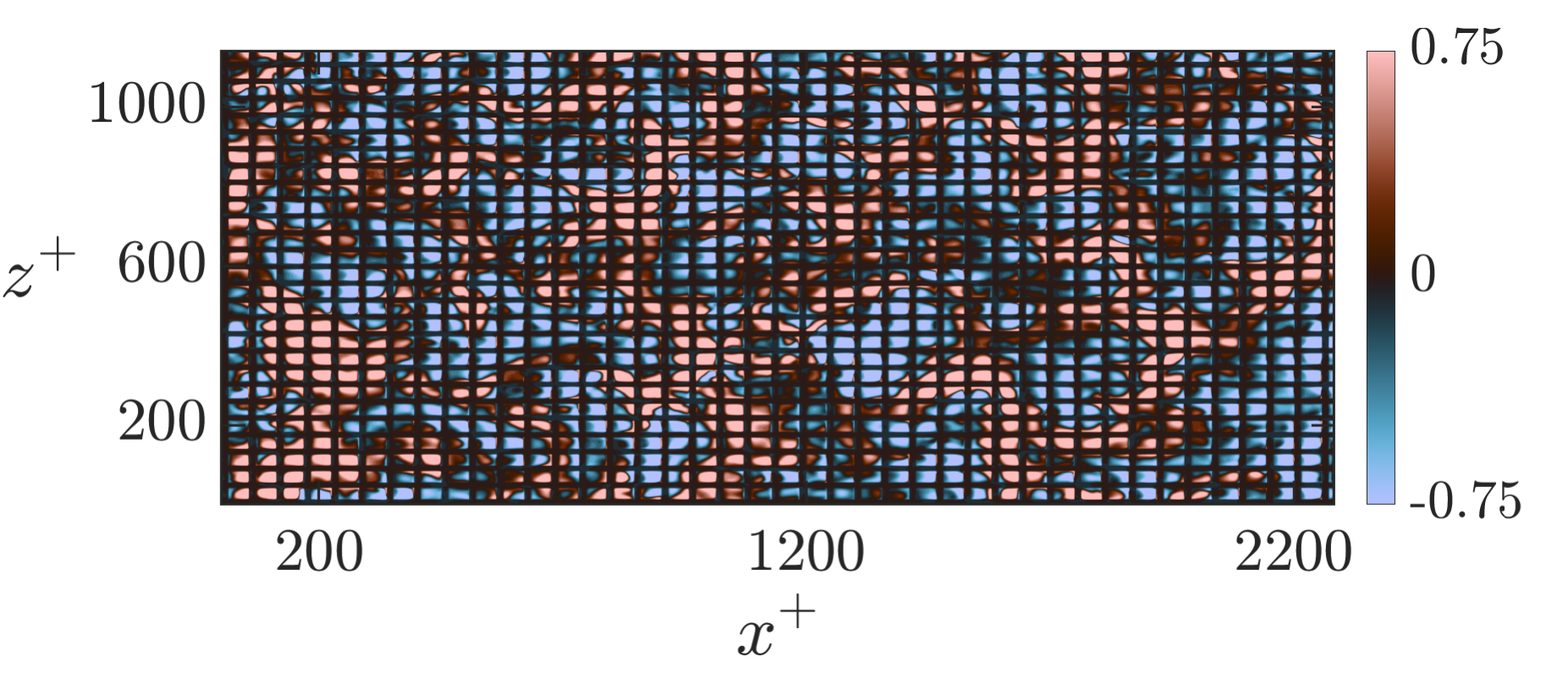}
    \end{subfigure}%
    \hspace*{-2mm}
    \begin{subfigure}[tbp]{.40\textwidth}
        {\captionsetup{position=bottom, labelfont=it,textfont=it,singlelinecheck=false,justification=raggedright,labelformat=parens}
        \caption{$KP2^{\prime\prime}$}\label{fig:v_velocity_contour_y_0:KP2-2}}
        \vspace*{-0.8mm}
        \includegraphics[width=1\linewidth]{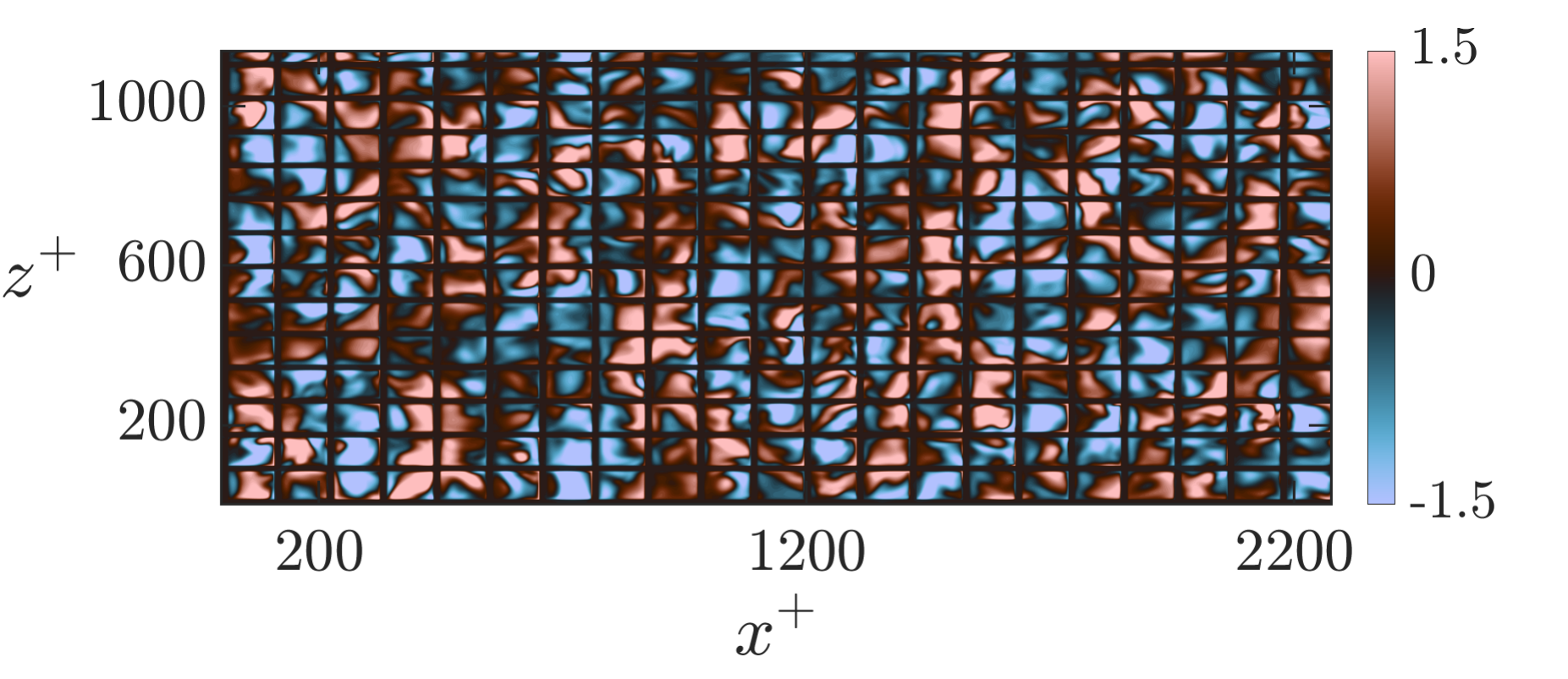}
    \end{subfigure}
    \hspace*{-13mm}
    \begin{subfigure}[tbp]{.40\textwidth}
        {\captionsetup{position=bottom, labelfont=it,textfont=it,singlelinecheck=false,justification=raggedright,labelformat=parens}
        \caption{KP3}\label{fig:v_velocity_contour_y_0:KP3}}
        \vspace*{-0.8mm}
        \includegraphics[width=1\linewidth]{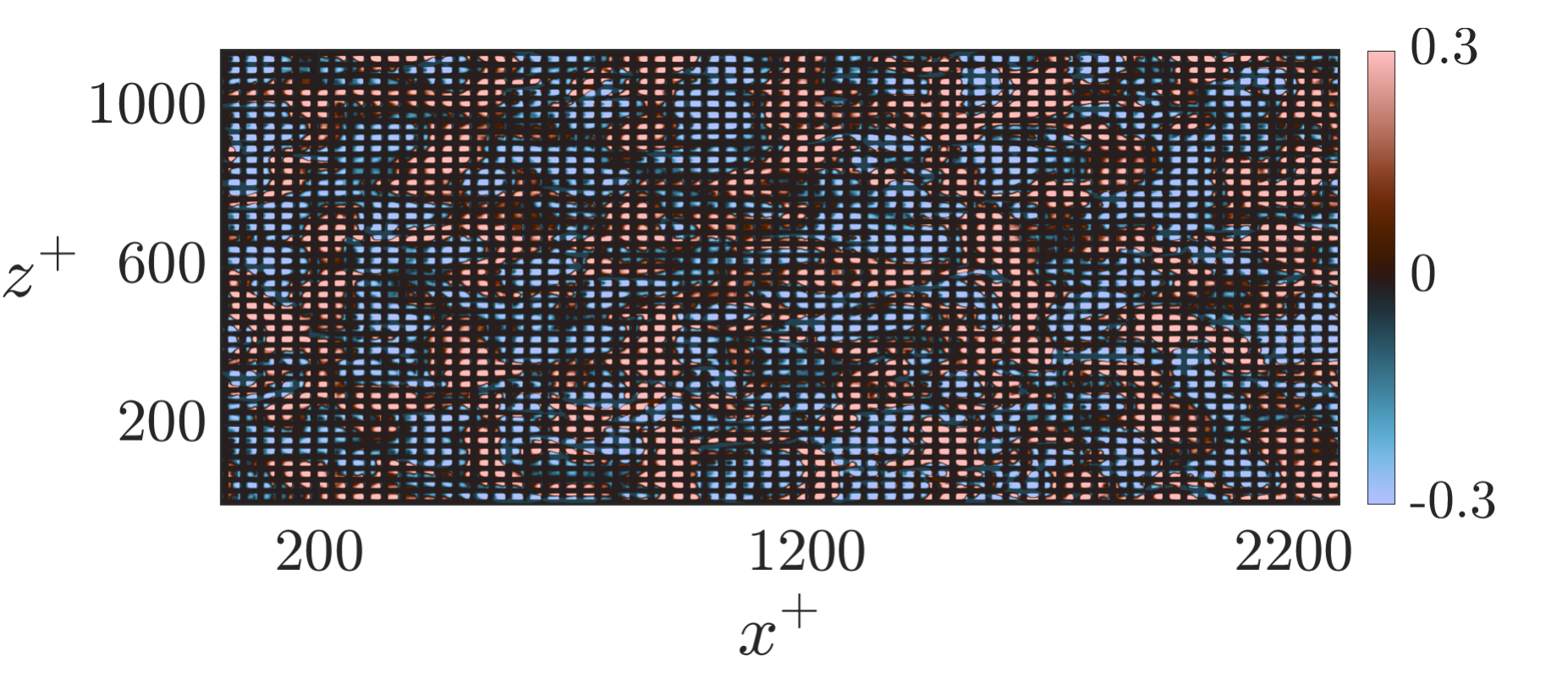}
    \end{subfigure}%
    \hspace*{-2mm}
    \begin{subfigure}[tbp]{.40\textwidth}
        {\captionsetup{position=bottom, labelfont=it,textfont=it,singlelinecheck=false,justification=raggedright,labelformat=parens}
        \caption{$KP3^{\prime}$}\label{fig:v_velocity_contour_y_0:KP3-1}}
        \vspace*{-0.8mm}
        \includegraphics[width=1\linewidth]{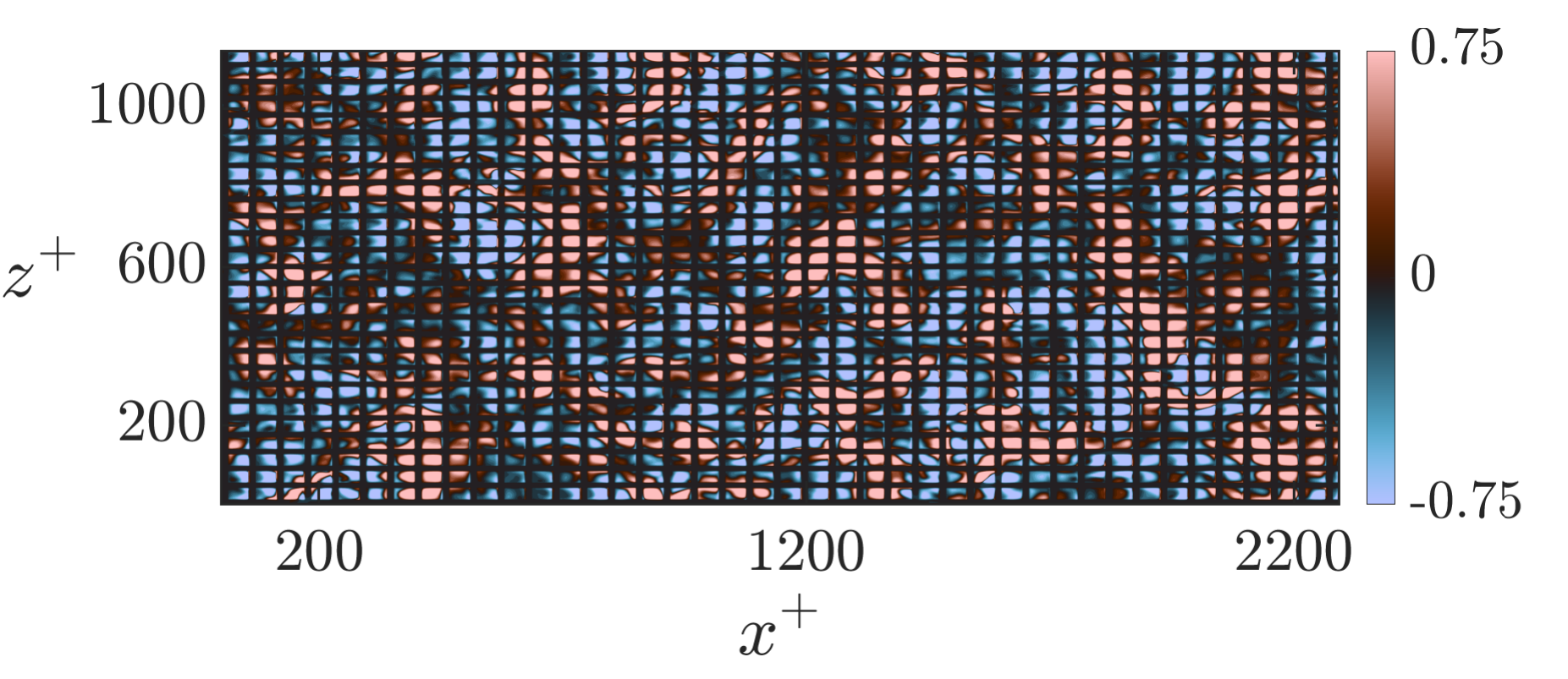}
    \end{subfigure}%
    \hspace*{-2mm}
    \begin{subfigure}[tbp]{.40\textwidth}
        {\captionsetup{position=bottom, labelfont=it,textfont=it,singlelinecheck=false,justification=raggedright,labelformat=parens}
        \caption{$KP3^{\prime\prime}$}\label{fig:v_velocity_contour_y_0:KP3-2}}
        \vspace*{-0.8mm}
        \includegraphics[width=1\linewidth]{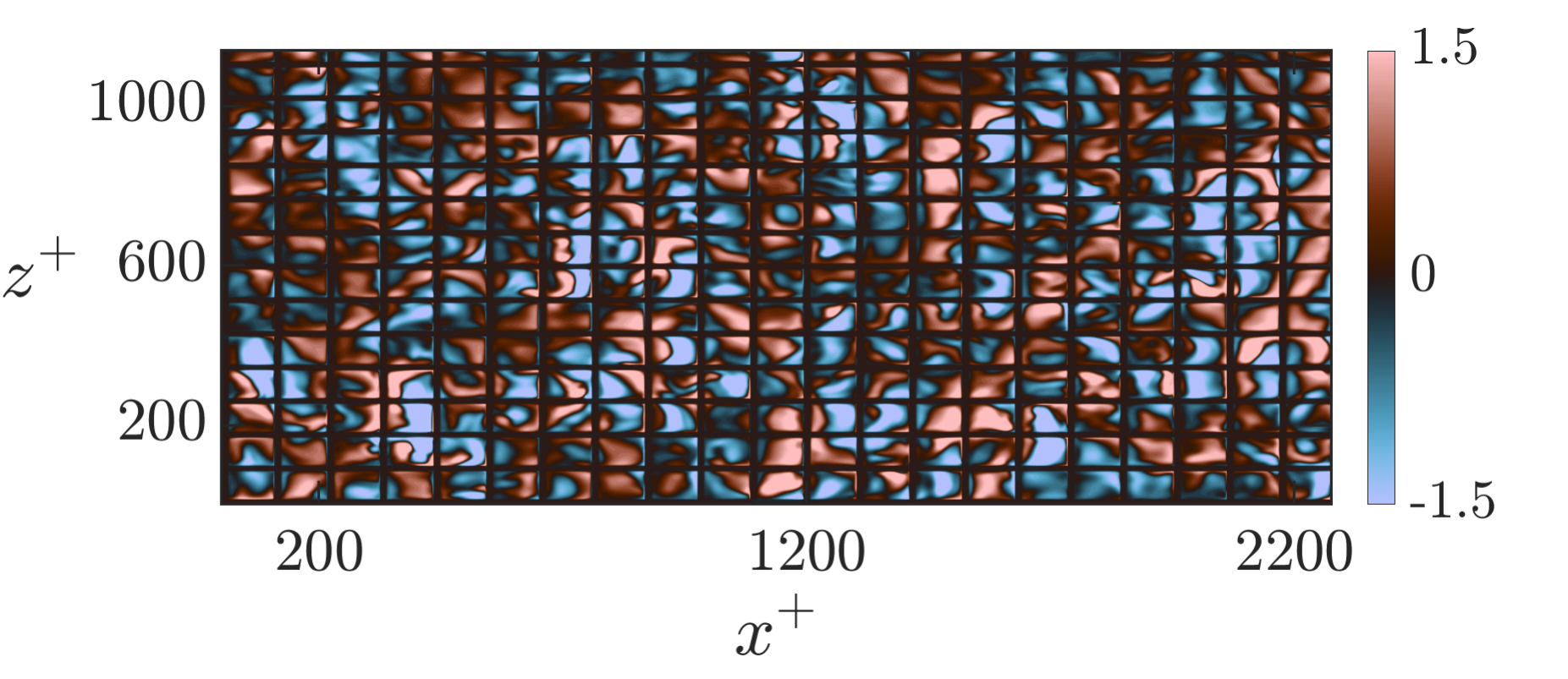}
    \end{subfigure}
    \hspace*{0mm}
    \begin{subfigure}[tbp]{.40\textwidth}
        {\captionsetup{position=bottom, labelfont=it,textfont=it,singlelinecheck=false,justification=raggedright,labelformat=parens}
        \caption{SP}\label{fig:v_velocity_contour_y_0:SP}}
        \vspace*{-0.8mm}
        \includegraphics[width=1\linewidth]{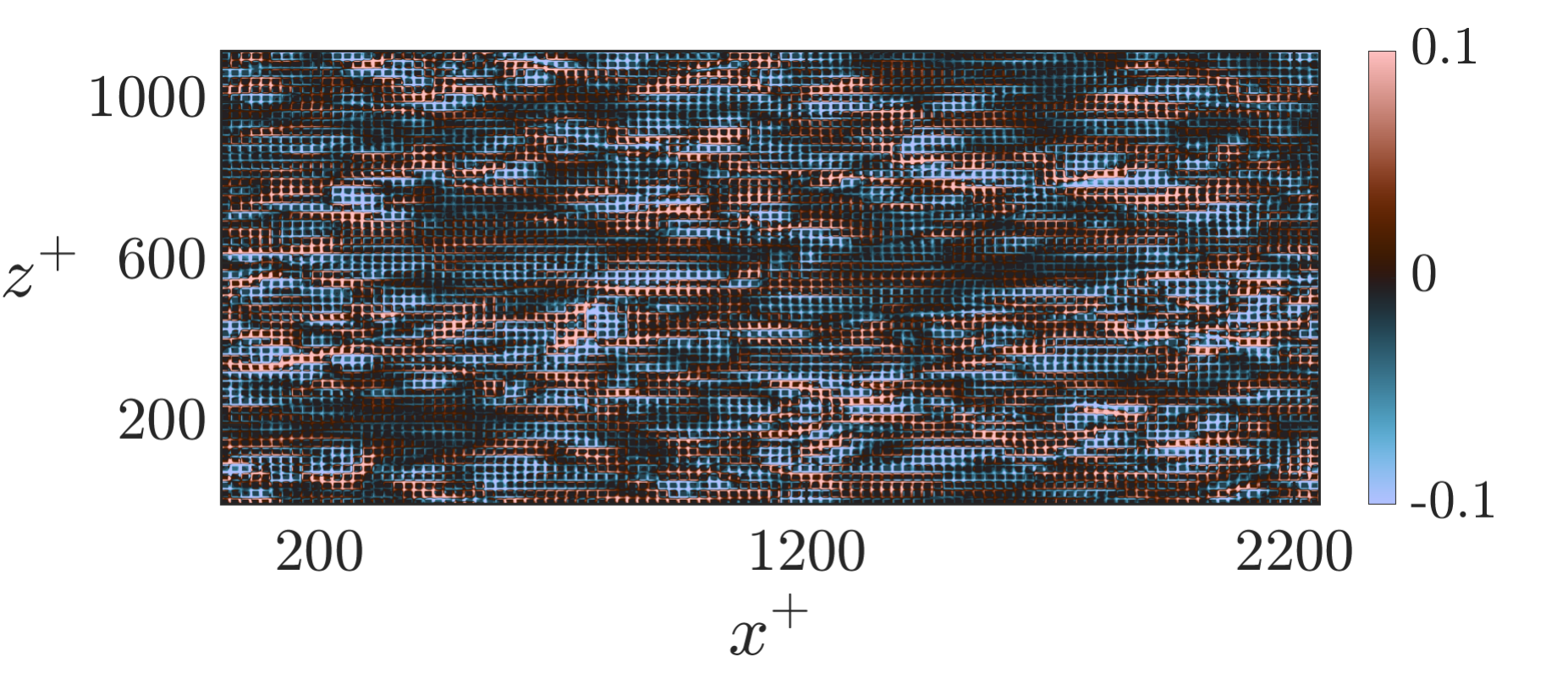}
    \end{subfigure}%
    \hspace*{-1mm}
    \begin{subfigure}[tbp]{.40\textwidth}
        {\captionsetup{position=bottom, labelfont=it,textfont=it,singlelinecheck=false,justification=raggedright,labelformat=parens}
        \caption{TP}\label{fig:v_velocity_contour_y_0:TP}}
        \vspace*{-0.8mm}
        \includegraphics[width=1\linewidth]{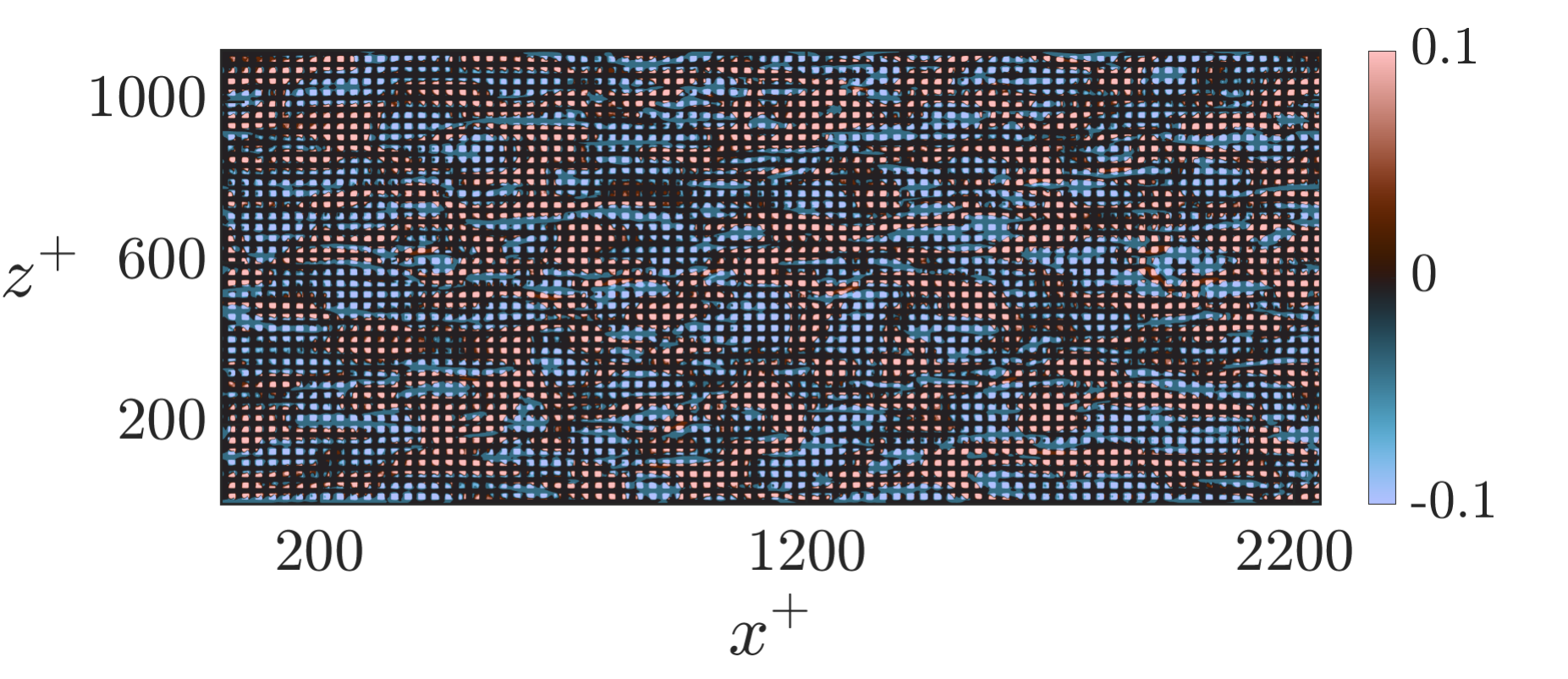}
    \end{subfigure}
    \vspace*{-2mm}
    \captionsetup{width=1\textwidth, justification=justified}
    \caption{Instantaneous snapshots of wall-normal velocity fluctuations at $y=0$ normalized by $u_{\tau}$.}
    \label{fig:v_velocity_contours_y_0}
 \end{center}
\end{figure}

 The changes in flow turbulence structure arising between the different cases can be visually examined in the instantaneous flow-field snapshots of figures \hyperref[fig:u_velocity_contours_y_0]{\ref*{fig:u_velocity_contours_y_0}} and \hyperref[fig:v_velocity_contours_y_0]{\ref*{fig:v_velocity_contours_y_0}}. \Cref{fig:u_velocity_contours_y_0} depicts streamwise velocity fluctuations at the surface ($y=0$). The low permeability case $SP$ retains the characteristics of smooth-wall flow, with streamwise elongated streaky patches visible in its velocity-field (\cref{fig:u_velocity_contour_y_0:SP}). As the porous wall is made more permeable, the streaky structure becomes disrupted and changes to groupings of smaller patches, such as those in figures \hyperref[fig:u_velocity_contour_y_0:KP2]{\ref*{fig:u_velocity_contour_y_0:KP2}} and \hyperref[fig:u_velocity_contour_y_0:KP1]{\ref*{fig:u_velocity_contour_y_0:KP1}}. These patches have a degree of spanwise alignment and get convected downstream.
 The change is even more notable when examining the wall-normal velocity in figures \hyperref[fig:v_velocity_contour_y_0:KP2]{\ref*{fig:v_velocity_contour_y_0:KP2}} and \hyperref[fig:v_velocity_contour_y_0:KP1]{\ref*{fig:v_velocity_contour_y_0:KP1}}, where lengthy spanwise-oriented patches have emerged. The increased horizontal pore dimensions in walls $KP\langle\rangle^{\prime\prime}$ by almost a factor of three leads to approximately a quadrupling of the effective wall-normal permeability, $\sqrt{K_{y\phantom{y}}^+}$. The effect that this has on the flow structure compared to the $KP\langle\rangle$ walls is larger but more disorganized patches at the surface, such as those visible in figures \hyperref[fig:u_velocity_contour_y_0:KP1-2]{\ref*{fig:u_velocity_contour_y_0:KP1-2}}, \hyperref[fig:u_velocity_contour_y_0:KP2-2]{\ref*{fig:u_velocity_contour_y_0:KP2-2}}, \hyperref[fig:v_velocity_contour_y_0:KP2-2]{\ref*{fig:v_velocity_contour_y_0:KP2-2}}, \hyperref[fig:v_velocity_contour_y_0:KP2-2]{\ref*{fig:v_velocity_contour_y_0:KP2-2}}.

\begin{figure}
 \begin{center}
    \hspace*{-8mm}
    \begin{subfigure}[tbp]{.38\textwidth}
        {\captionsetup{position=bottom, labelfont=it,textfont=it,size=scriptsize,singlelinecheck=false,justification=centering,labelformat=parens}
        \caption{$KP1^{\prime\prime},\;$ $k_x\,k_z\,E_{uu}$}\label{fig:u_spectra_y_0:KP1-2}}
        \vspace*{-0.8mm}
        \includegraphics[width=1\linewidth]{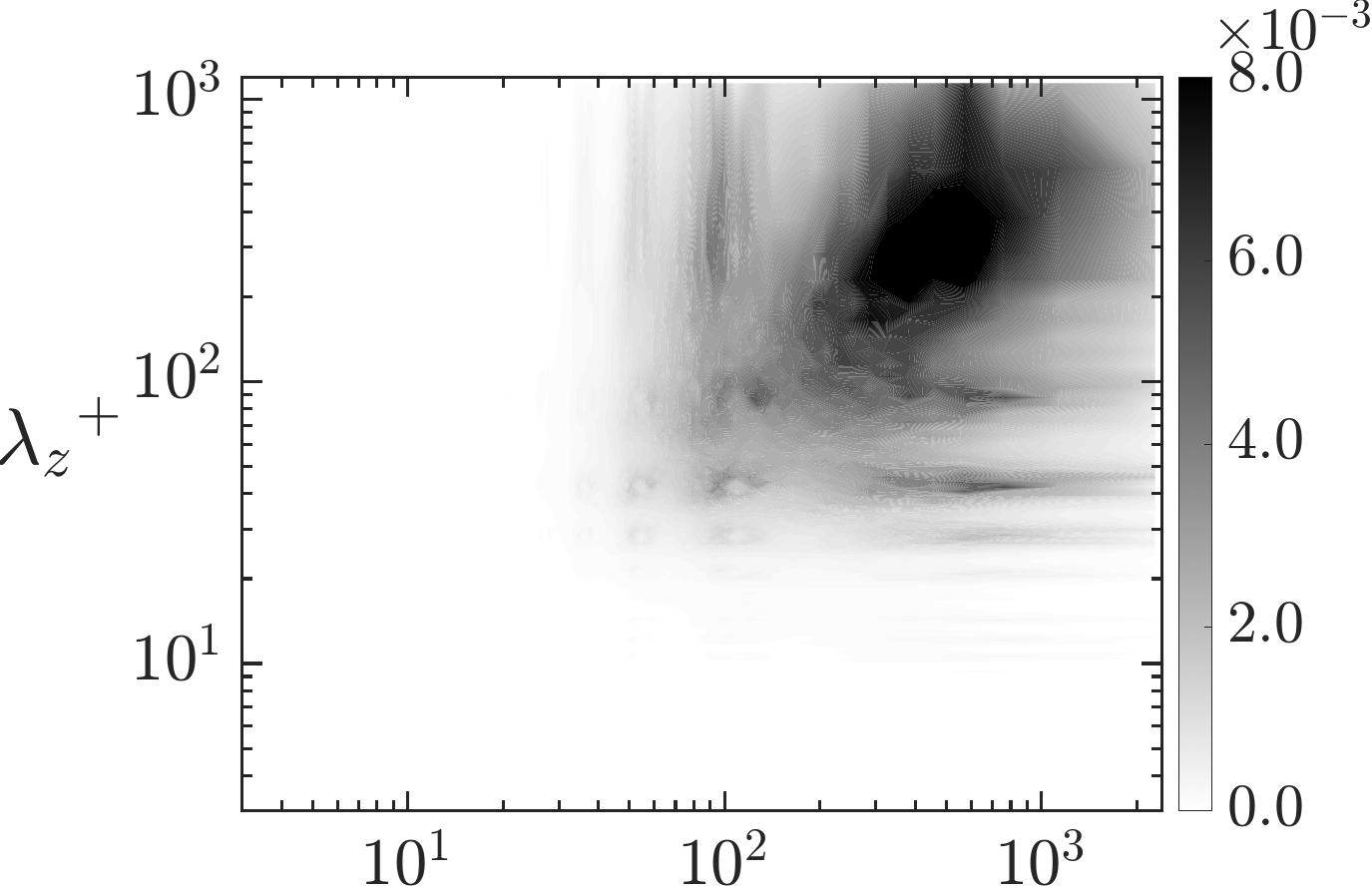}
    \end{subfigure}%
    \hspace*{2mm}
    \begin{subfigure}[tbp]{.34\textwidth}
        {\captionsetup{position=bottom, labelfont=it,textfont=it,size=scriptsize,singlelinecheck=false,justification=centering,labelformat=parens}
        \caption{$KP1^{\prime\prime},\;$ $k_x\,k_z\,E_{vv}$}\label{fig:v_spectra_y_0:KP1-2}}
        \vspace*{-0.8mm}
        \includegraphics[width=1\linewidth]{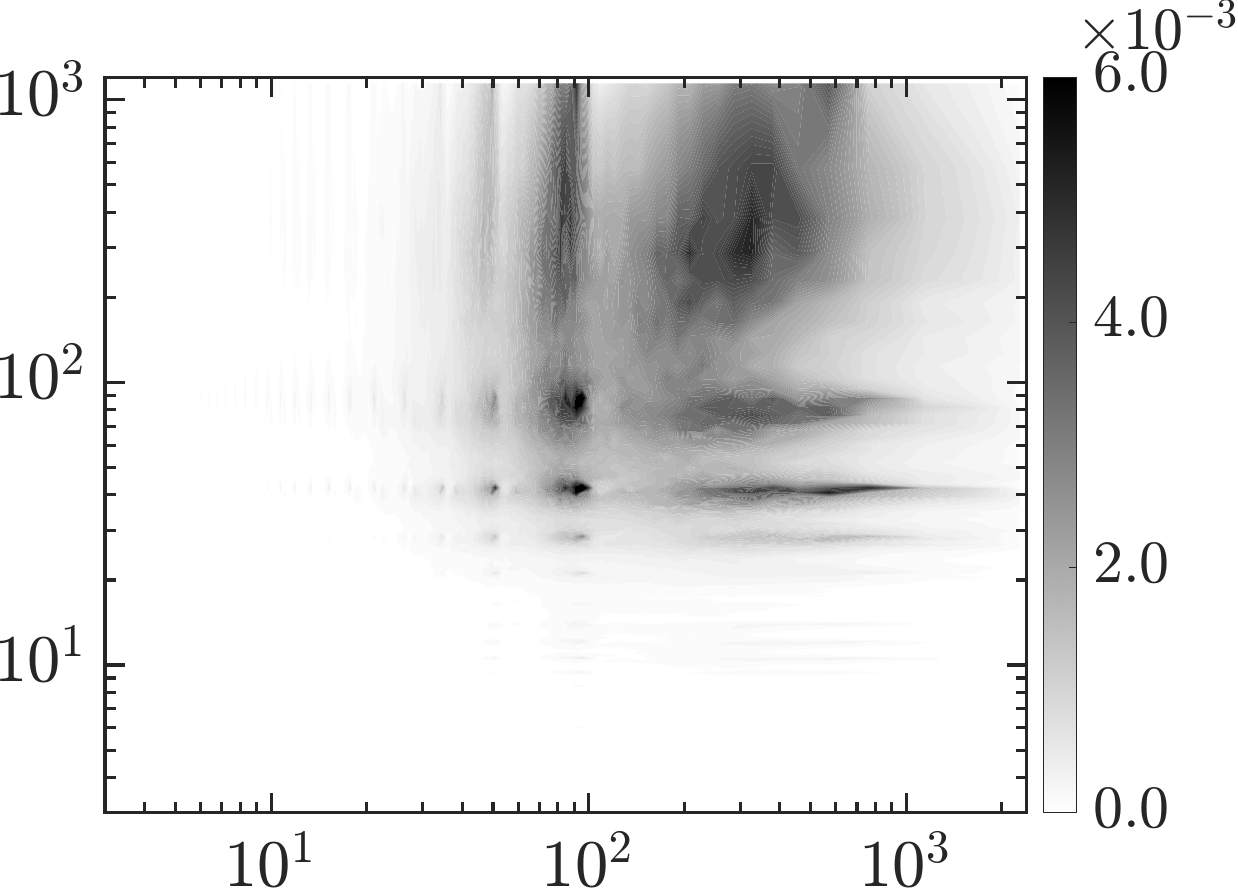}
    \end{subfigure}%
    \hspace*{2mm}
    \begin{subfigure}[tbp]{.34\textwidth}
        {\captionsetup{position=bottom, labelfont=it,textfont=it,size=scriptsize,singlelinecheck=false,justification=centering,labelformat=parens}
        \caption{$KP1^{\prime\prime},\;$ $k_x\,k_z\,E_{uv}$}\label{fig:uv_spectra_y_0:KP1-2}}
        \vspace*{-0.8mm}
        \includegraphics[width=1\linewidth]{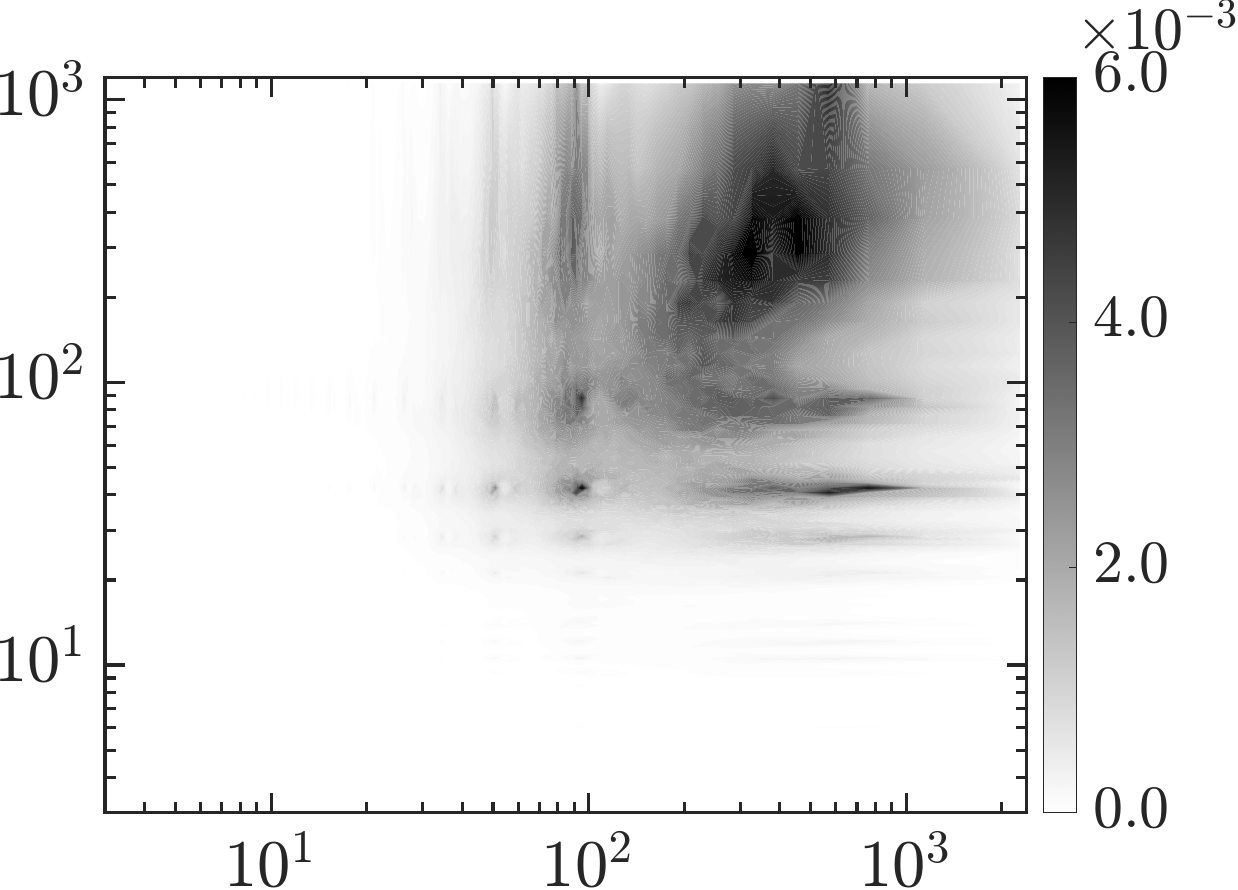}
    \end{subfigure}
    \hspace*{-8mm}
    \begin{subfigure}[tbp]{.38\textwidth}
        {\captionsetup{position=bottom, labelfont=it,textfont=it,size=scriptsize,singlelinecheck=false,justification=centering,labelformat=parens}
        \caption{$KP1^{\prime},\;$ $k_x\,k_z\,E_{uu}$}\label{fig:u_spectra_y_0:KP1-1}}
        \vspace*{-0.8mm}
        \includegraphics[width=1\linewidth]{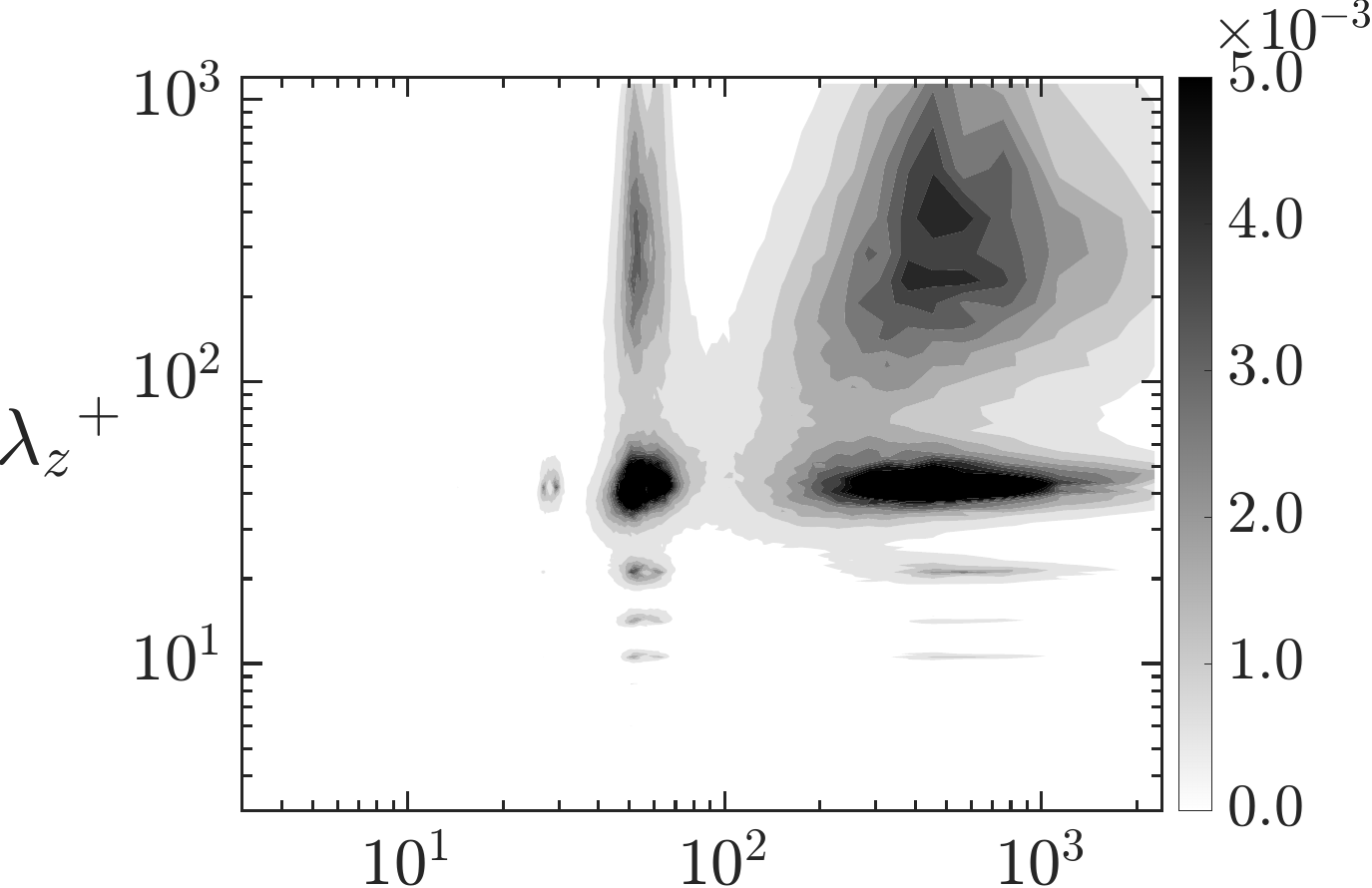}
    \end{subfigure}%
    \hspace*{2mm}
    \begin{subfigure}[tbp]{.34\textwidth}
        {\captionsetup{position=bottom, labelfont=it,textfont=it,size=scriptsize,singlelinecheck=false,justification=centering,labelformat=parens}
        \caption{$KP1^{\prime},\;$ $k_x\,k_z\,E_{vv}$}\label{fig:v_spectra_y_0:KP1-1}}
        \vspace*{-0.8mm}
        \includegraphics[width=1\linewidth]{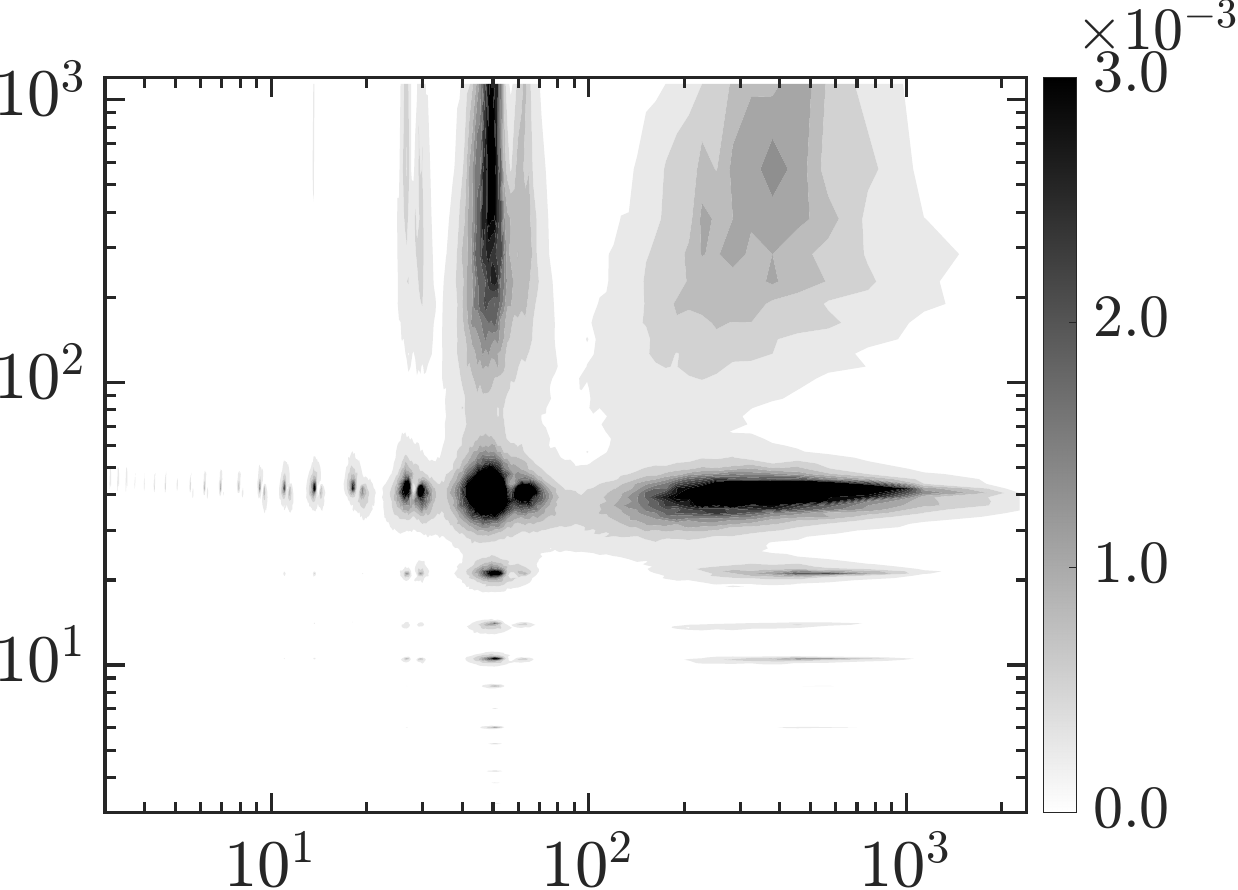}
    \end{subfigure}%
    \hspace*{2mm}
    \begin{subfigure}[tbp]{.34\textwidth}
        {\captionsetup{position=bottom, labelfont=it,textfont=it,size=scriptsize,singlelinecheck=false,justification=centering,labelformat=parens}
        \caption{$KP1^{\prime},\;$ $k_x\,k_z\,E_{uv}$}\label{fig:uv_spectra_y_0:KP1-1}}
        \vspace*{-0.8mm}
        \includegraphics[width=1\linewidth]{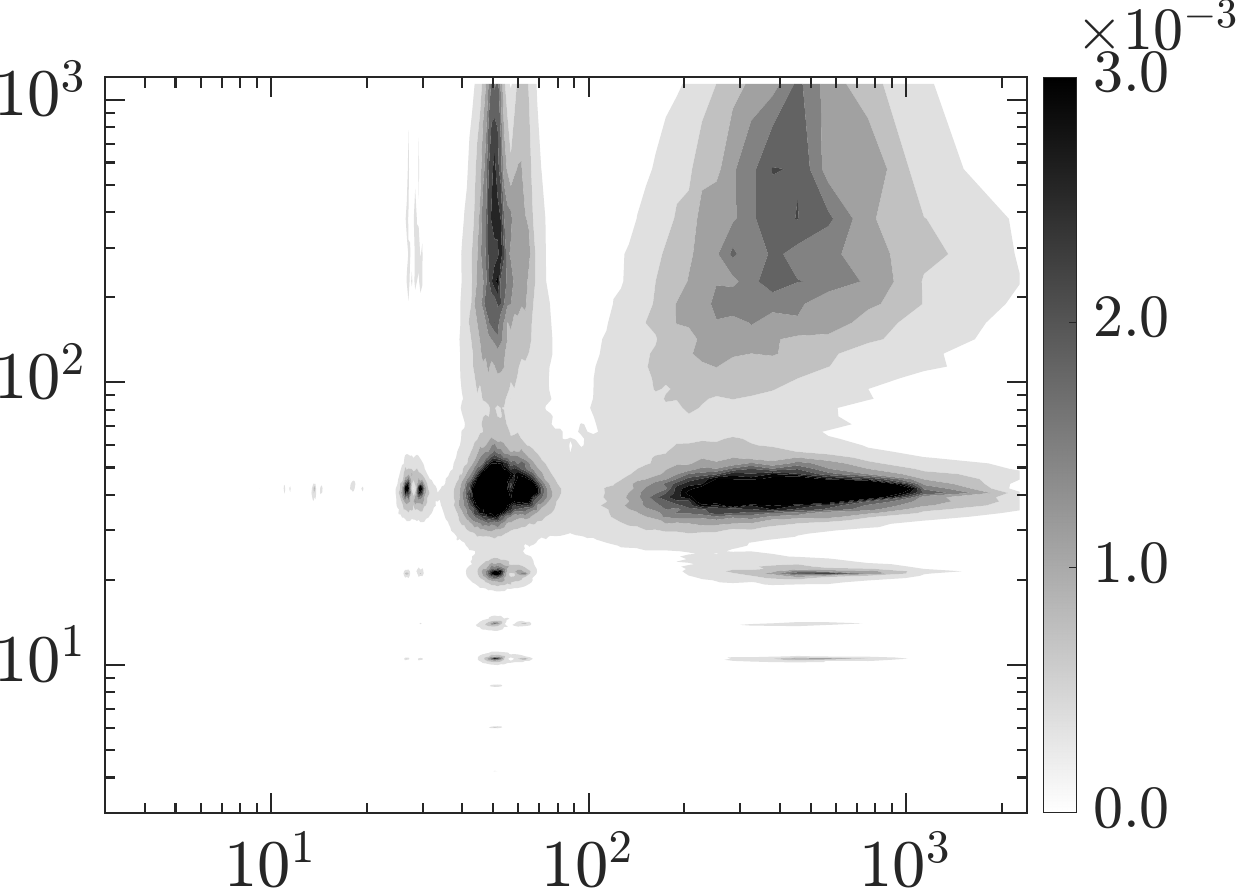}
    \end{subfigure}
    \hspace*{-8mm}
    \begin{subfigure}[tbp]{.38\textwidth}
        {\captionsetup{position=bottom, labelfont=it,textfont=it,size=scriptsize,singlelinecheck=false,justification=centering,labelformat=parens}
        \caption{$KP1,\;$ $k_x\,k_z\,E_{uu}$}\label{fig:u_spectra_y_0:KP1}}
        \vspace*{-0.8mm}
        \includegraphics[width=1\linewidth]{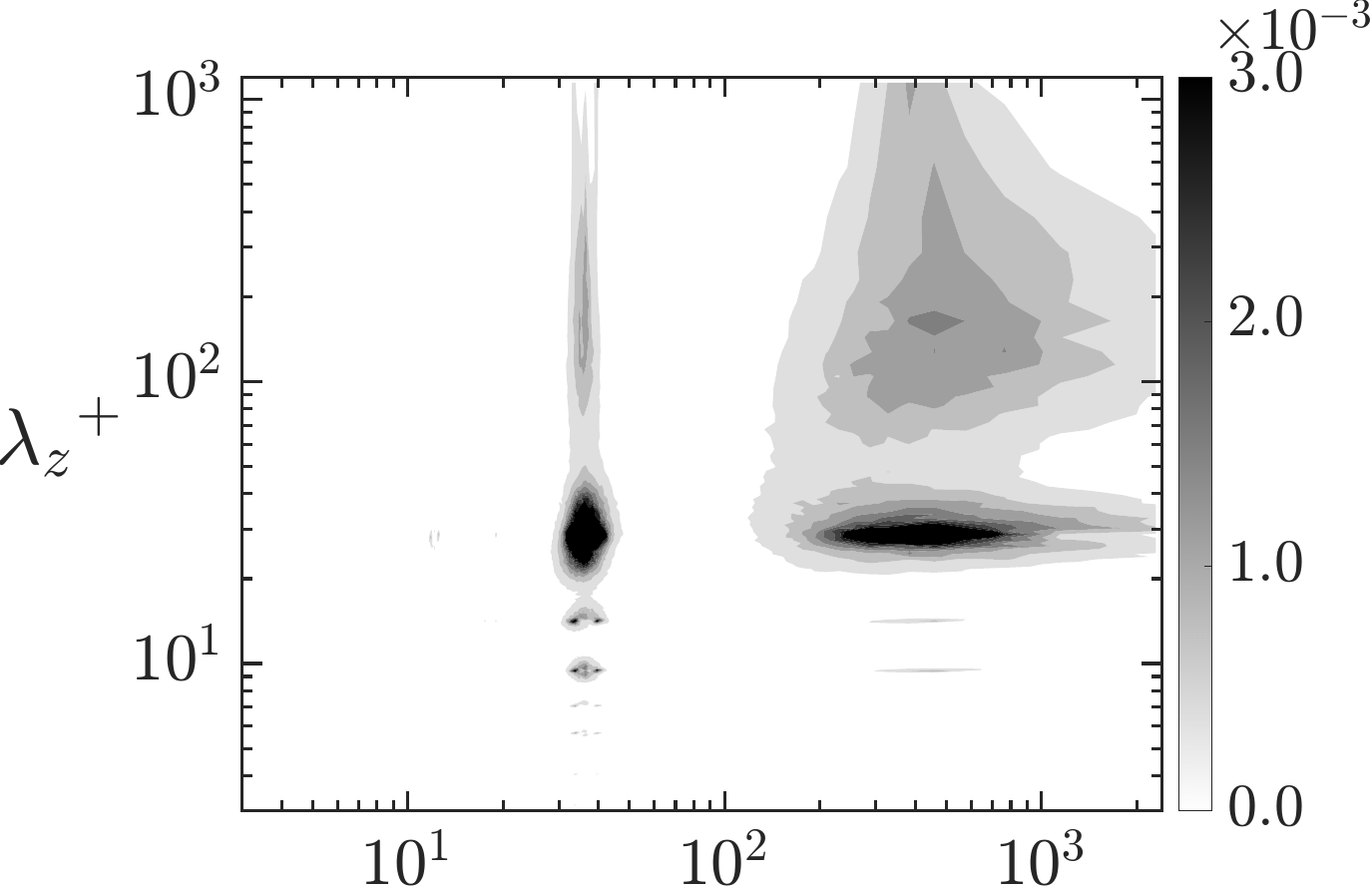}
    \end{subfigure}%
    \hspace*{2mm}
    \begin{subfigure}[tbp]{.34\textwidth}
        {\captionsetup{position=bottom, labelfont=it,textfont=it,size=scriptsize,singlelinecheck=false,justification=centering,labelformat=parens}
        \caption{$KP1,\;$ $k_x\,k_z\,E_{vv}$}\label{fig:v_spectra_y_0:KP1}}
        \vspace*{-0.8mm}
        \includegraphics[width=1\linewidth]{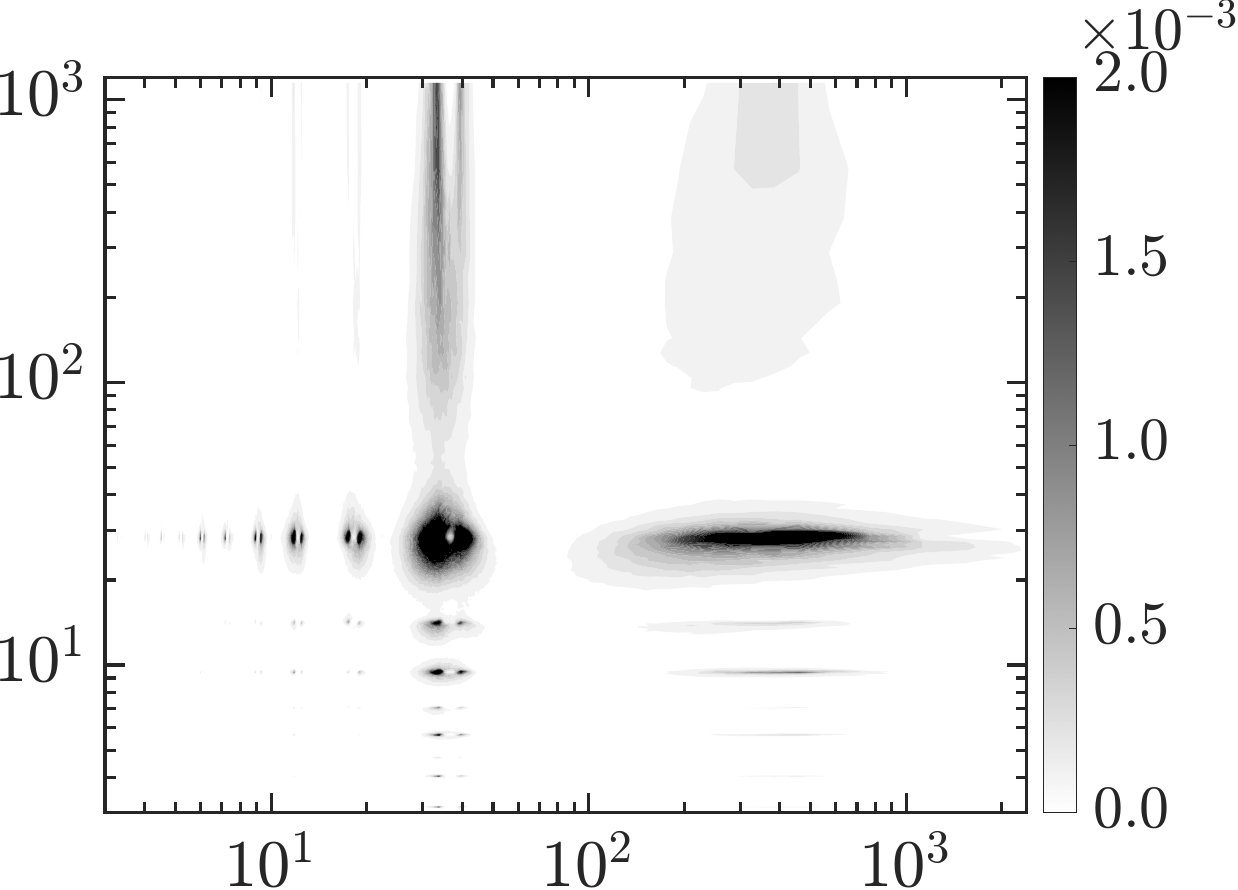}
    \end{subfigure}%
    \hspace*{2mm}
    \begin{subfigure}[tbp]{.34\textwidth}
        {\captionsetup{position=bottom, labelfont=it,textfont=it,size=scriptsize,singlelinecheck=false,justification=centering,labelformat=parens}
        \caption{$KP1,\;$ $k_x\,k_z\,E_{uv}$}\label{fig:uv_spectra_y_0:KP1}}
        \vspace*{-0.8mm}
        \includegraphics[width=1\linewidth]{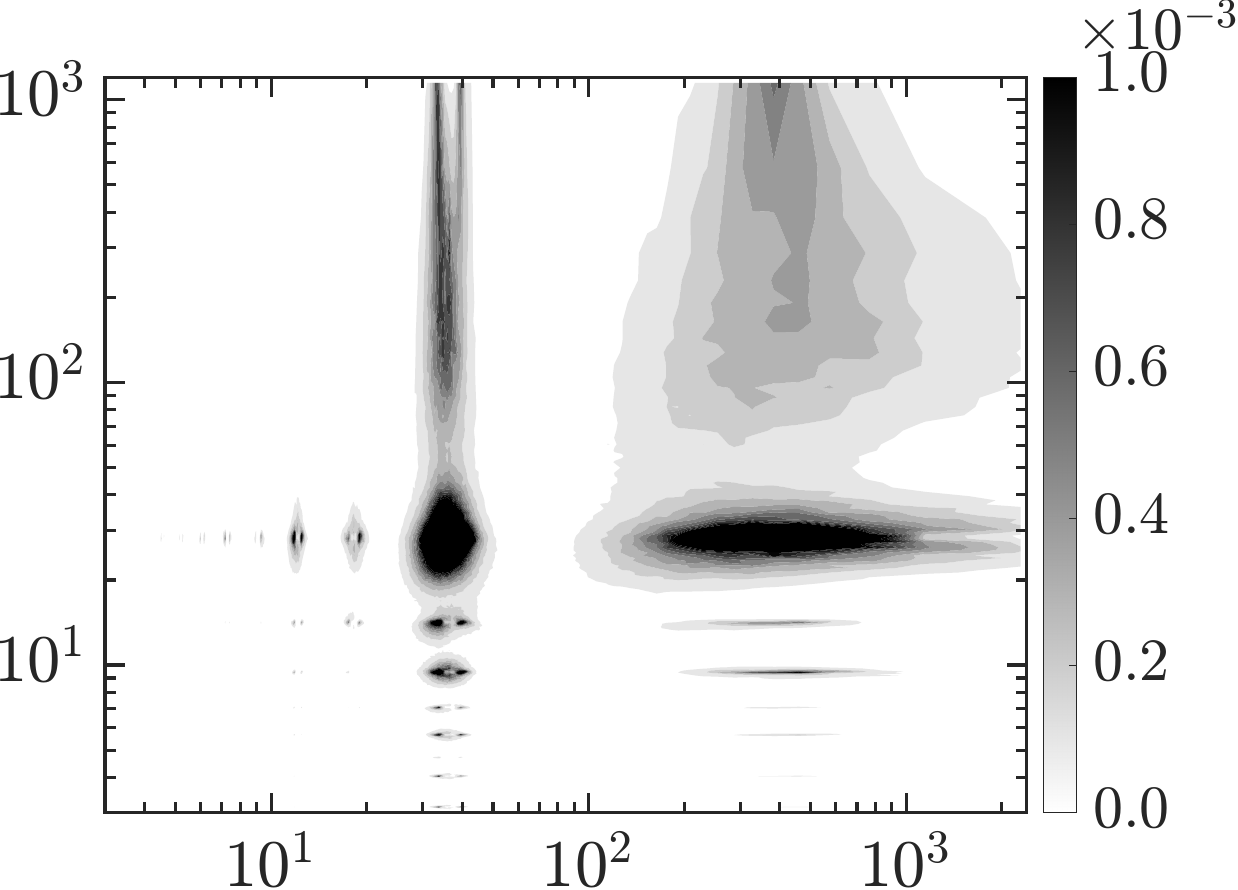}
    \end{subfigure}
    \hspace*{-8mm}
    \begin{subfigure}[tbp]{.38\textwidth}
        {\captionsetup{position=bottom, labelfont=it,textfont=it,size=scriptsize,singlelinecheck=false,justification=centering,labelformat=parens}
        \caption{$SP,\;$ $k_x\,k_z\,E_{uu}$}\label{fig:u_spectra_y_0:SP}}
        \vspace*{-0.8mm}
        \includegraphics[width=1\linewidth]{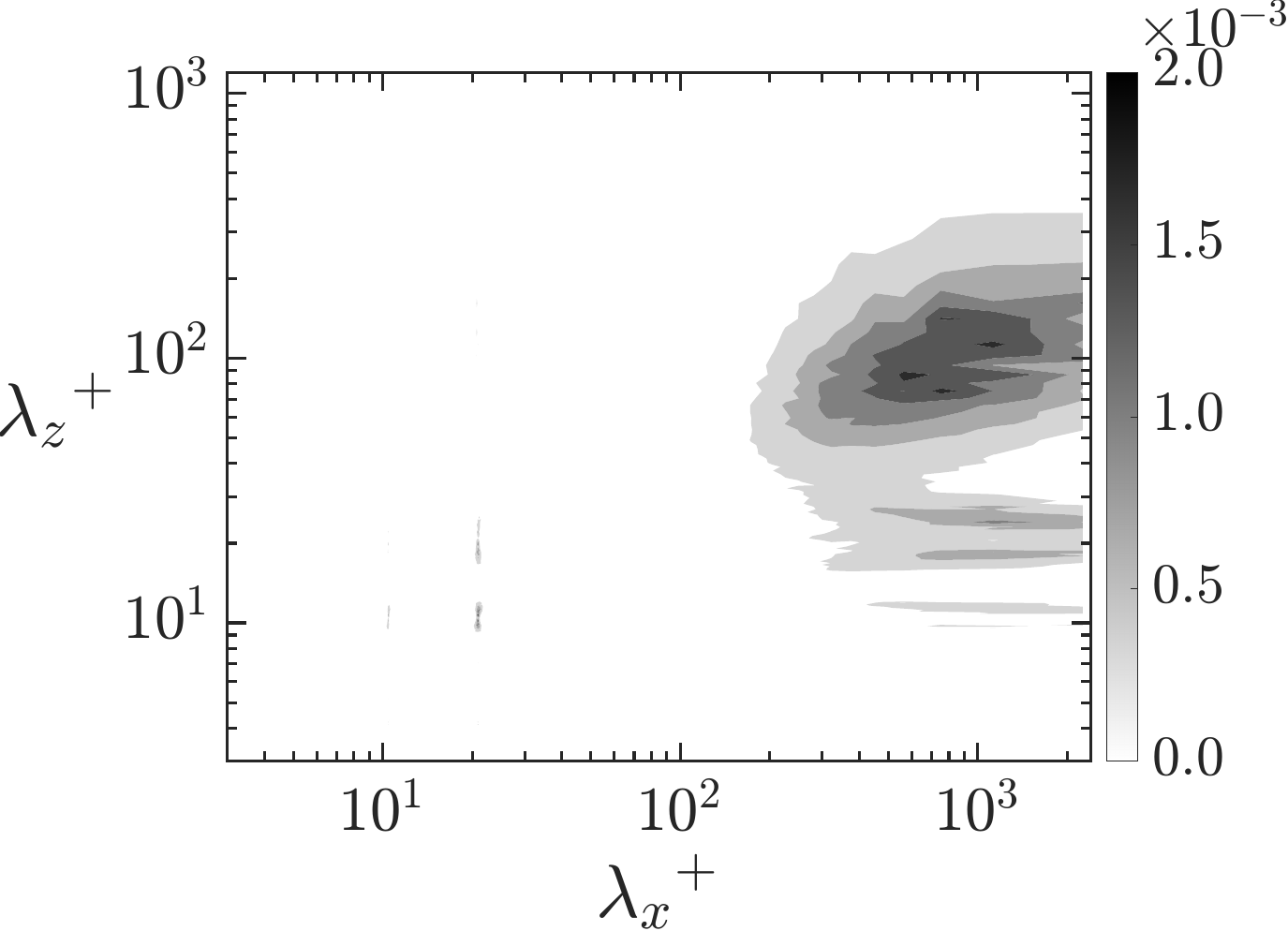}
    \end{subfigure}%
    \hspace*{2mm}
    \begin{subfigure}[tbp]{.34\textwidth}
        {\captionsetup{position=bottom, labelfont=it,textfont=it,size=scriptsize,singlelinecheck=false,justification=centering,labelformat=parens}
        \caption{$SP,\;$ $k_x\,k_z\,E_{vv}$}\label{fig:v_spectra_y_0:SP}}
        \vspace*{-0.8mm}
        \includegraphics[width=1\linewidth]{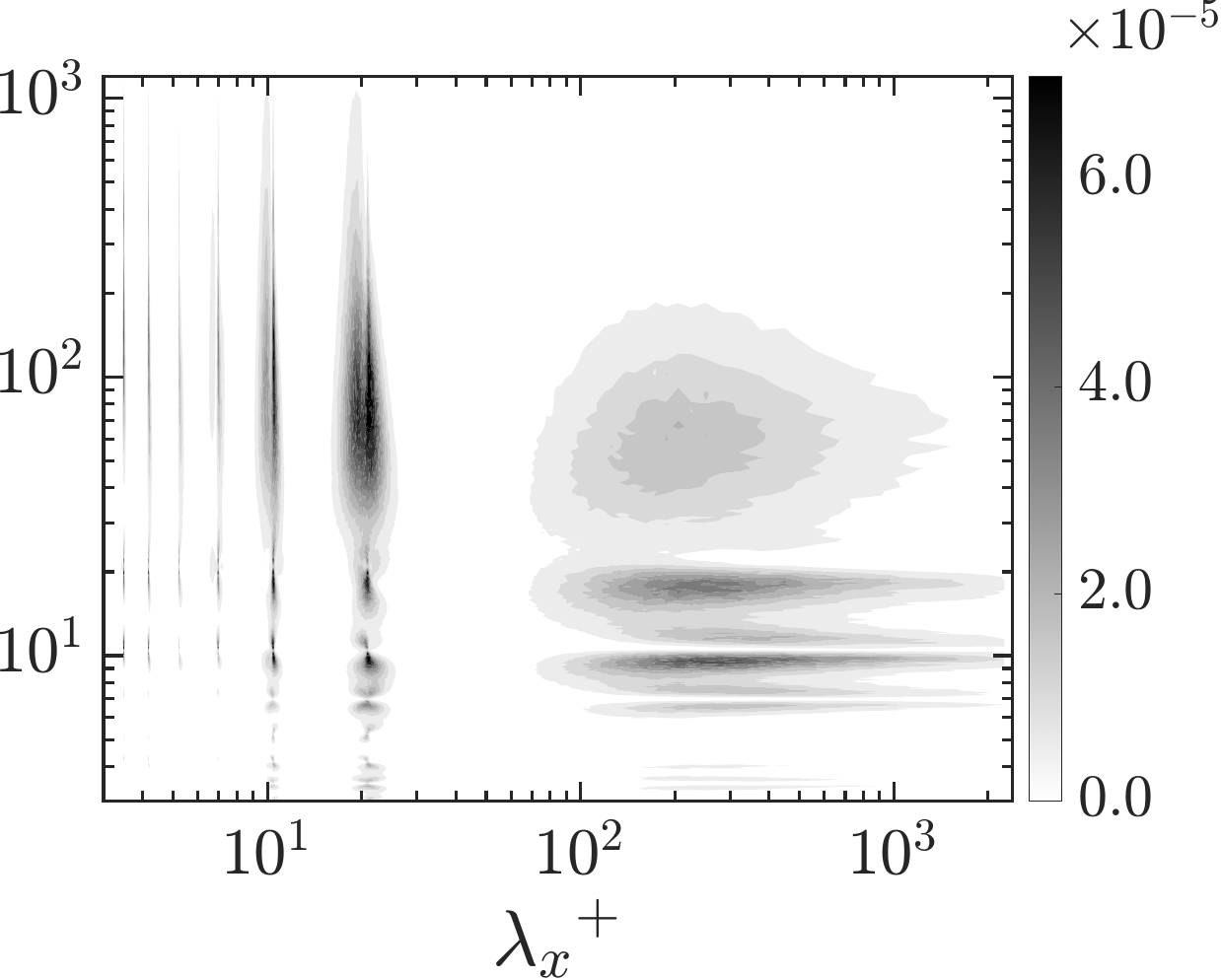}
    \end{subfigure}%
    \hspace*{2mm}
    \begin{subfigure}[tbp]{.34\textwidth}
        {\captionsetup{position=bottom, labelfont=it,textfont=it,size=scriptsize,singlelinecheck=false,justification=centering,labelformat=parens}
        \caption{$SP,\;$ $k_x\,k_z\,E_{uv}$}\label{fig:uv_spectra_y_0:SP}}
        \vspace*{-0.8mm}
        \includegraphics[width=1\linewidth]{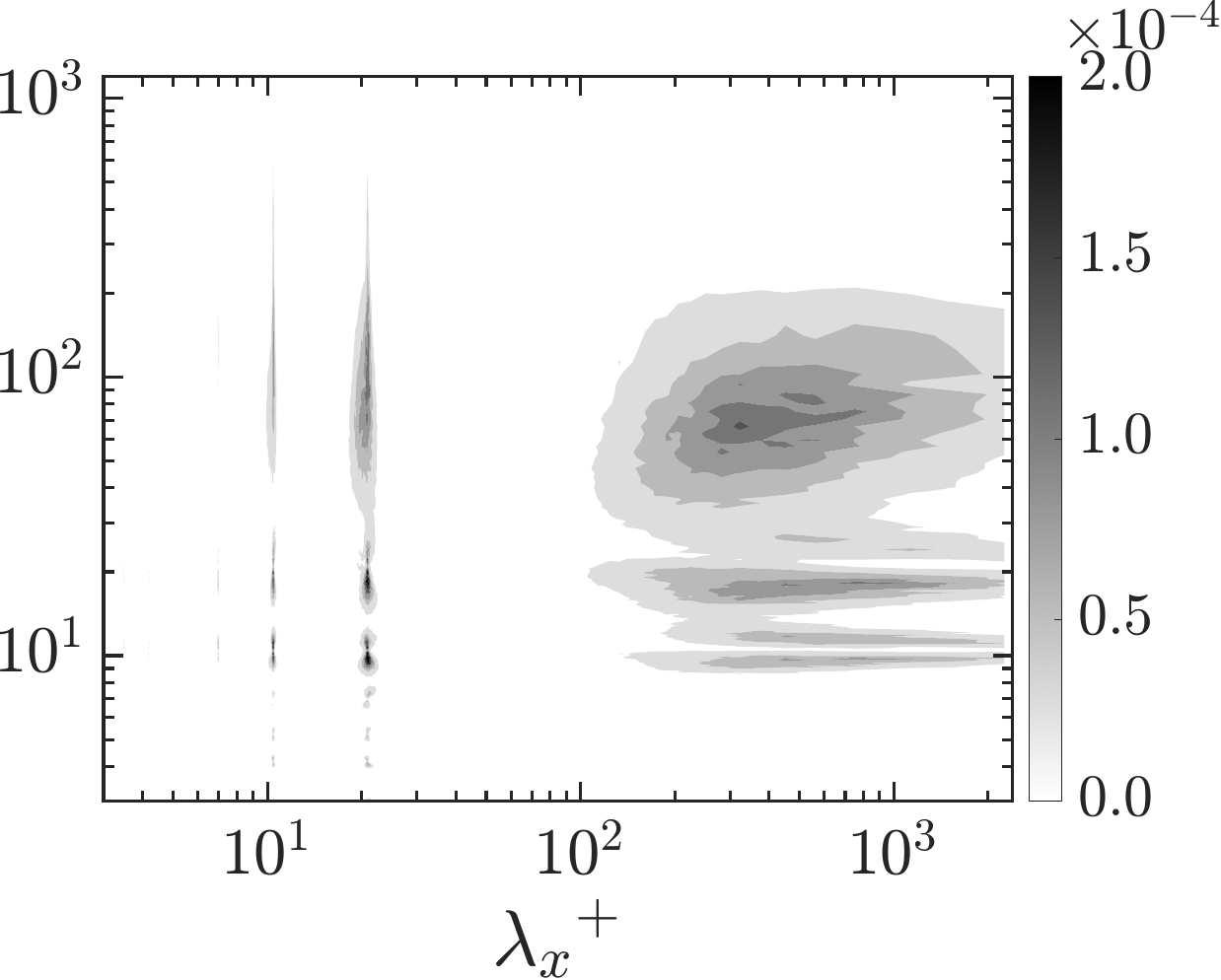}
    \end{subfigure}
    \vspace*{-1mm}
    \captionsetup{width=0.95\textwidth, justification=justified}
    \caption{Premultiplied spectral energies $k_x\,k_z\,E_{uu}$, $k_x\,k_z\,E_{vv}$, and $k_x\,k_z\,E_{uv}$ at $y=0$: (\emph{a-c}) $KP1^{\prime\prime}$; (\emph{d-f}) $KP1^{\prime}$; (\emph{g-i}) $KP1$; (\emph{j-l}) $SP$.}
    \label{fig:velocity_spectra_y_0}
 \end{center}
\end{figure}

 As a final assessment of the how the flow differs over the different walls, and to tie the observations made in this section to the description of porous-wall turbulent flows given in \cref{sec:porous_wall_flow}, the spectral energies of the velocity fluctuations at the surface ($y=0$) are examined in \cref{fig:velocity_spectra_y_0}. The smooth-wall-like $SP$ case clearly shows a different spectral signature across $u$, $v$ and $uv$ in figures \hyperref[fig:u_spectra_y_0:SP]{\ref*{fig:u_spectra_y_0:SP}}, \hyperref[fig:v_spectra_y_0:SP]{\ref*{fig:v_spectra_y_0:SP}}, and \hyperref[fig:uv_spectra_y_0:SP]{\ref*{fig:uv_spectra_y_0:SP}} compared to those of $KP1$ in figures \hyperref[fig:u_spectra_y_0:KP1]{\ref*{fig:u_spectra_y_0:KP1}}, \hyperref[fig:v_spectra_y_0:KP1]{\ref*{fig:v_spectra_y_0:KP1}}, and \hyperref[fig:uv_spectra_y_0:KP1]{\ref*{fig:uv_spectra_y_0:KP1}}. The latter exhibits energetic scales over large spanwise length-scales which are absent in $SP$. These scales belong to the K-H-like structures that were described in \cref{sec:porous_wall_flow}. Energetic regions belonging to the texture-coherent flow as described in \cref{fig:pore_coherent_flow_illustration} are visible for both $SP$ and $KP1$, but are more energetic in the latter case. By increasing the streamwise and spanwise pitches ($s_x^+, s_z^+$) of the pores in case $KP1^{\prime}$, these energetic regions move closer to the region belonging to the ambient turbulence and lead to direct interactions between them. Further increasing the pitches in case $KP1^{\prime\prime}$ results in significant scale overlap, which consequently causes the texture-coherent and ambient turbulence scales to no longer be clearly distinguishable from one another. The energy is distributed across a wide range of scales more evenly, although not uniformly. Cases $KP2$ and $KP3$ follow the same trend as $KP1$ with regards to the change in flow structure when going to their $KP\langle\rangle^{\prime}$ and $KP\langle\rangle^{\prime\prime}$ counterparts, which are not shown here for the sake of brevity.
%%%%%%%%%%%%%%%%%%%%%%% u contours at y+ = 0 %%%%%%%%%%%%%%%%%%%%%%%%%%%%%%%%%%%
%%%%%%%%%%%%%%%%%%%%%%%%%%%%%%%%%%%%%%%%%%%%%%%%%%%%%%%%%%%%%%%%%%%%%%%%%%%%%%%%
%%%%%%%%%%%%%%%%%%%%%%%%%%%%%%%%%%%%%%%%%%%%%%%%%%%%%%%%%%%%%%%%%%%%%%%%%%%%%%%%
%%%%%%%%%%%%%%%%%%%%%%%%%%%%%%%%%%%%%%%%%%%%%%%%%%%%%%%%%%%%%%%%%%%%%%%%%%%%%%%%
%%%%%%%%%%%%%%%%%%%%%%%%%%%%%%%%%%%%%%%%%%%%%%%%%%%% spectral energies of velocities at y+ = 0 %%%%%%%%%%%%%%%%%%%%%%%%%%%%%%%%%%%%%%%%%%%%%%%%%%%%%%%%%%%%%%%%%%%%%%%%
%%%%%%%%%%%%%%%%%%%%%%%%%%%%%%%%%%%%%%%%%%%%%%%%%%%%%%%%%%%%%%%%%%%%%%%%%%%%%%%%%%%%%%%%%%%%%%%%%%%%%%%%%%%%%%%%%%%%%%%%%%%%%%%%%%%%%%%%%%%%%%%%%%%%%%

 Having established the differences in flow structure by examining the velocity fields both statistically and spectrally, the temperature field and heat transfer of the cases can now be examined. It is once again emphasized that the content of this section was merely to serve as a basis for the heat transfer analysis and not to be an exhaustive analysis of porous-wall turbulence itself. For more thorough details regarding such flows, the references provided in \cref{sec:porous_wall_flow} can be referred to.

\subsection{Temperature field statistics}\label{subsec:heat_stats}

 \begin{figure}
    \begin{center}
    \hspace*{-10pt}
    \begin{subfigure}{.45\textwidth}
        {\captionsetup{labelfont=it,textfont=normalfont,singlelinecheck=false,justification=raggedright,labelformat=parens}\caption{}\label{fig:mean_temp_channel}}%
        \includegraphics[width=1\linewidth]{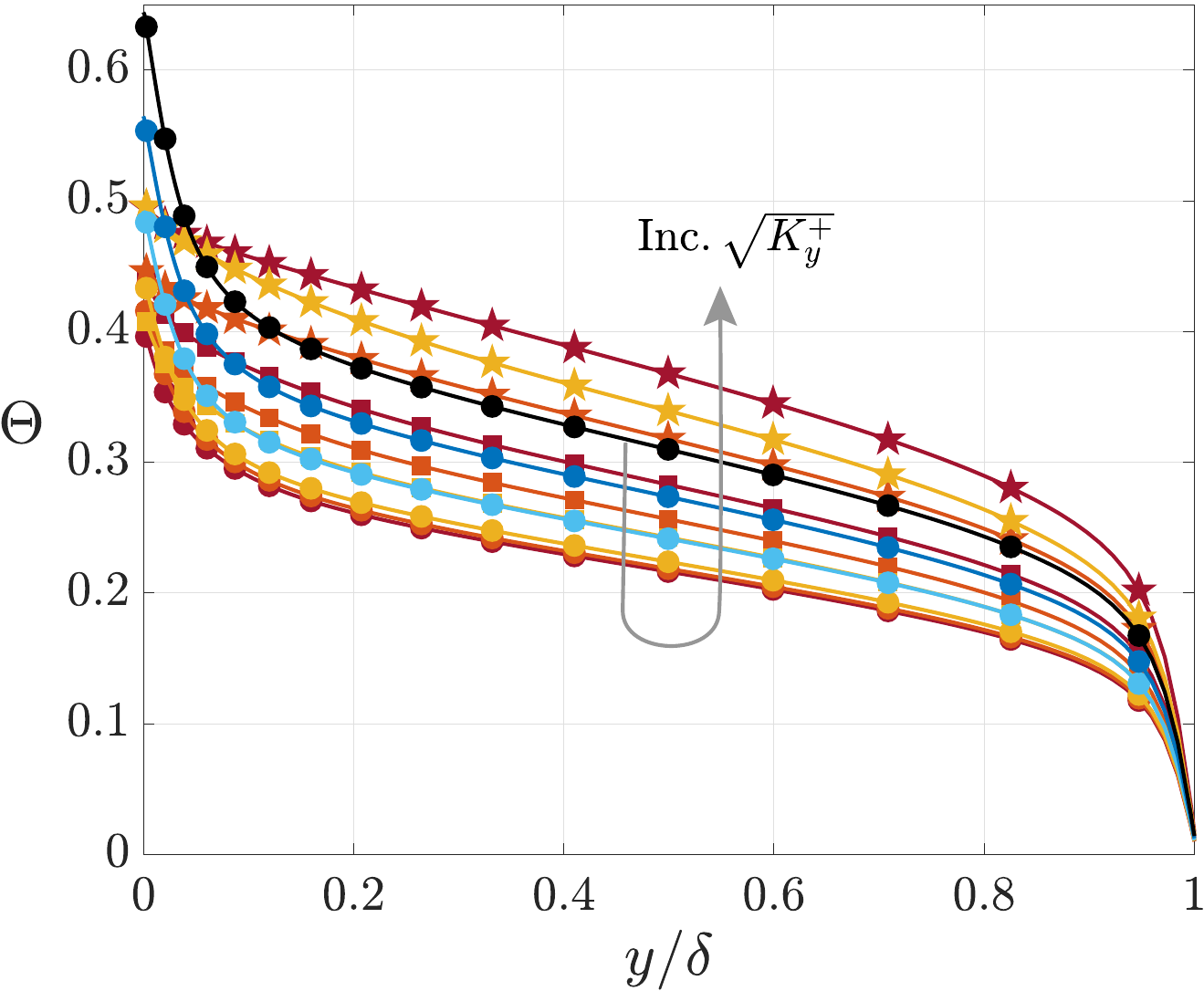}
    \end{subfigure}%
    \hspace*{10pt}
    \begin{subfigure}{.45\textwidth}
        {\captionsetup{labelfont=it,textfont=normalfont,singlelinecheck=false,justification=raggedright,labelformat=parens}\caption{}\label{fig:mean_temp_porous}}%
        \includegraphics[width=1\linewidth]{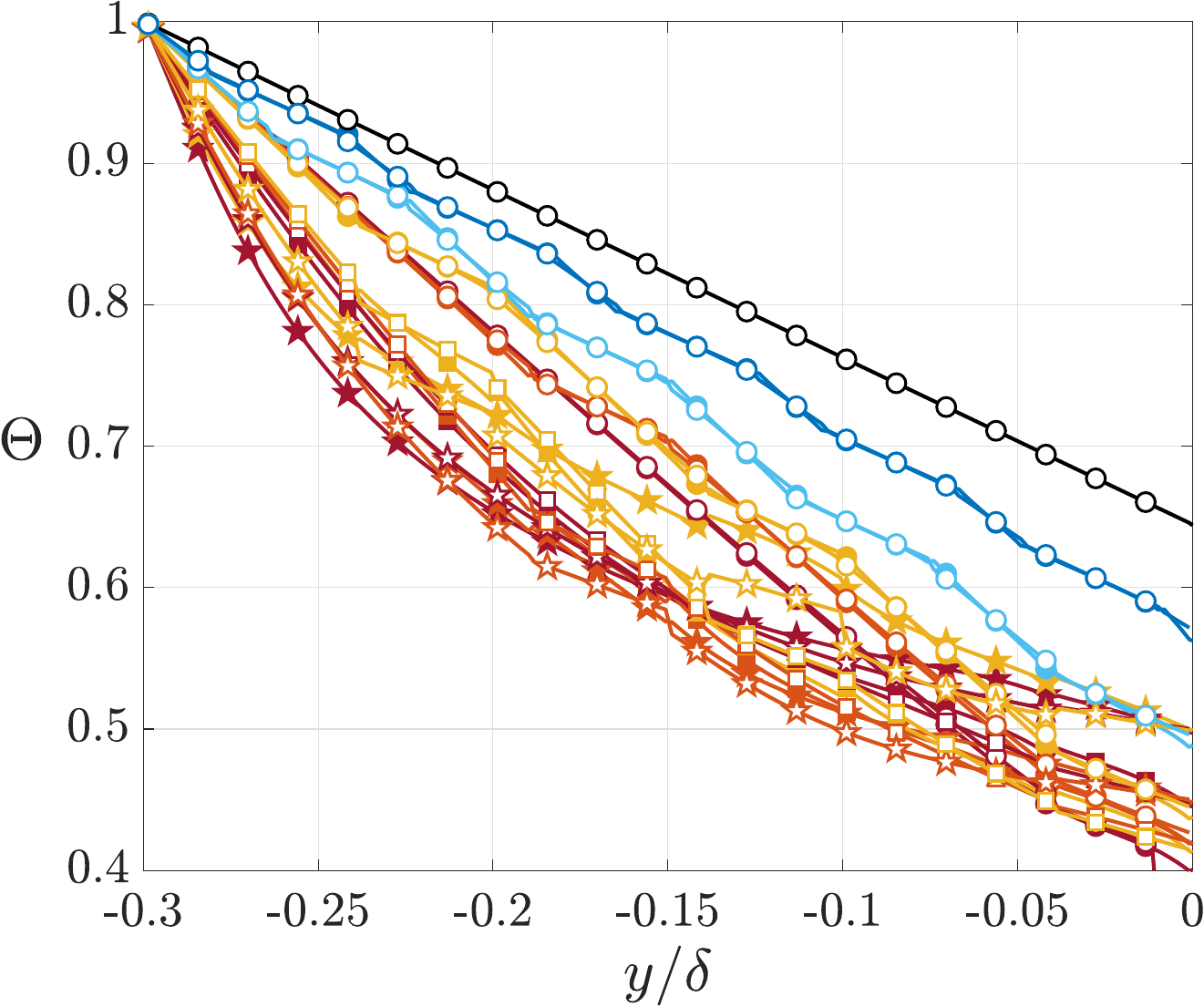}
    \end{subfigure}
    \captionsetup{width=0.95\textwidth, justification=justified}
    \caption{Mean temperature profiles in (\emph{a}) channel region and (\emph{b}) porous wall region. Filled symbols are the fluid-phase-averaged temperature and open symbols are the solid-phase-averaged temperature. The direction of the arrow in (\emph{a}) indicates increasing wall-normal permeability.}
    \label{fig:mean_temperature_profiles}
    \end{center}
 \end{figure}

 We start by examining the mean temperature profiles in \cref{fig:mean_temperature_profiles}. Initially, as the wall is made porous, the overall temperature throughout the channel region (\hyperref[fig:mean_temp_channel]{figure \ref*{fig:mean_temp_channel}}) reduces compared to case $ZP$ where the wall is an impenetrable solid block. This initial cooling down of the fluid temperature suggests that the rate at which heat gets extracted from the wall becomes diminished, and is likely due to the reduction in the overall conductivity of the wall as it is made permeable and high conductivity solid material becomes replaced with lower conductivity fluid material. This trend of temperature reduction persists until the $KP\langle\rangle^{\prime}$ cases, after which the temperature in the channel region begins to rise as the wall-normal permeability, $\sqrt{K_{y}^+}$, becomes larger. This rise in temperature continues until cases $KP1^{\prime\prime}$, $KP2^{\prime\prime}$, and $KP3^{\prime\prime}$, where it exceeds that of $ZP$. These are the cases with the highest wall-normal permeabilities ($\sqrt{K_{y}^+}>10$) and where the texture-coherent and ambient turbulence scales overlap, as was shown earlier in \cref{fig:velocity_spectra_y_0}. Whether or not these are the only cases that demonstrate an improvement in net heat transfer will be conclusively assessed later by measuring the heat flux.
 Other notable changes in the temperature distribution is the diminishment of their diffusive quality in the vicinity of the surface as $\sqrt{K_{y}^+}$ becomes larger. This causes the overall profile to become more uniform and can be attributed to the intensified turbulent mixing occurring in the vicinity of the porous wall, as previously seen in \cref{fig:Reynolds_stress_channel} for the Reynolds shear stresses, and which reduces the difference between the fluid temperature close to the wall and away from it.
 
 Moving into the porous region in \cref{fig:mean_temp_porous}, for the impermeable case of $ZP$ a linear variation of the temperature profile with respect to the $y$ coordinate is observed since heat is transferred purely through conduction. As the wall is made more permeable, it becomes cooler relative to the purely conductive $ZP$ case, and the temperature distribution gradually develops a non-linear quality. Unlike in the channel region, the cooling trend in the wall region is consistent. For the highly permeable $KP\langle\rangle^{\prime\prime}$ walls, the temperature distribution takes on a slightly more complex quality. In the shallow part of these walls, the temperature smoothly rises from its value above the surface, whereas for the other cases there is a rapid temperature rise in the surface vicinity.
 Note that the temperature distribution in both the solid and fluid phases are very similar, with some differences being visible for the $KP\langle\rangle^{\prime\prime}$ walls, but they are not significant.

 \begin{figure}
    \begin{center}
    \begin{subfigure}{.35\textwidth}
        {\captionsetup{labelfont=it,textfont=normalfont,singlelinecheck=false,justification=raggedright,labelformat=parens}\caption{}\label{fig:rms_temp_channel}}%
        \hspace{-24pt}\includegraphics[width=1\linewidth]{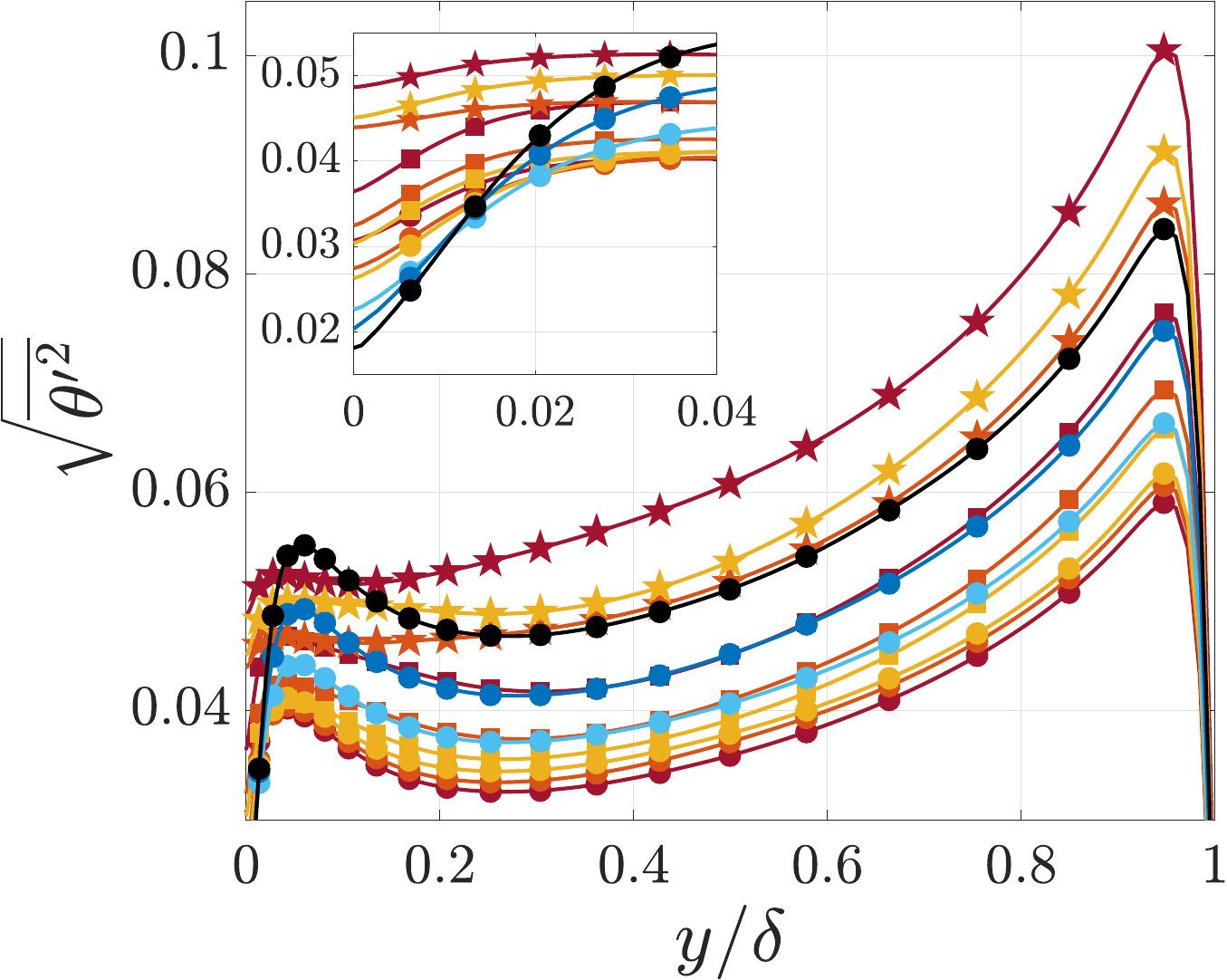}
    \end{subfigure}%
    \begin{subfigure}{.35\textwidth}
        {\captionsetup{labelfont=it,textfont=normalfont,singlelinecheck=false,justification=raggedright,labelformat=parens}\caption{}\label{fig:rms_temp_porous_fluid}}%
        \hspace{-18pt}\includegraphics[width=1\linewidth]{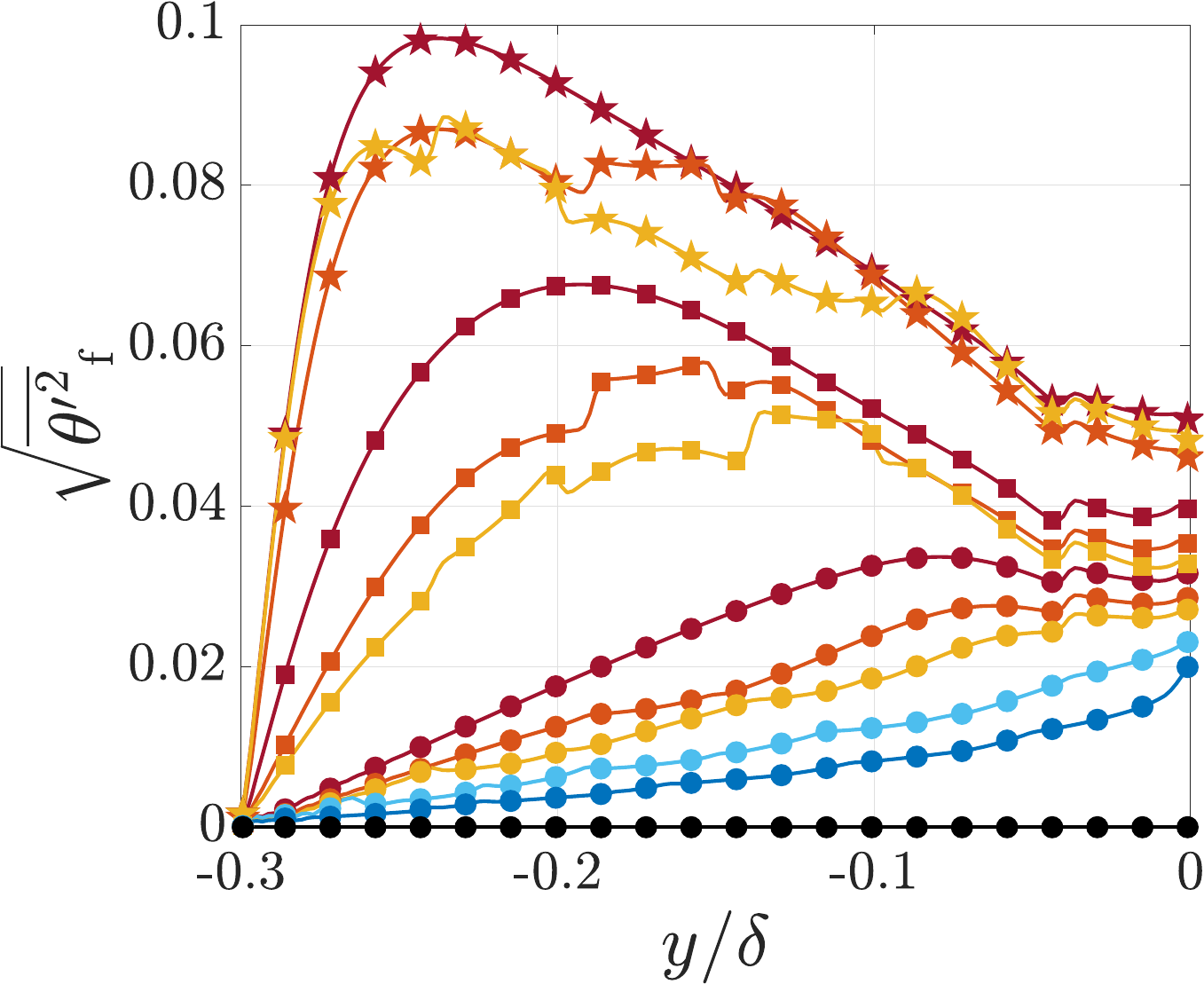}
    \end{subfigure}%
    \begin{subfigure}{.35\textwidth}
        {\captionsetup{labelfont=it,textfont=normalfont,singlelinecheck=false,justification=raggedright,labelformat=parens}\caption{}\label{fig:rms_temp_porous_solid}}%
        \hspace{-12pt}\includegraphics[width=1\linewidth]{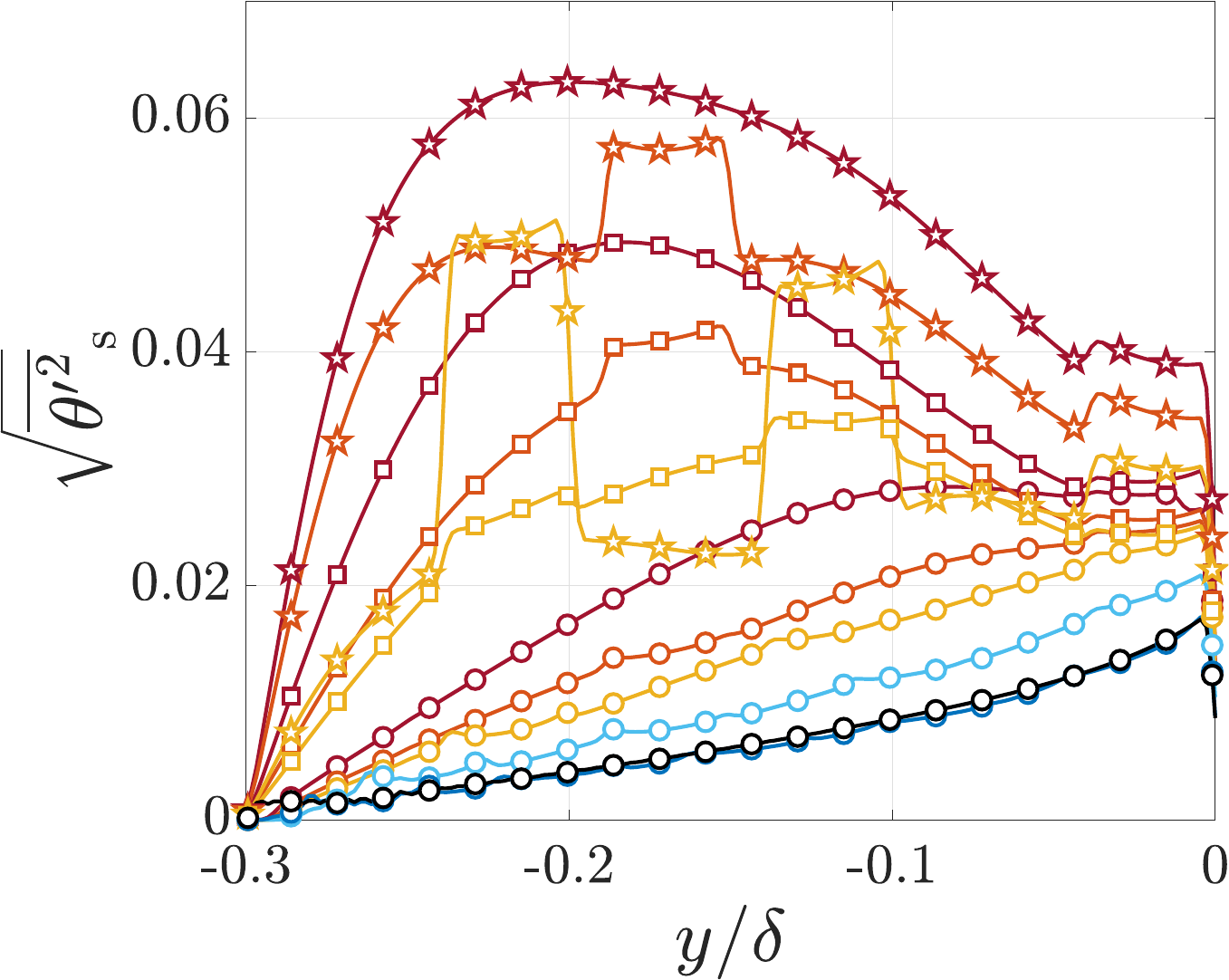}
    \end{subfigure}
    \captionsetup{width=0.95\textwidth, justification=justified}
    \caption{root-mean-square temperature fluctuation profiles: (\emph{a}) profiles in the channel region; (\emph{b}) fluid-phase-averaged profiles in the porous region; (\emph{c}) solid-phase-averaged profiles in the porous region.}
    \label{fig:rms_profiles}
    \end{center}
 \end{figure}

 \Cref{fig:rms_profiles} shows the r.m.s. temperature fluctuations throughout the domain. In the channel region (\hyperref[fig:rms_temp_channel]{figure \ref*{fig:rms_temp_channel}}), the changes in temperature fluctuations mirror those of the mean velocity profiles of \cref{fig:mean_temp_channel}. The intensity of the fluctuations reduces when going from $ZP$ to $KP3$, then $KP2$, and finally $KP1$. The fluctuations then subsequently increase when going from $KP1$ to $KP1^{\prime\prime}$ in rising order of $\sqrt{K_{y}^+}$ ($KP1 \rightarrow KP3^\prime \rightarrow KP2^\prime \rightarrow KP1^\prime \rightarrow KP3^{\prime\prime} \rightarrow KP2^{\prime\prime} \rightarrow KP1^{\prime\prime}$), with only the $KP\langle\rangle^{\prime\prime}$ cases having stronger fluctuations than $ZP$. The region immediately above the surface is an exception. There, the intensity of the fluctuations has a consistent increasing trend as shown in the inset plot of \cref{fig:rms_temp_channel}. The near-surface peak also moves closer to the surface as the wall is made more permeable. The increased turbulent flux of heat at the surface causes this, as it leads to a greater production of temperature variance \citep{leonardi_orlandi_djenidi_antonia_2015}. Case $ZP$ however registers the strongest near-surface peak, which is attributable to the strong temperature gradient of its mean temperature profile in this region.

 Figures \hyperref[fig:rms_temp_porous_fluid]{\ref*{fig:rms_temp_porous_fluid}} and \hyperref[fig:rms_temp_porous_solid]{\ref*{fig:rms_temp_porous_solid}} show a greater distinction in the r.m.s. temperature fluctuations of the $KP\langle\rangle^{\prime}$ and $KP\langle\rangle^{\prime\prime}$ cases from the others in the wall region. To begin with, the fluid-phase-averaged profiles in \cref{fig:rms_temp_porous_fluid} show a steady increase in fluctuation intensity as the wall grows more preamble ($ZP \rightarrow KP1$, $\sqrt{K_{y}^+}=0 \rightarrow \sqrt{K_{y}^+}\approx4$), with a peak beginning to emerge in the shallow part of the wall above $y=-0.1$. The fluctuations show an overall steady decay toward the bottom of the wall and are skewed toward the surface.
 Moving to cases of higher permeability ($KP3^{\prime} \rightarrow KP2^{\prime} \rightarrow KP1^{\prime}$, $\sqrt{K_{y}^+}\approx5.8 \rightarrow \sqrt{K_{y}^+}\approx6.5 \rightarrow \sqrt{K_{y}^+}=8.2$), the intensity of the fluctuations increases notably and the peak moves toward the lower\nobreakdash-half of the wall. Consequently, the fluctuations become skewed toward the bottom and grow stronger further down into the wall. This trend persists for the highest permeability cases ($KP3^{\prime\prime} \rightarrow KP2^{\prime\prime} \rightarrow KP1^{\prime\prime}$, $\sqrt{K_{y}^+}\approx12.7 \rightarrow \sqrt{K_{y}^+}\approx15.3 \rightarrow \sqrt{K_{y}^+}\approx18.9$), with the fluctuation intensity growing steadily towards the bottom of the wall, where it first attains a peak and then subsequently decays to zero due to the imposed temperature boundary condition. For these latter cases, the position of the peak is $y^+\approx-0.25$, placing it very close to the bottom of the wall. Also notable, is that for $KP1^{\prime\prime}$, the peak within the wall is almost equal to its peak in the channel region.
 
 The observations made for the fluid-phase r.m.s. profiles in \cref{fig:rms_temp_porous_fluid} also hold for the solid-phase profiles in \cref{fig:rms_temp_porous_solid}, with the difference being an overall lower fluctuation intensity than those in the fluid-phase, but which are otherwise similar. Notable jumps occur in the profile of $KP3^{\prime\prime}$ as it passes from regions with less solid to regions with more solid where the conductive capacity of transferring heat becomes higher. While these jumps are also visible in some of the other profiles, specifically those belonging to the higher permeability walls, they are not as pronounced as in $KP3^{\prime\prime}$. This could be a consequence of the higher anisotropy in $KP3^{\prime\prime}$ compared to the other cases (${\Phi_{xy}}\approx0.6$).

%%%%%%%%%%%%%%%%%%%%%%% u contours at y+ = 0 %%%%%%%%%%%%%%%%%%%%%%%%%%%%%%%%%%%
%%%%%%%%%%%%%%%%%%%%%%%%%%%%%%%%%%%%%%%%%%%%%%%%%%%%%%%%%%%%%%%%%%%%%%%%%%%%%%%%
\begin{figure}
\vspace*{-20mm}
 \begin{center}
    \hspace*{-13mm}
    \begin{subfigure}[tbp]{.40\textwidth}
        {\captionsetup{position=bottom, labelfont=it, textfont=it, singlelinecheck=false, justification=raggedright, labelformat=parens}
        \caption{KP1}\label{fig:tmp_velocity_contour_y_0:KP1}}
        \vspace*{-1mm}
        \includegraphics[width=1\linewidth]{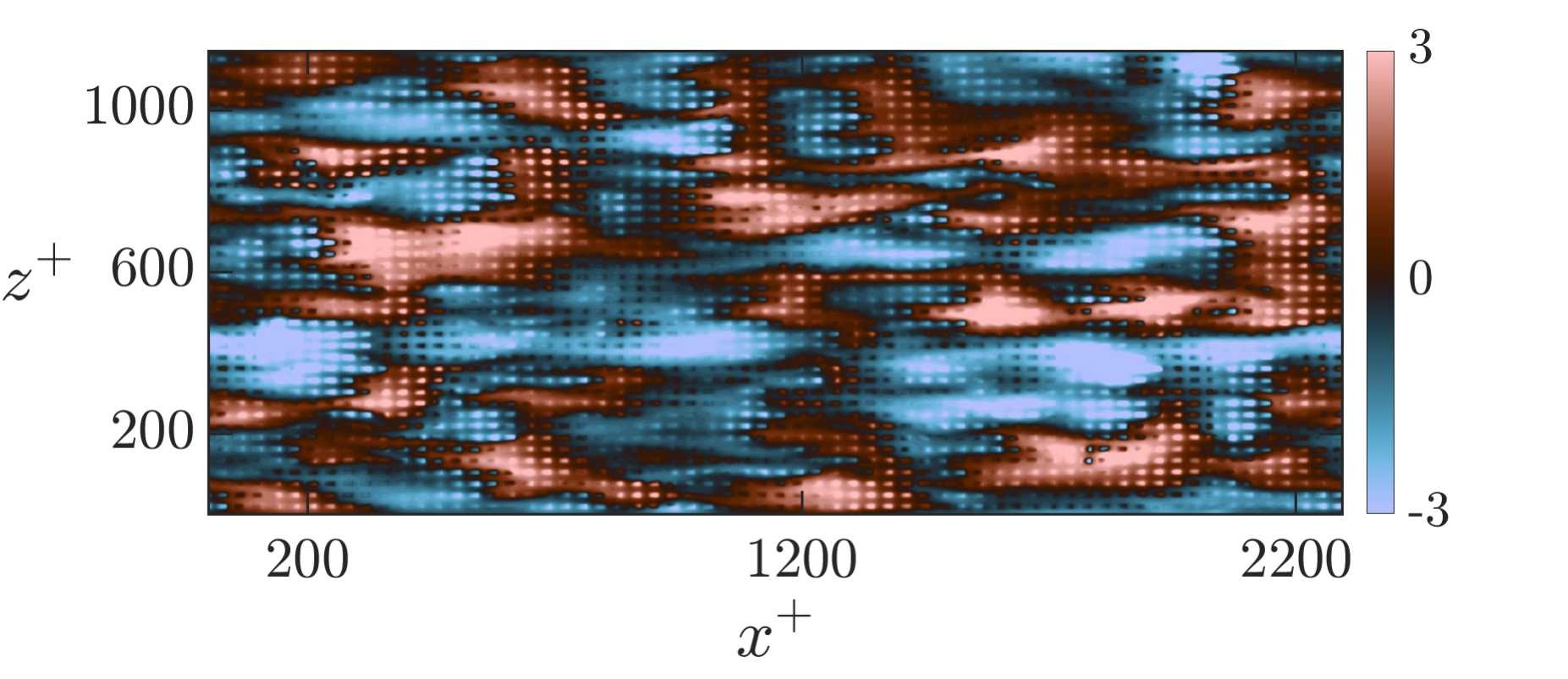}
    \end{subfigure}%
    \hspace*{-2mm}
    \begin{subfigure}[tbp]{.40\textwidth}
        {\captionsetup{position=bottom, labelfont=it,textfont=it,singlelinecheck=false,justification=raggedright,labelformat=parens}
        \caption{$KP1^\prime$}\label{fig:tmp_velocity_contour_y_0:KP1-1}}
        \vspace*{-1mm}
        \includegraphics[width=1\linewidth]{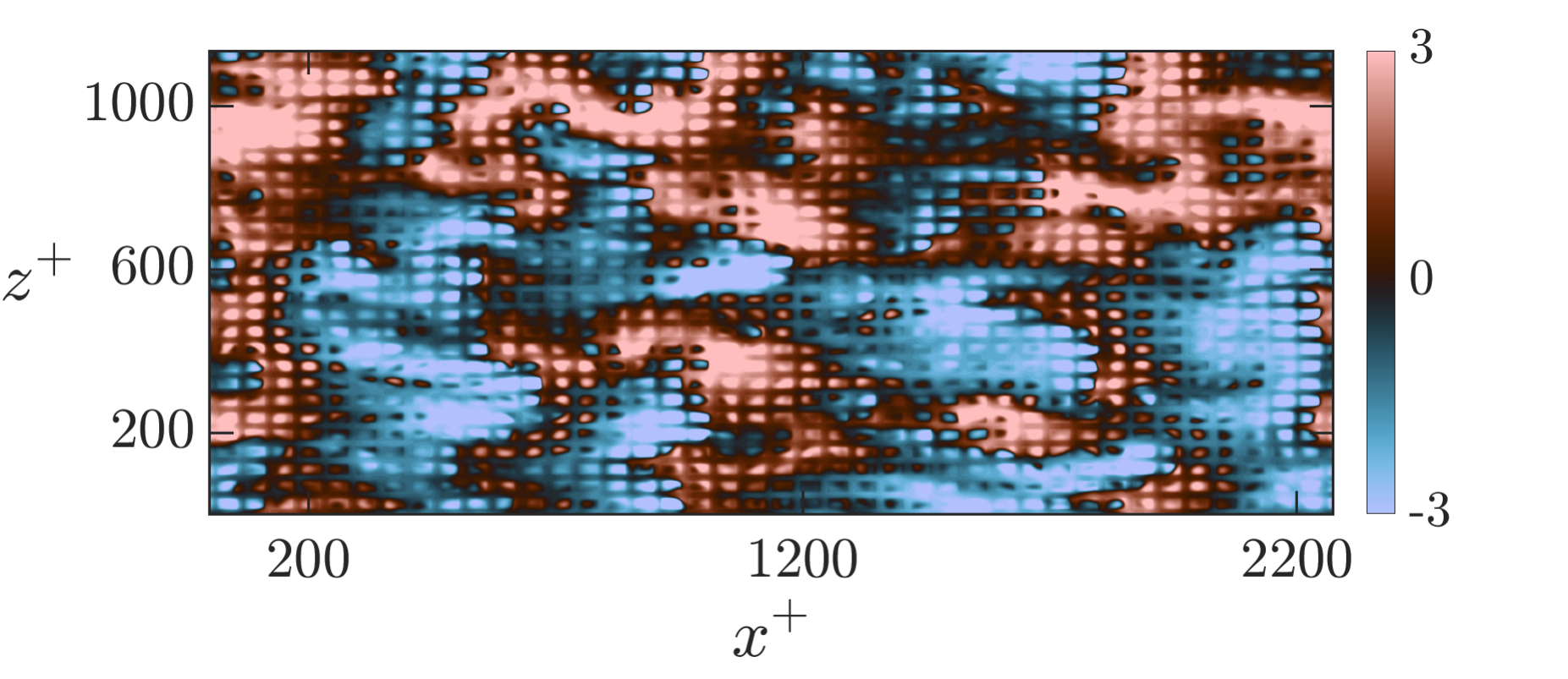}
    \end{subfigure}%
    \hspace*{-2mm}
    \begin{subfigure}[tbp]{.40\textwidth}
        {\captionsetup{position=bottom, labelfont=it,textfont=it,singlelinecheck=false,justification=raggedright,labelformat=parens}
        \caption{$KP1^{\prime\prime}$}\label{fig:tmp_velocity_contour_y_0:KP1-2}}
        \vspace*{-1mm}
        \includegraphics[width=1\linewidth]{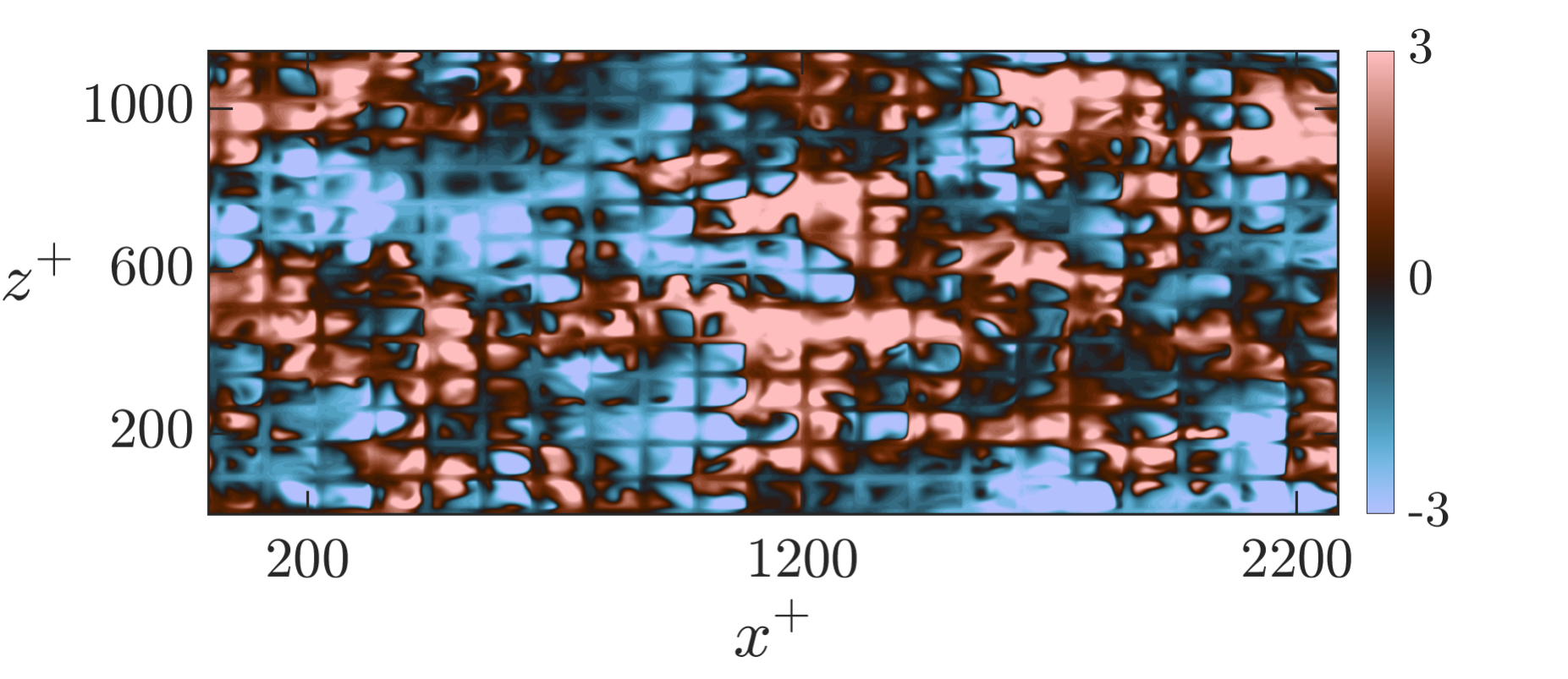}
    \end{subfigure}
    \hspace*{0mm}
    \begin{subfigure}[tbp]{.40\textwidth}
        {\captionsetup{position=bottom, labelfont=it,textfont=it,singlelinecheck=false,justification=raggedright,labelformat=parens}
        \caption{SP}\label{fig:tmp_velocity_contour_y_0:SP}}
        \vspace*{-0.8mm}
        \includegraphics[width=1\linewidth]{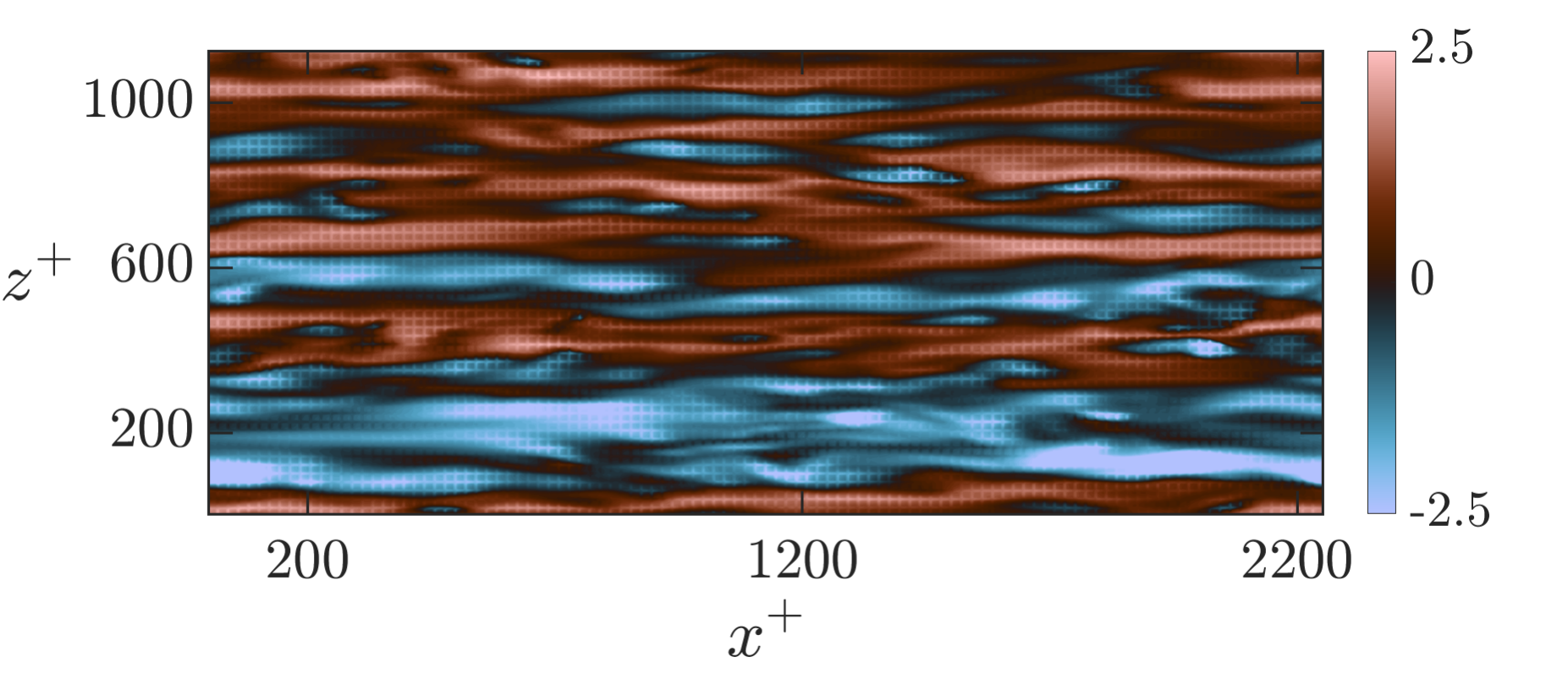}
    \end{subfigure}%
    \hspace*{-1mm}
    \begin{subfigure}[tbp]{.40\textwidth}
        {\captionsetup{position=bottom, labelfont=it,textfont=it,singlelinecheck=false,justification=raggedright,labelformat=parens}
        \caption{TP}\label{fig:tmp_velocity_contour_y_0:TP}}
        \vspace*{-0.8mm}
        \includegraphics[width=1\linewidth]{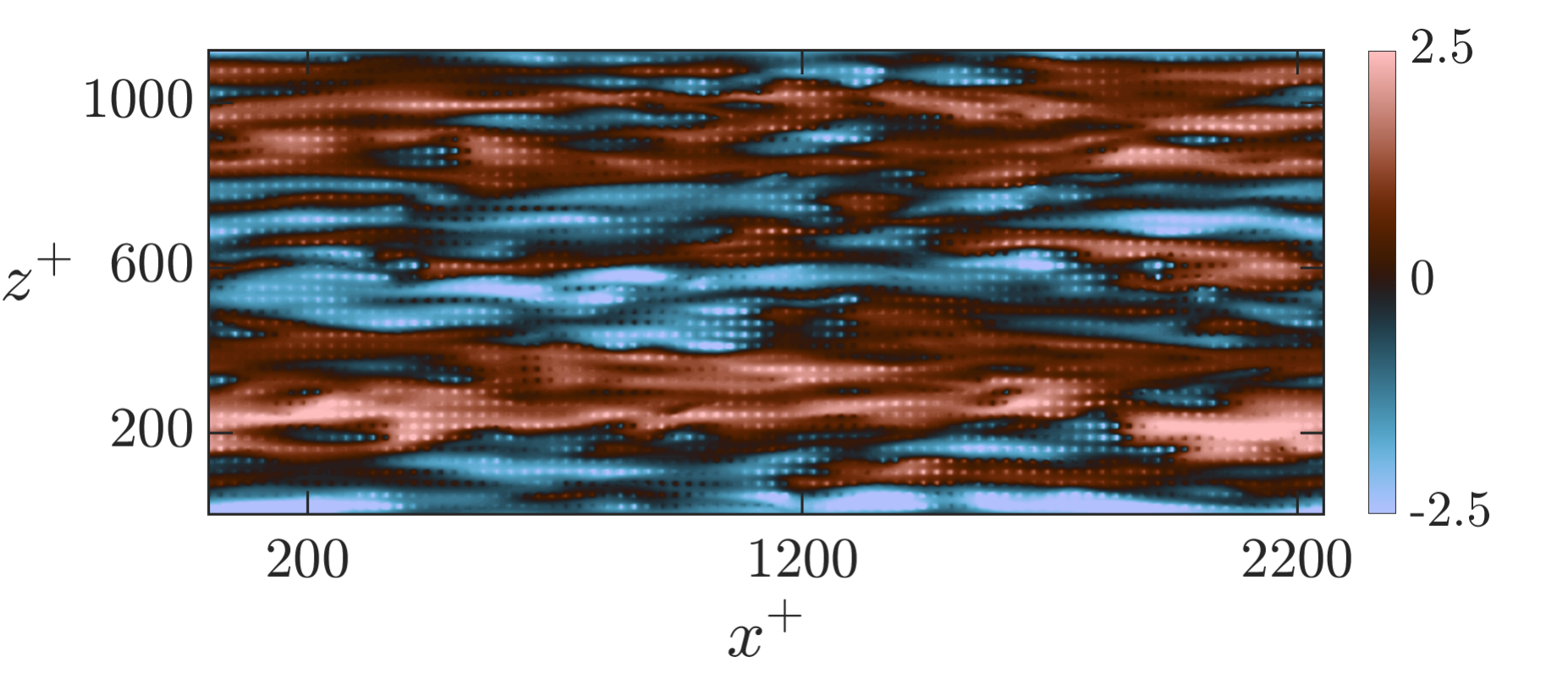}
    \end{subfigure}
    \vspace*{-2mm}
    \captionsetup{width=1\textwidth, justification=justified}
    \caption{Instantaneous snapshots of temperature fluctuations at $y^+=0$ for select cases of \cref{tab:DNS}.}
    \label{fig:tmp_velocity_contours_y_0}
 \end{center}
\end{figure}
%%%%%%%%%%%%%%%%%%%%%%%%%%%%%%%%%%%%%%%%%%%%%%%%%%%%  temeprature spectra at y = 0 %%%%%%%%%%%%%%%%%%%%%%%%%%%%%%%%%%%%%%%%%%%%%%%%%%%%%%%%%%%%%%%%%%%%%%%%
\begin{figure}
 \begin{center}
    \hspace*{-4mm}
    \begin{subfigure}[tbp]{.38\textwidth}
        {\captionsetup{position=bottom, labelfont=it,textfont=it,size=scriptsize,singlelinecheck=false,justification=centering,labelformat=parens}
        \caption{$KP1,\;$ $k_x\,k_z\,E_{\theta \theta}$}\label{fig:tmp_spectra_y_0:KP1}}
        \vspace*{-0.8mm}
        \includegraphics[width=1\linewidth]{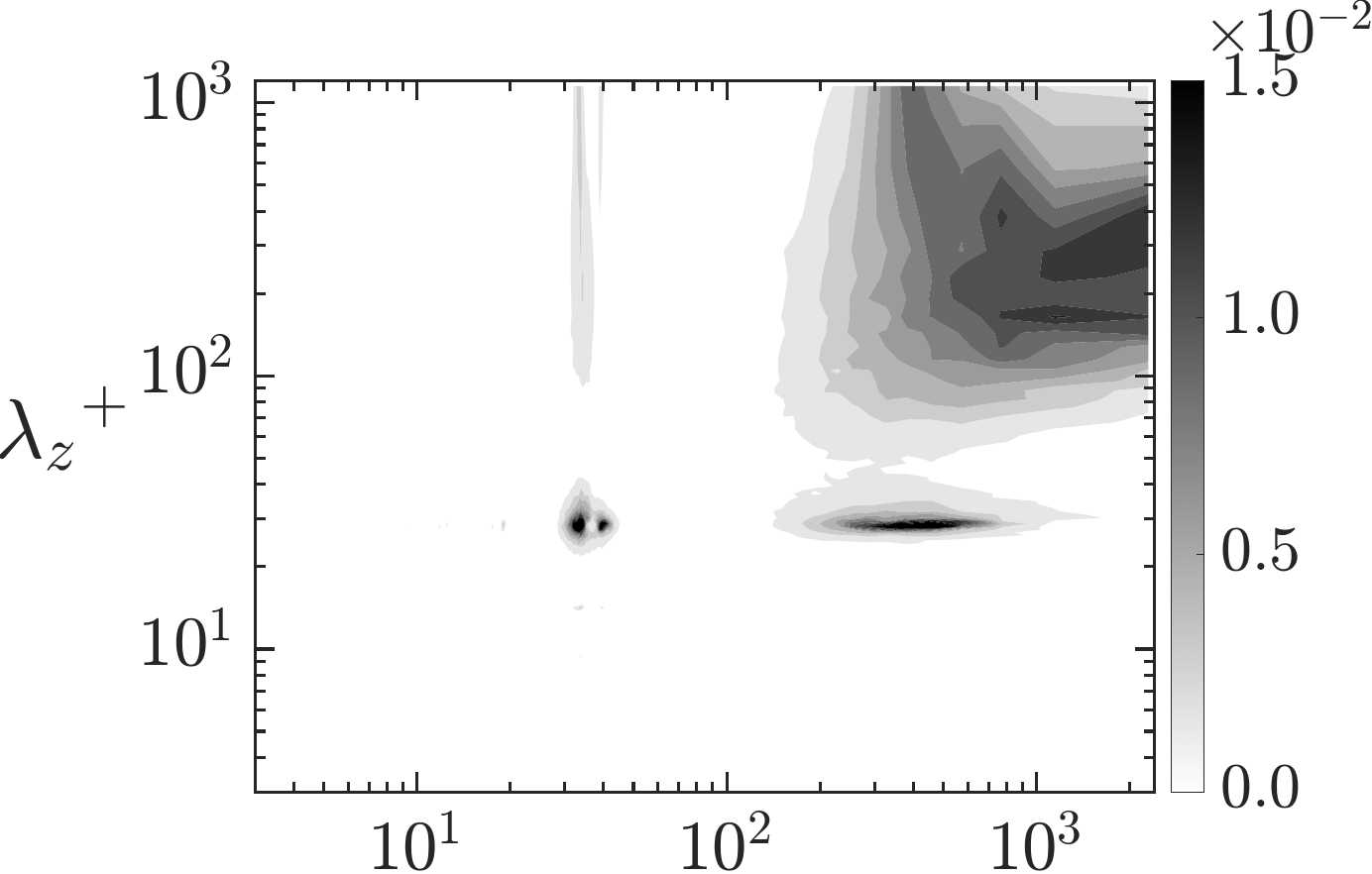}
    \end{subfigure}%
    \hspace*{0mm}
    \begin{subfigure}[tbp]{.34\textwidth}
        {\captionsetup{position=bottom, labelfont=it,textfont=it,size=scriptsize,singlelinecheck=false,justification=centering,labelformat=parens}
        \caption{$KP1^{\prime},\;$ $k_x\,k_z\,E_{\theta \theta}$}\label{fig:tmp_spectra_y_0:KP1-1}}
        \vspace*{-0.8mm}
        \includegraphics[width=1\linewidth]{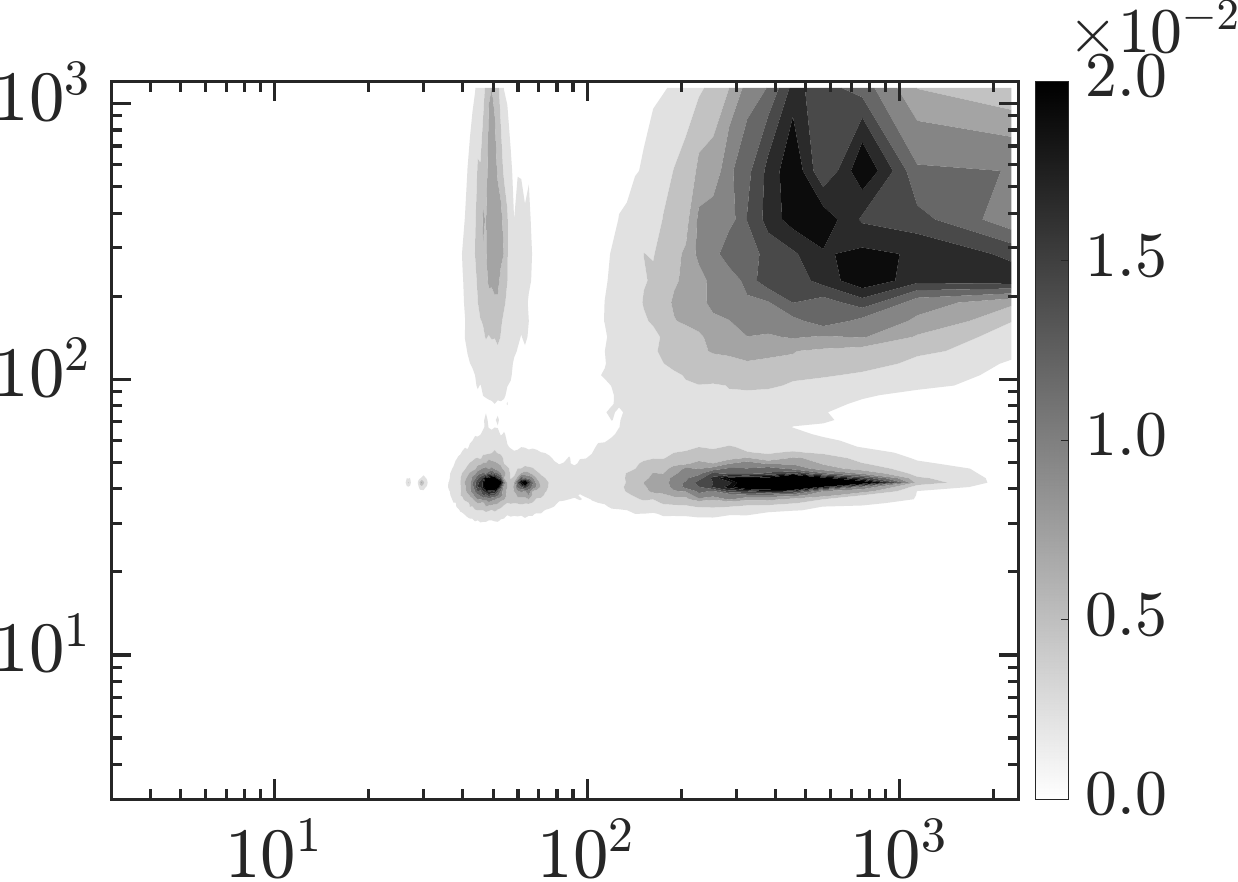}
    \end{subfigure}%
    \hspace*{0mm}
    \begin{subfigure}[tbp]{.34\textwidth}
        {\captionsetup{position=bottom, labelfont=it,textfont=it,size=scriptsize,singlelinecheck=false,justification=centering,labelformat=parens}
        \caption{$KP1^{\prime\prime},\;$ $k_x\,k_z\,E_{\theta \theta}$}\label{fig:tmp_spectra_y_0:KP1-2}}
        \vspace*{-0.8mm}
        \includegraphics[width=1\linewidth]{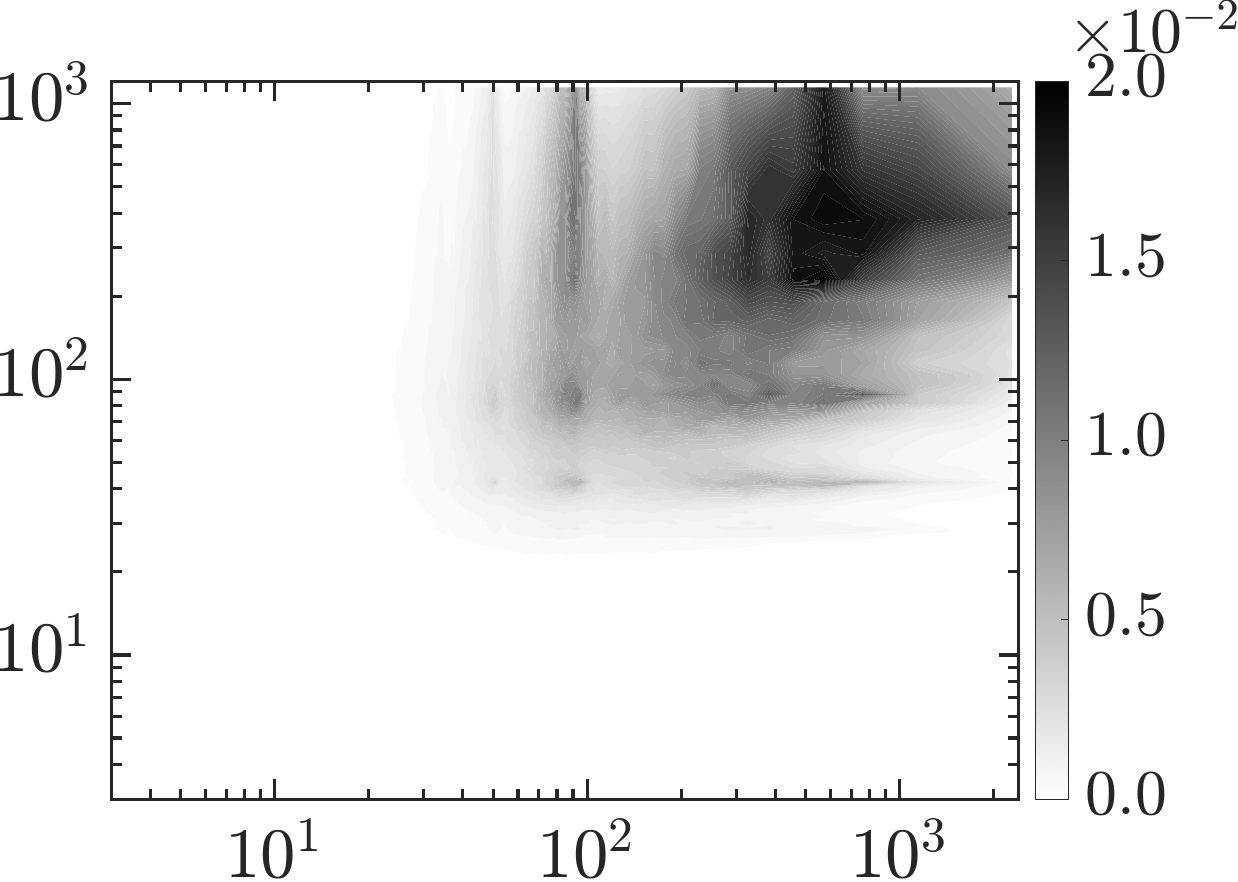}
    \end{subfigure}
    \hspace*{-4mm}
    \begin{subfigure}[tbp]{.38\textwidth}
        {\captionsetup{position=bottom, labelfont=it,textfont=it,size=scriptsize,singlelinecheck=false,justification=centering,labelformat=parens}
        \caption{$KP1,\;$ $k_x\,k_z\,E_{\theta v}$}\label{fig:tmpv_spectra_y_0:KP1}}
        \vspace*{-0.8mm}
        \includegraphics[width=1\linewidth]{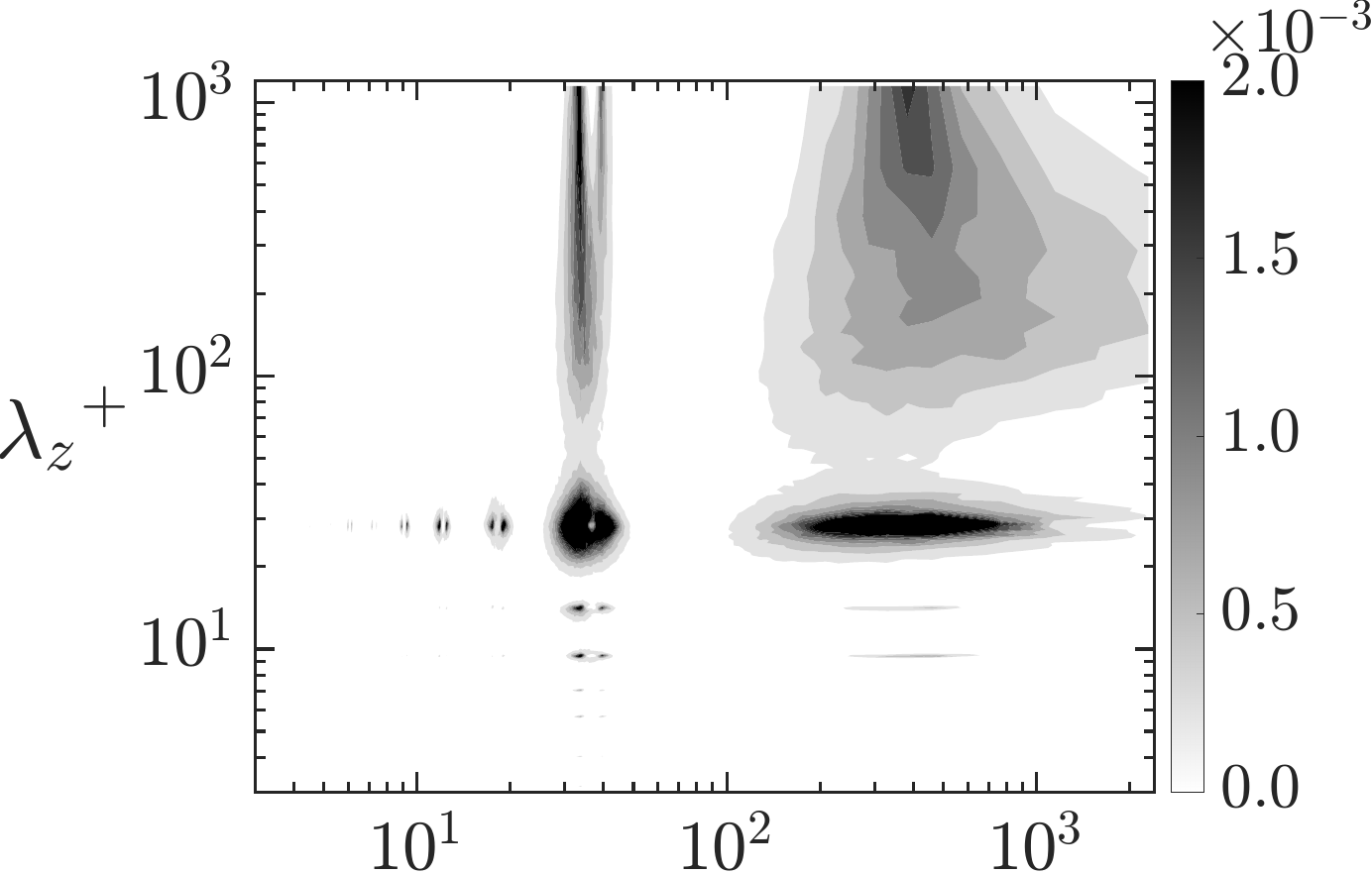}
    \end{subfigure}%
    \hspace*{0mm}
    \begin{subfigure}[tbp]{.34\textwidth}
        {\captionsetup{position=bottom, labelfont=it,textfont=it,size=scriptsize,singlelinecheck=false,justification=centering,labelformat=parens}
        \caption{$KP1^{\prime},\;$ $k_x\,k_z\,E_{\theta v}$}\label{fig:tmpv_spectra_y_0:KP1-1}}
        \vspace*{-0.8mm}
        \includegraphics[width=1\linewidth]{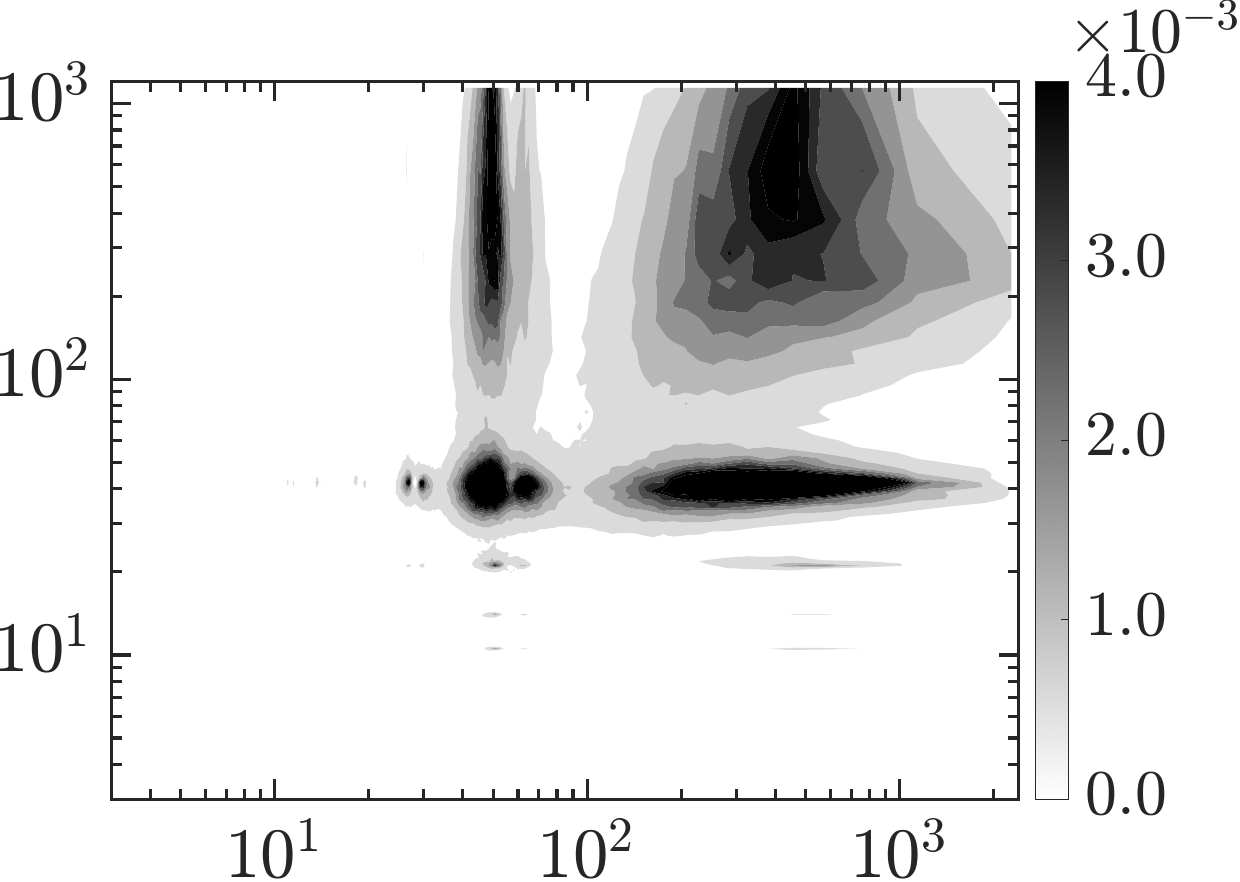}
    \end{subfigure}%
    \hspace*{0mm}
    \begin{subfigure}[tbp]{.34\textwidth}
        {\captionsetup{position=bottom, labelfont=it,textfont=it,size=scriptsize,singlelinecheck=false,justification=centering,labelformat=parens}
        \caption{$KP1^{\prime\prime},\;$ $k_x\,k_z\,E_{\theta v}$}\label{fig:tmpv_spectra_y_0:KP1-2}}
        \vspace*{-0.8mm}
        \includegraphics[width=1\linewidth]{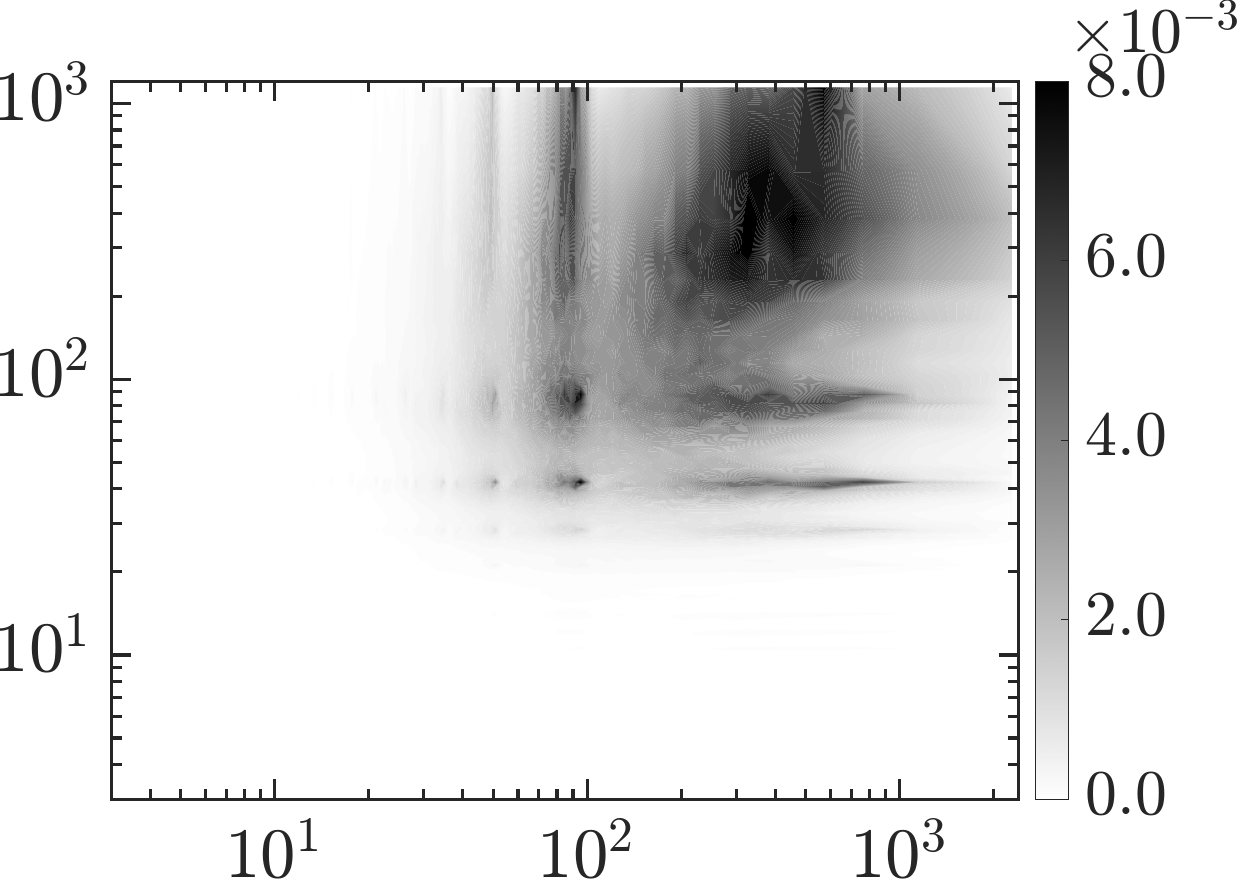}
    \end{subfigure}
    \hspace*{0mm}
    \begin{subfigure}[tbp]{.38\textwidth}
        {\captionsetup{position=bottom, labelfont=it,textfont=it,size=scriptsize,singlelinecheck=false,justification=centering,labelformat=parens}
        \caption{$SP,\;$ $k_x\,k_z\,E_{\theta \theta}$}\label{fig:tmp_spectra_y_0:SP}}
        \vspace*{-0.8mm}
        \includegraphics[width=1\linewidth]{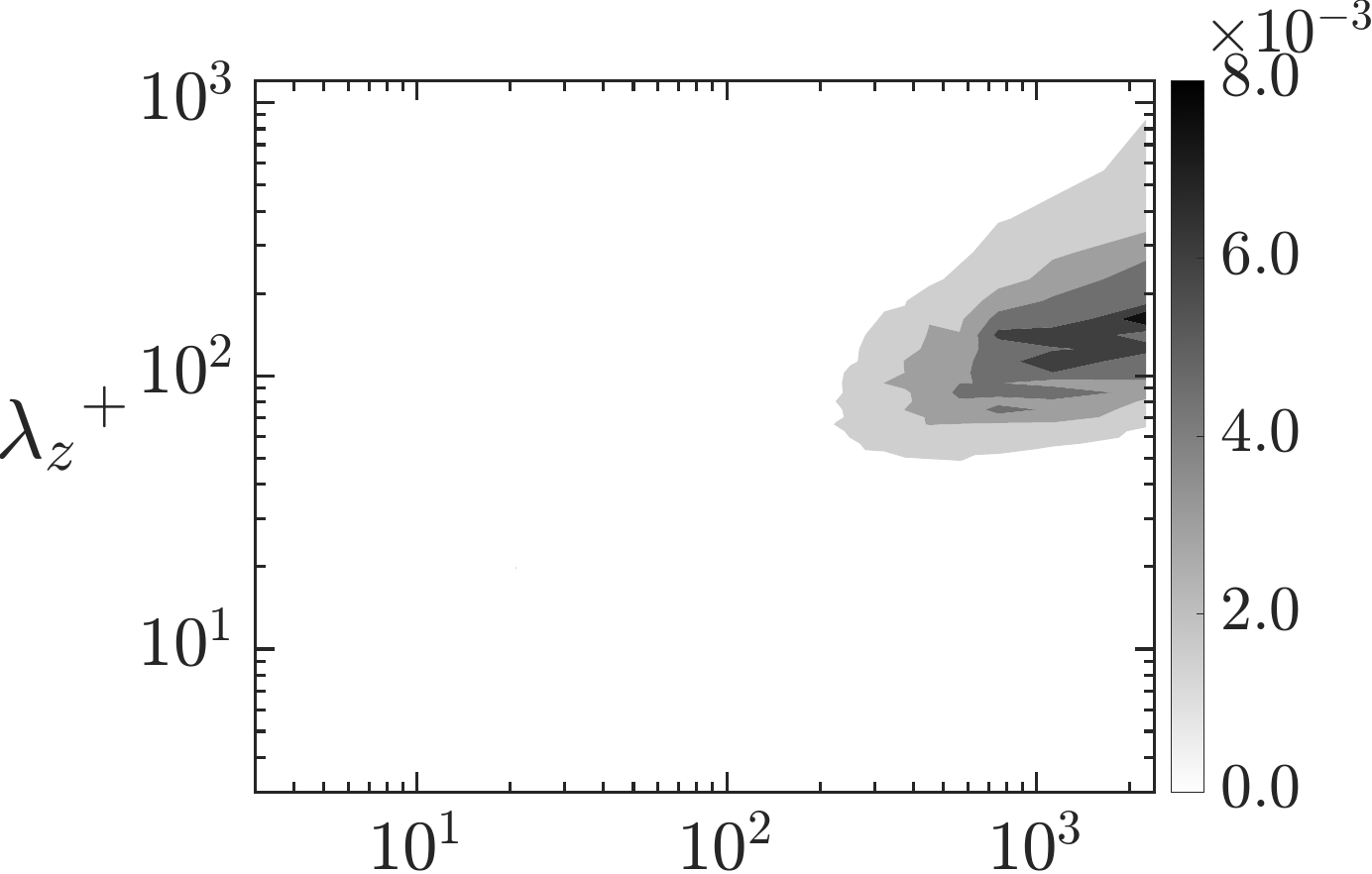}
    \end{subfigure}%
    \hspace*{0mm}
    \begin{subfigure}[tbp]{.34\textwidth}
        {\captionsetup{position=bottom, labelfont=it,textfont=it,size=scriptsize,singlelinecheck=false,justification=centering,labelformat=parens}
        \caption{$TP,\;$ $k_x\,k_z\,E_{\theta \theta}$}\label{fig:tmp_spectra_y_0:TP}}
        \vspace*{-0.8mm}
        \includegraphics[width=1\linewidth]{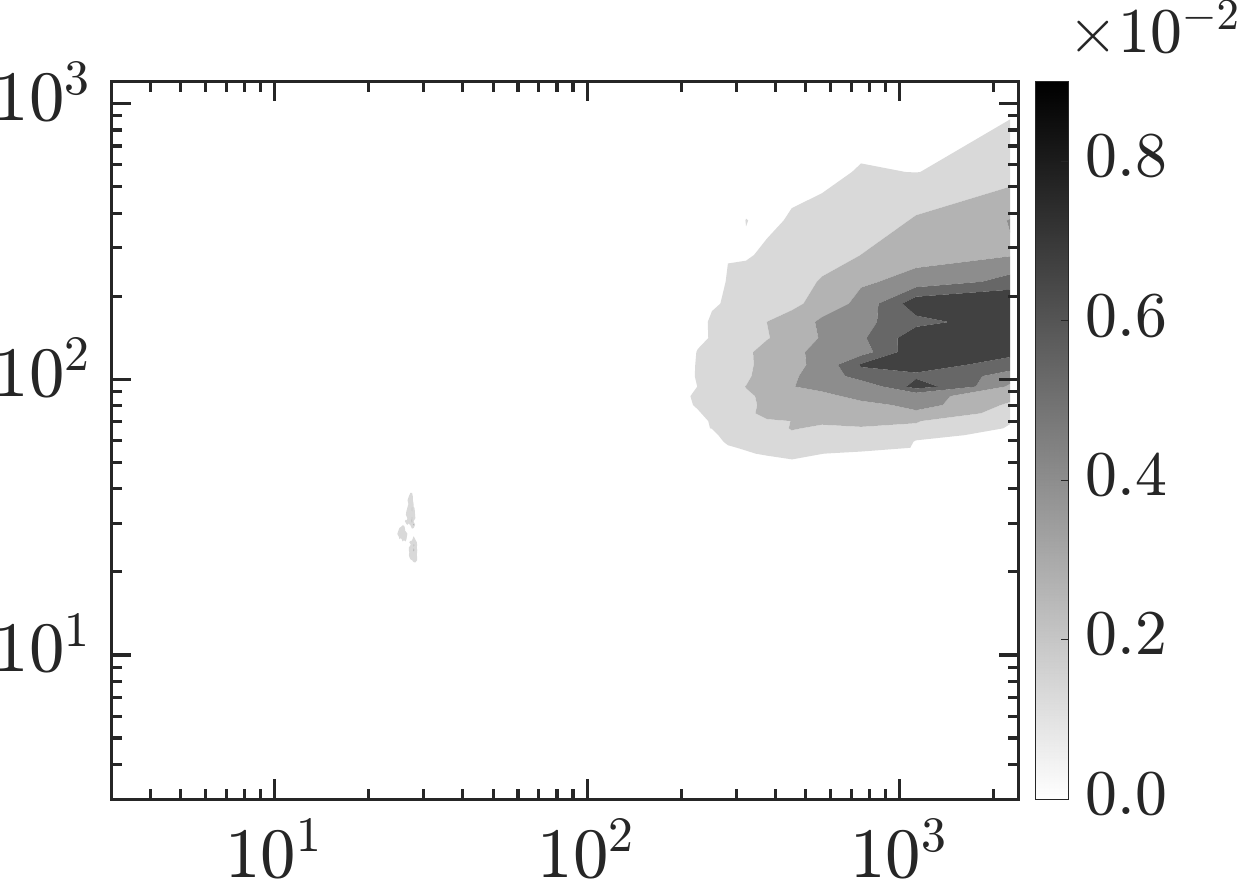}
    \end{subfigure}
    \hspace*{0mm}
    \begin{subfigure}[tbp]{.38\textwidth}
        {\captionsetup{position=bottom, labelfont=it,textfont=it,size=scriptsize,singlelinecheck=false,justification=centering,labelformat=parens}
        \caption{$SP,\;$ $k_x\,k_z\,E_{\theta v}$}\label{fig:tmpv_spectra_y_0:SP}}
        \vspace*{-0.8mm}
        \includegraphics[width=1\linewidth]{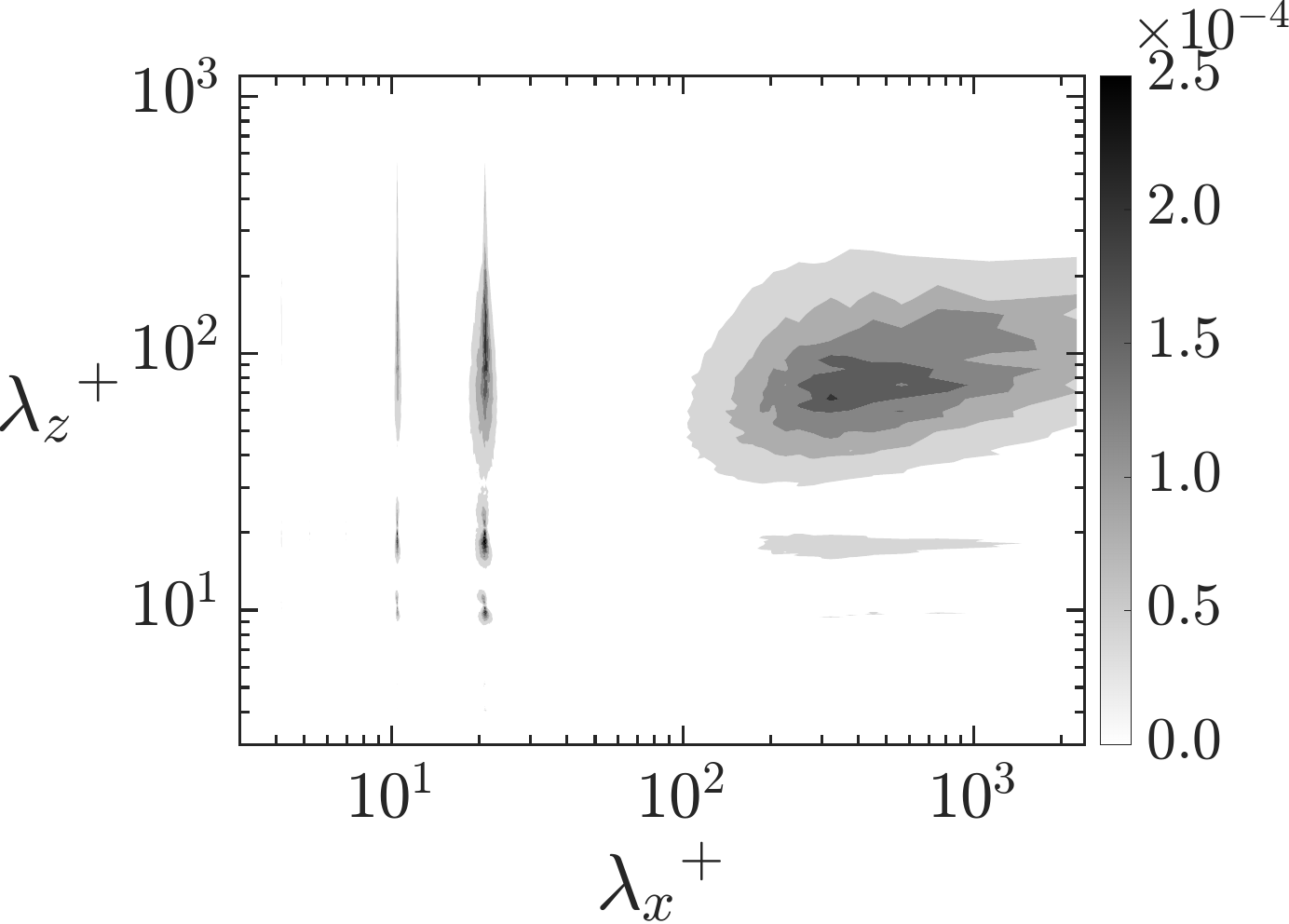}
    \end{subfigure}%
    \hspace*{0mm}
    \begin{subfigure}[tbp]{.34\textwidth}
        {\captionsetup{position=bottom, labelfont=it,textfont=it,size=scriptsize,singlelinecheck=false,justification=centering,labelformat=parens}
        \caption{$TP,\;$ $k_x\,k_z\,E_{\theta v}$}\label{fig:tmpv_spectra_y_0:TP}}
        \vspace*{-0.8mm}
        \includegraphics[width=1\linewidth]{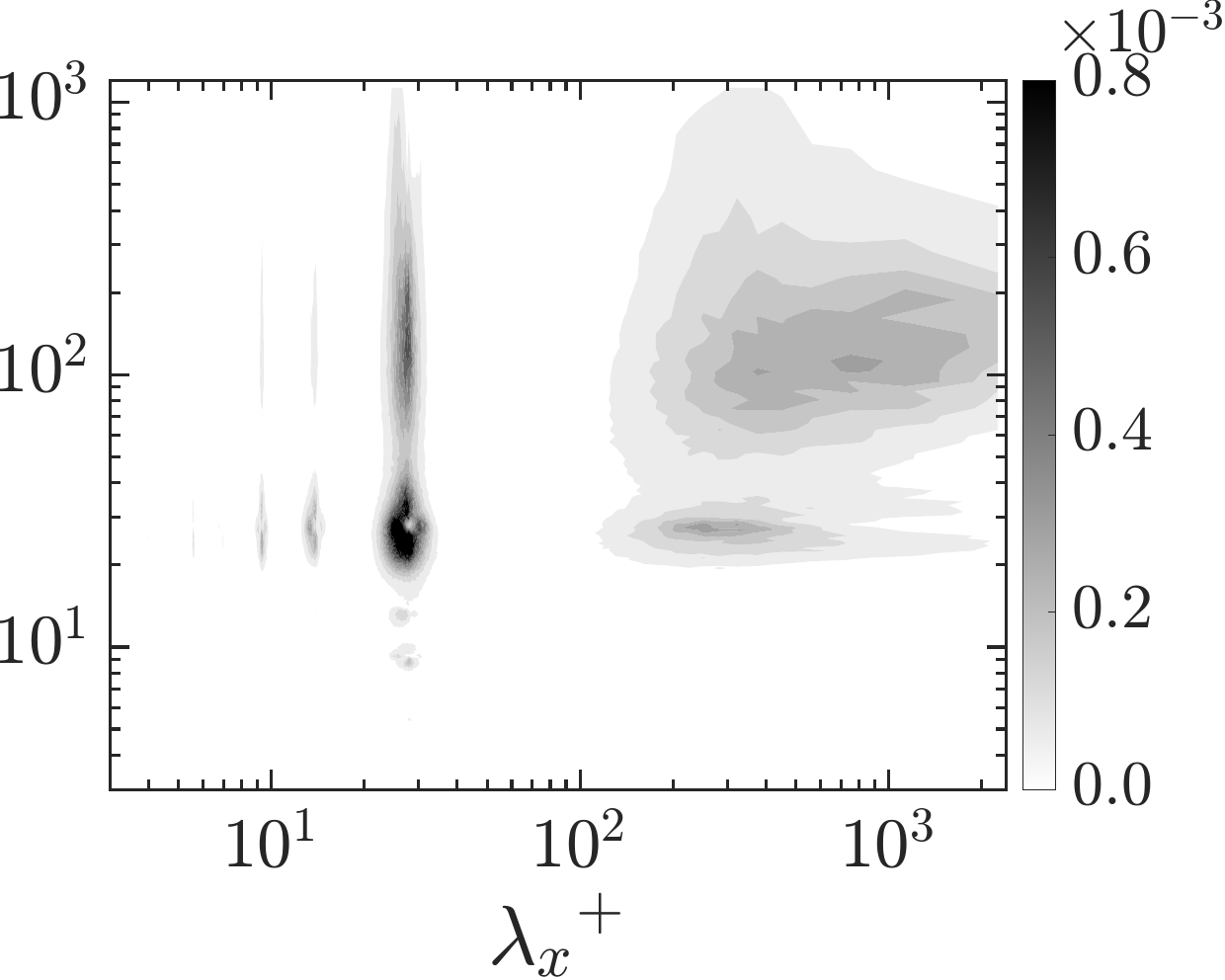}
    \end{subfigure}
    \vspace*{-1mm}
    \captionsetup{width=1\textwidth, justification=justified}
    \caption{Premultiplied spectral energies $k_x\,k_z\,E_{\theta \theta}$ and $k_x\,k_z\,E_{\theta v}$ at $y^+=0$ for the cases shown in \cref{fig:tmp_velocity_contours_y_0}.}
    \label{fig:temperature_spectra_y_0}
 \end{center}
\end{figure}
%%%%%%%%%%%%%%%%%%%%%%%%%%%%%%%%%%%%%%%%%%%%%%%%%%%%%%%%%%%%%%%%%%%%%%%%%%%%%%%%%%%%%%%%%%%%%%%%%%%%%%%%%%%%%%%%%%%%%%%%%%%%%%%%%%%%%%%%%%%%%%%%%%%%%%
%%%%%%%%%%%%%%%%%%%%%%%%%%%%%%%%%%%%%%%%%%%%%%%%%%%%%%%%%%%%%%%%%%%%%%%%%%%%%%%%
%%%%%%%%%%%%%%%%%%%%%%%%%%%%%%%%%%%%%%%%%%%%%%%%%%%%%%%%%%%%%%%%%%%%%%%%%%%%%%%%

 The overall picture of the temperature fluctuations is reflective of the velocity fluctuations that were observed in \cref{fig:velocity_statistics_channel}. Thermal activity becomes intensified in the vicinity of the surface in \cref{fig:rms_temp_channel} as it becomes more permeable. This intensification is attributable to the enhanced mixing caused by the K-H-like rollers, which are energetic at large wavelengths (\cref{fig:velocity_spectra_y_0}) and cause high levels of Reynolds shear stress (\cref{fig:Reynolds_stress_channel}).
 Examining the temperature field at $y=0$ in \cref{fig:tmp_velocity_contours_y_0}, we can observe the same evolution in its structure as we previously did for the streamwise velocity field in \cref{fig:u_velocity_contours_y_0}. The initial streak-like patterns in the temperature field become disrupted and broken down into shorter patches as the wall becomes more permeable. However, for the highest permeability case of $KP1^{\prime\prime}$ in \cref{fig:tmp_spectra_y_0:KP1-2}, the patches still demonstrate a greater level of streamwise coherence compared to those of the streamwise velocity in \cref{fig:u_velocity_contour_y_0:KP1-2}. This streamwise coherence is also reflected in the spectra of the temperature fluctuations for cases $KP1$, $KP1^{\prime}$, and $KP1^{\prime\prime}$ shown in figures \hyperref[fig:tmp_spectra_y_0:KP1]{\ref*{fig:tmp_spectra_y_0:KP1}}, \hyperref[fig:tmp_spectra_y_0:KP1-1]{\ref*{fig:tmp_spectra_y_0:KP1-1}}, \hyperref[fig:tmp_spectra_y_0:KP1-2]{\ref*{fig:tmp_spectra_y_0:KP1-2}}, where high levels of energy remain concentrated at large streamwise wavelengths.   
 Additionally, the presence of the signature of the K-H-like structures at long spanwise wavelengths in both the spectra of the temperature fluctuations and the wall-normal turbulent heat flux shown in figures \hyperref[fig:tmpv_spectra_y_0:KP1]{\ref*{fig:tmpv_spectra_y_0:KP1}}, \hyperref[fig:tmpv_spectra_y_0:KP1-1]{\ref*{fig:tmpv_spectra_y_0:KP1-1}}, \hyperref[fig:tmpv_spectra_y_0:KP1-2]{\ref*{fig:tmpv_spectra_y_0:KP1-2}} is evident. The intense temperature fluctuations deep inside the walls for cases with large permeabilities is indicative of strong levels of turbulent heat flux penetrating into them, which is what will be examined next.
 
 \begin{figure}
    \begin{center}
    \begin{subfigure}{.49\textwidth}
        {\captionsetup{labelfont=it,textfont=normalfont,singlelinecheck=false,justification=raggedright,labelformat=parens}\caption{}
        \label{fig:Total_verticle_heat_flux}}%
        \includegraphics[width=1\linewidth]{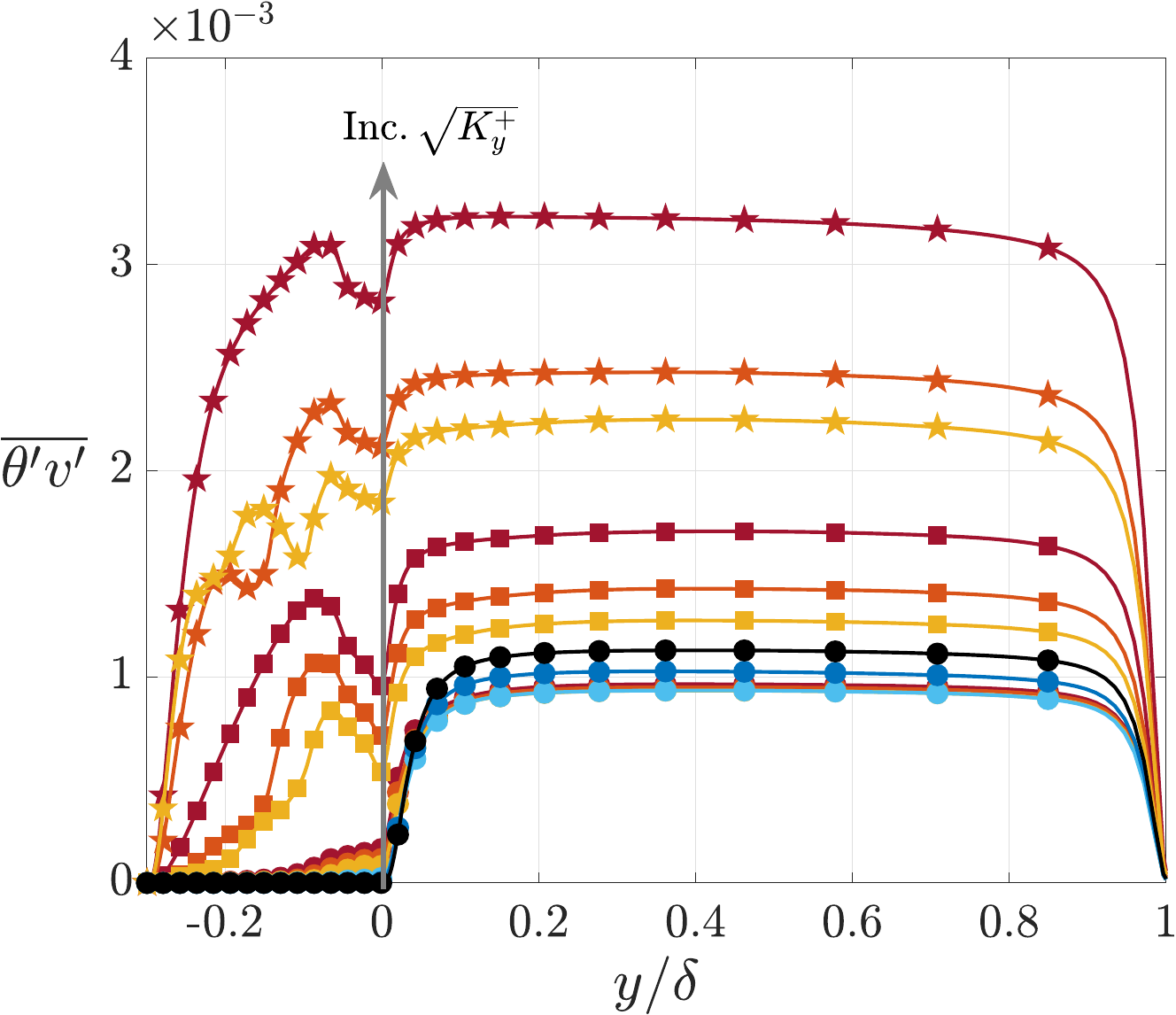}
    \end{subfigure}
    \begin{subfigure}{.48\textwidth}
        {\captionsetup{labelfont=it,textfont=normalfont,singlelinecheck=false,justification=raggedright,labelformat=parens}\caption{}
        \label{fig:Dispersive_verticle_heat_flux_decomp}}%
        \hspace*{2.5pt}\includegraphics[width=1\linewidth]{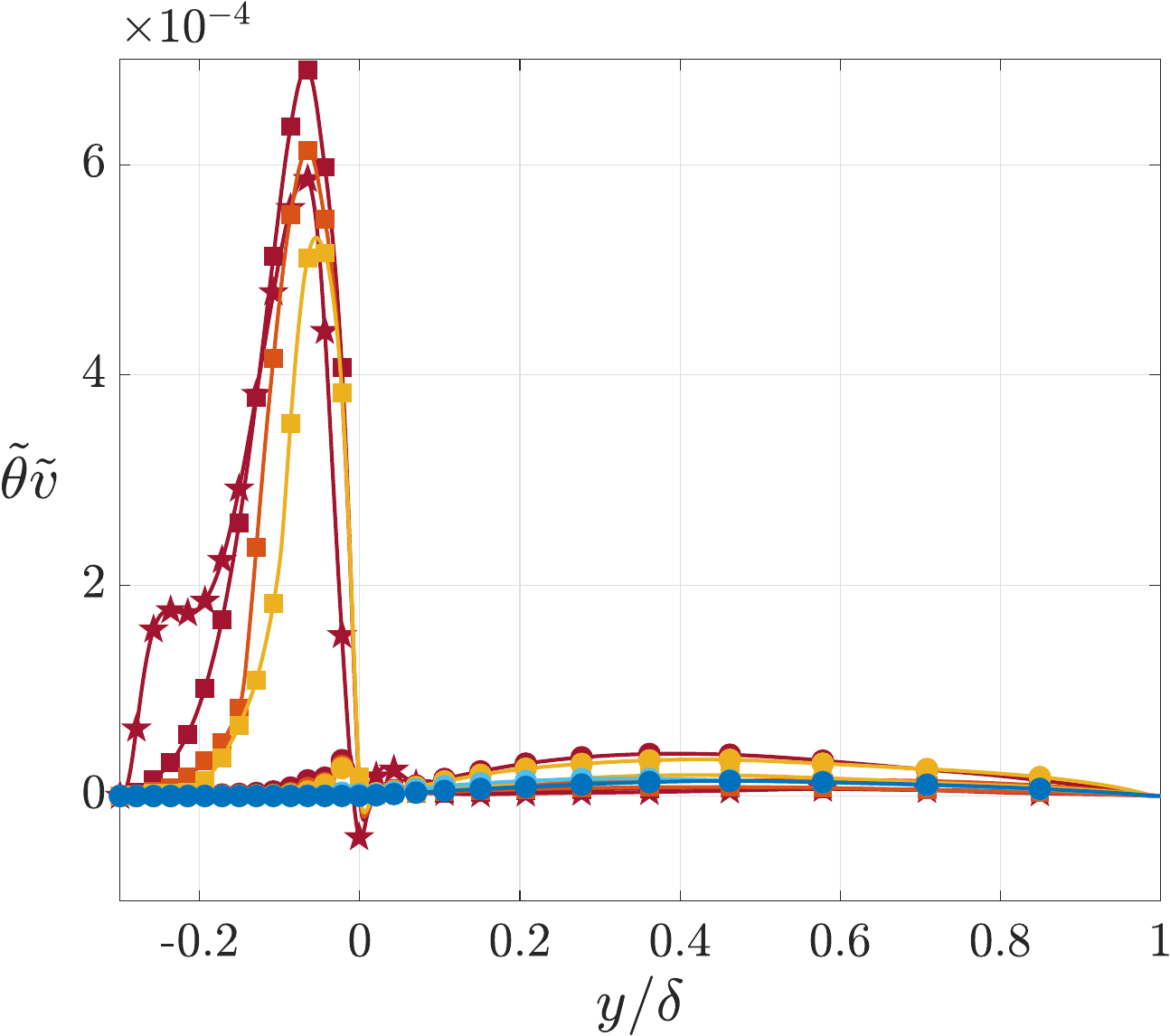}
    \end{subfigure}
    \captionsetup{width=0.95\textwidth, justification=justified}
    \caption{Wall-normal heat flux throughout the whole domain: \emph{(a)} total wall-normal heat flux and \emph{(b)} dispersive wall-normal heat flux. Direction of the gray arrow in \emph{(a)} indicates increasing wall-normal permeability.}
    \label{fig:turbulent_heat_flux}
 \end{center}
 \end{figure}
 \begin{figure}
    \begin{center}
    \begin{subfigure}{.49\textwidth}
        {\captionsetup{labelfont=it,textfont=normalfont,singlelinecheck=false,justification=raggedright,labelformat=parens}\caption{}
        \label{fig:heat_flux_decomp}}%
        \hspace*{-2pt}\includegraphics[width=1\linewidth]{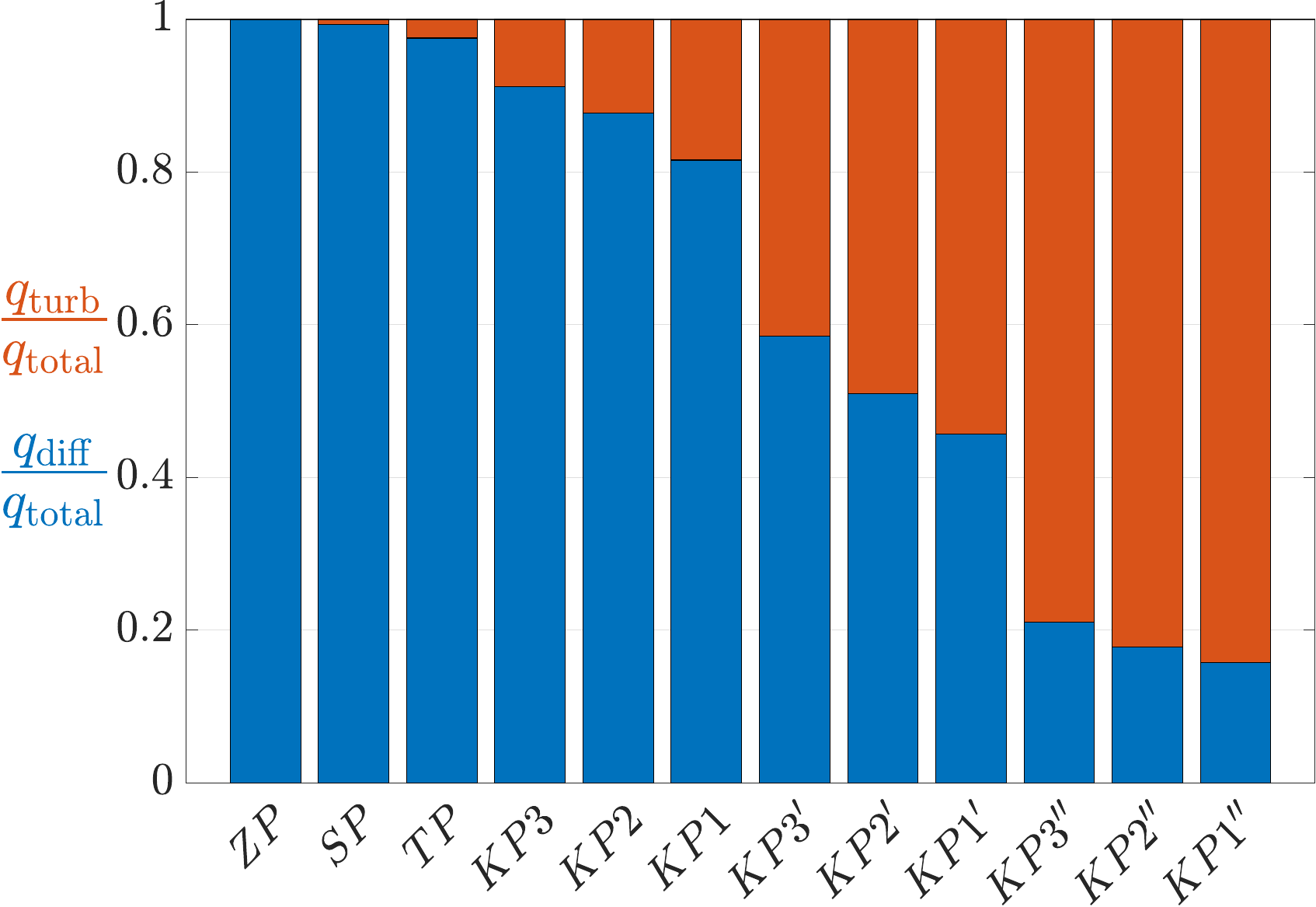}
        \vspace*{1pt}
    \end{subfigure}
    \begin{subfigure}{.49\textwidth}
        {\captionsetup{labelfont=it,textfont=normalfont,singlelinecheck=false,justification=raggedright,labelformat=parens}\caption{}
        \label{fig:Nu}}%
        \includegraphics[width=1\linewidth]{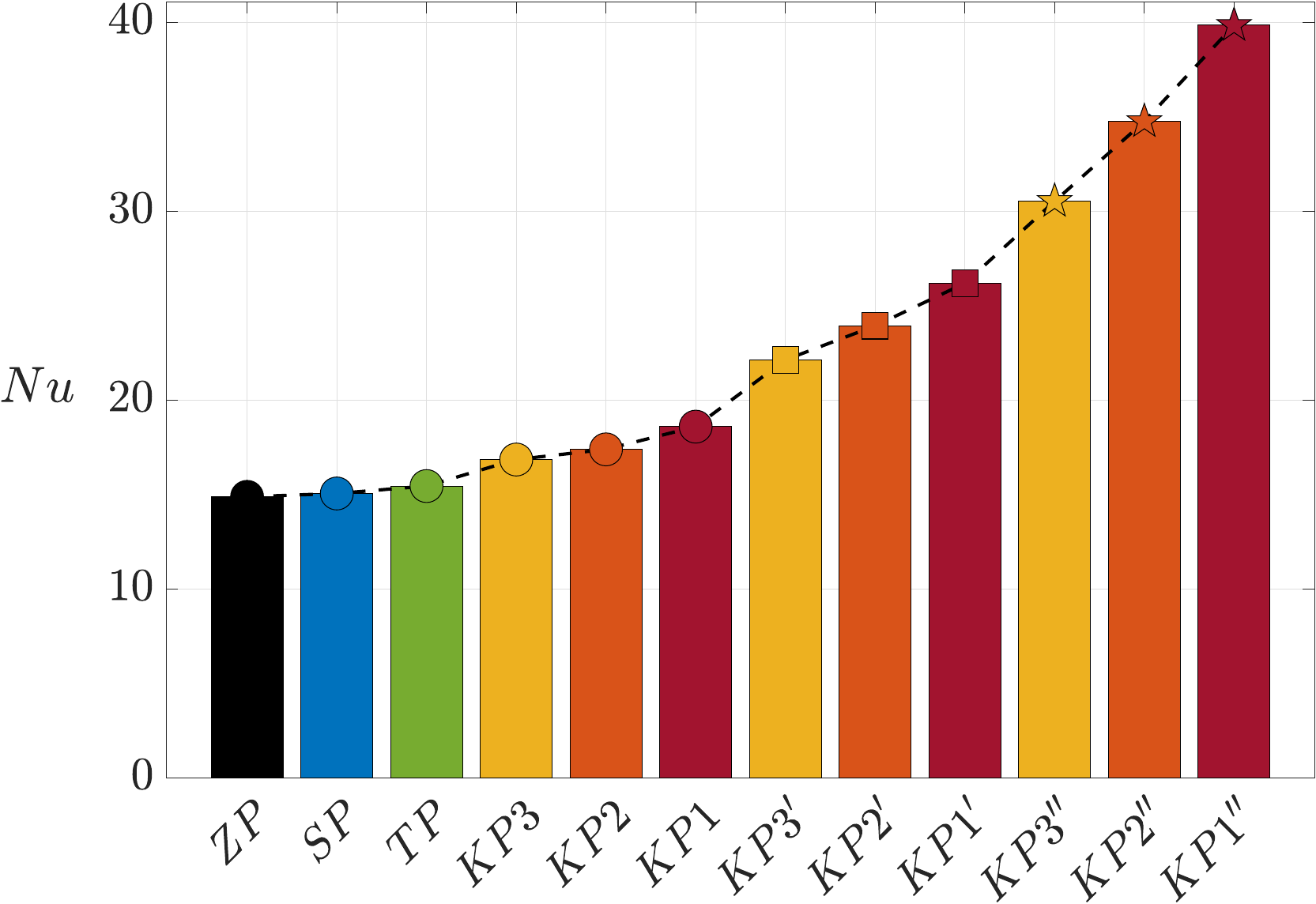}
    \end{subfigure}
    \captionsetup{width=0.98\textwidth, justification=justified}
    \caption{(\emph{a}) Diffusive (blue) and turbulent (orange) fractions of the heat flux at the surface ($y=0$), (\emph{b}) Nusselt number for the cases of \cref{tab:DNS}.}
    \label{fig:Nu_and_heat_flux}
    \end{center}
 \end{figure}

 The total wall-normal heat flux, $\overline{\theta^{\prime}v^{\prime}}$, is shown in \cref{fig:Total_verticle_heat_flux}. It is clear that as $\sqrt{K_{y}^+}$ becomes larger, the heat flux grows proportionally stronger in the vicinity of the surface. The low-permeability cases $SP$ and $TP$ show small gains in $\overline{\theta^{\prime}v^{\prime}}$ close to the surface. After the onset of the K-H-like instability (beyond $TP$), the gains in $\overline{\theta^{\prime}v^{\prime}}$ relative to $\sqrt{K_{y}^+}$ become greater. Cases $KP3^{\prime\prime}$, $KP2^{\prime\prime}$, and $KP1^{\prime\prime}$ with their very high permeabilities ($\sqrt{K_{y}^+}\approx12.7, \sqrt{K_{y}^+}\approx15.3, \sqrt{K_{y}^+}\approx18.9$) show significant levels of $\overline{\theta^{\prime}v^{\prime}}$ at the surface. This is why the mean temperature distribution of these cases lacked a diffusive characteristic in the inner part of their profiles in \cref{fig:mean_temp_channel}, because the diffusive heat flux component close to the surface becomes diminished and overtaken by the wall-normal heat flux. The change in overall $\overline{\theta^{\prime}v^{\prime}}$ throughout the channel region ($y/\delta>0$) mirrors that of the temperature fluctuations (\cref{fig:rms_temp_channel}). The overall magnitude initially falls when going from $ZP$ to $KP1$ ($ZP \rightarrow SP \rightarrow TP \rightarrow KP3 \rightarrow KP2 \rightarrow KP1$), but grows when going from $KP1$ to $KP1^{\prime\prime}$ ($KP1 \rightarrow KP3^\prime \rightarrow KP2^\prime \rightarrow KP1^\prime \rightarrow KP3^{\prime\prime} \rightarrow KP2^{\prime\prime} \rightarrow KP1^{\prime\prime}$). The $KP\langle\rangle^{\prime}$ and $KP\langle\rangle^{\prime\prime}$ cases have overall higher levels of $\overline{\theta^{\prime}v^{\prime}}$ compared to the baseline $ZP$ case.
 Similar to what was observed for the dispersive Reynolds shear stress in \cref{fig:Reynolds_stress_channel}, the dispersive wall-normal heat flux in \cref{fig:Dispersive_verticle_heat_flux_decomp} is also overall negligible. Therefore the wall-normal transport of heat is predominately attributable to the turbulence scales generated by the K-H-like rollers.
 
 In the porous wall region ($y/\delta<0$), negligible $\overline{\theta^{\prime}v^{\prime}}$ penetrates beneath the surface in cases $SP$ and $TP$. The stronger $\overline{\theta^{\prime}v^{\prime}}$ scales of $KP3$, $KP2$, and $KP1$ manage to penetrate deeper into the porous region, down to $y/\delta\approx-0.1$, before dissipating. Cases $KP\langle\rangle^{\prime}$ and $KP\langle\rangle^{\prime\prime}$ have $\overline{\theta^{\prime}v^{\prime}}$ distributions which undergo intensification in the shallow part of the wall ($-0.1<{y/\delta}<0$), developing a peak there, and then subsequently decaying toward the floor of the wall. The overall magnitude of $\overline{\theta^{\prime}v^{\prime}}$ for these cases is much higher compared to those of $KP\langle\rangle$ and penetrates far down into the porous wall, close to the floor of it ($y/\delta=-0.3$).
 Note that in the $KP\langle\rangle^{\prime\prime}$ cases, the peak of $\overline{\theta^{\prime}v^{\prime}}$ in the wall is practically on par with $\overline{\theta^{\prime}v^{\prime}}$ throughout the channel region. The peaks of the $KP\langle\rangle^{\prime}$ cases also come close to the value of $\overline{\theta^{\prime}v^{\prime}}$ throughout the channel. This quality is particular to these cases and is not observed for the other walls. Similarly, it is only these cases which register any notable dispersive heat flux activity in the porous wall region (\cref{fig:Dispersive_verticle_heat_flux_decomp}). The overall weak levels of dispersive heat flux not withstanding, the gulf in dispersive heat flux intensity throughout the porous wall region between these cases ($KP\langle\rangle^{\prime}$, $KP\langle\rangle^{\prime\prime}$) and the other cases is considerable.

 \subsection{Heat transfer and Reynolds analogy analysis}\label{subsec:analysis_heat_reynolds_analogy}
 
 It is now suitable to examine the components of the heat flux at the surface of the porous walls through which heat is transferred to the bulk flow in the channel region. To this end, the components of the total heat flux,
 \begin{gather}\label{eq:total_heat}
        q_{\mathrm{tot}} = \underbrace{\overline{\theta^{\prime}v^{\prime}}}_{q_{\mathrm{turb}}} - \overbrace{\alpha_{f}\cfrac{d\Theta}{dy}}^{q_{\mathrm{diff}}},
 \end{gather}
 were calculated at the surface ($y/\delta=0$) and are shown in \cref{fig:heat_flux_decomp}. Note that since the dispersive heat flux turned out to be negligible, we denote $\overline{\theta^{\prime}v^{\prime}}$ as $q_{\mathrm{turb}}$ (the turbulent heat flux) to indicate that it is due to turbulence alone. It is clear in \cref{fig:heat_flux_decomp} that for the $KP\langle\rangle^{\prime\prime}$ cases, the diffusive heat flux component accounts for less than $20\%$ of the total heat flux at the surface and the turbulent heat flux represents the dominant mechanism of heat transfer there. Since $\overline{\theta^{\prime}v^{\prime}}$ becomes considerably enhanced throughout the channel region (\cref{fig:turbulent_heat_flux}), the increase in convective heat transfer caused by this can be gauged using the Nusselt number, defined as
 \begin{gather}\label{eq:Nu}
        Nu = \cfrac{q_{\mathrm{tot}} \; \delta}{\alpha_{\mathrm{f}} \; (\Theta_{s} - \Theta_{b})},
 \end{gather}
 where $\Theta_{b} = {\int_0^\delta U \,\Theta \, dy} \Big/ {\int_0^\delta U \, dy}$ and $\Theta_{s} = \Theta({y=0})$ are the bulk mean temperature and the mean temperature at the surface, respectively. The Nusselt number values for all cases are gathered in \cref{fig:Nu}, which serves as a companion plot to \cref{fig:heat_flux_decomp} by showing how the growth of convective heat transfer mirrors the increasing share of the turbulent component in the overall surface heat flux.

 \begin{figure}
    \begin{center}
    \hspace*{-4mm}
    \begin{subfigure}[tbp]{.57\textwidth}
        {\captionsetup{labelfont=it,textfont=normalfont,singlelinecheck=false,justification=raggedright,labelformat=parens}\caption{}
        \label{fig:St_vs_KxKy}}
        \vspace*{1mm}
        \includegraphics[width=1\linewidth]{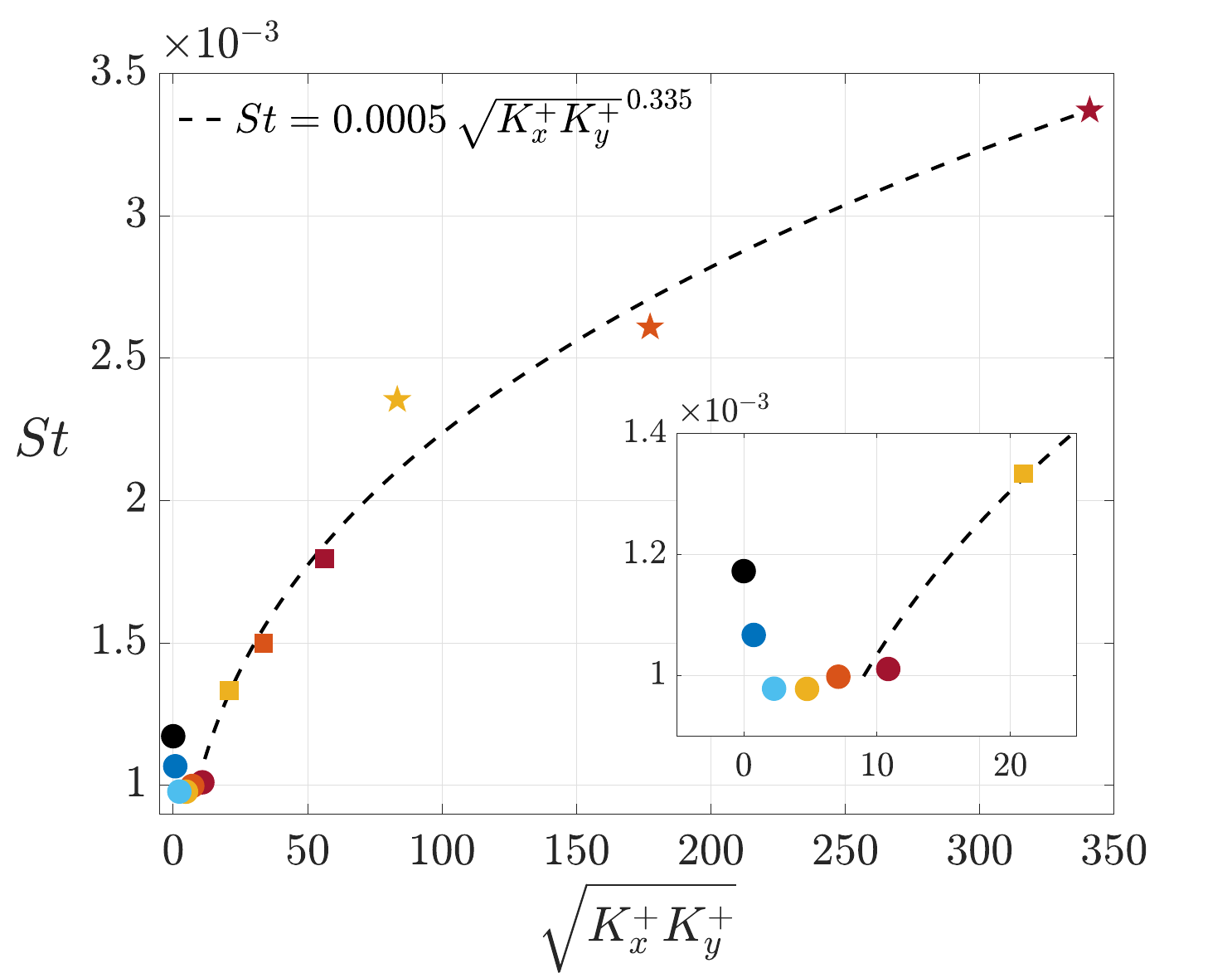}
    \end{subfigure}%
    \hspace*{-4mm}
    \begin{subfigure}[tbp]{.52\textwidth}
        {\captionsetup{labelfont=it,textfont=normalfont,singlelinecheck=false,justification=raggedright,labelformat=parens}\caption{}
        \label{fig:St_vs_cf}}
        \vspace*{-1mm}
        \includegraphics[width=1\linewidth]{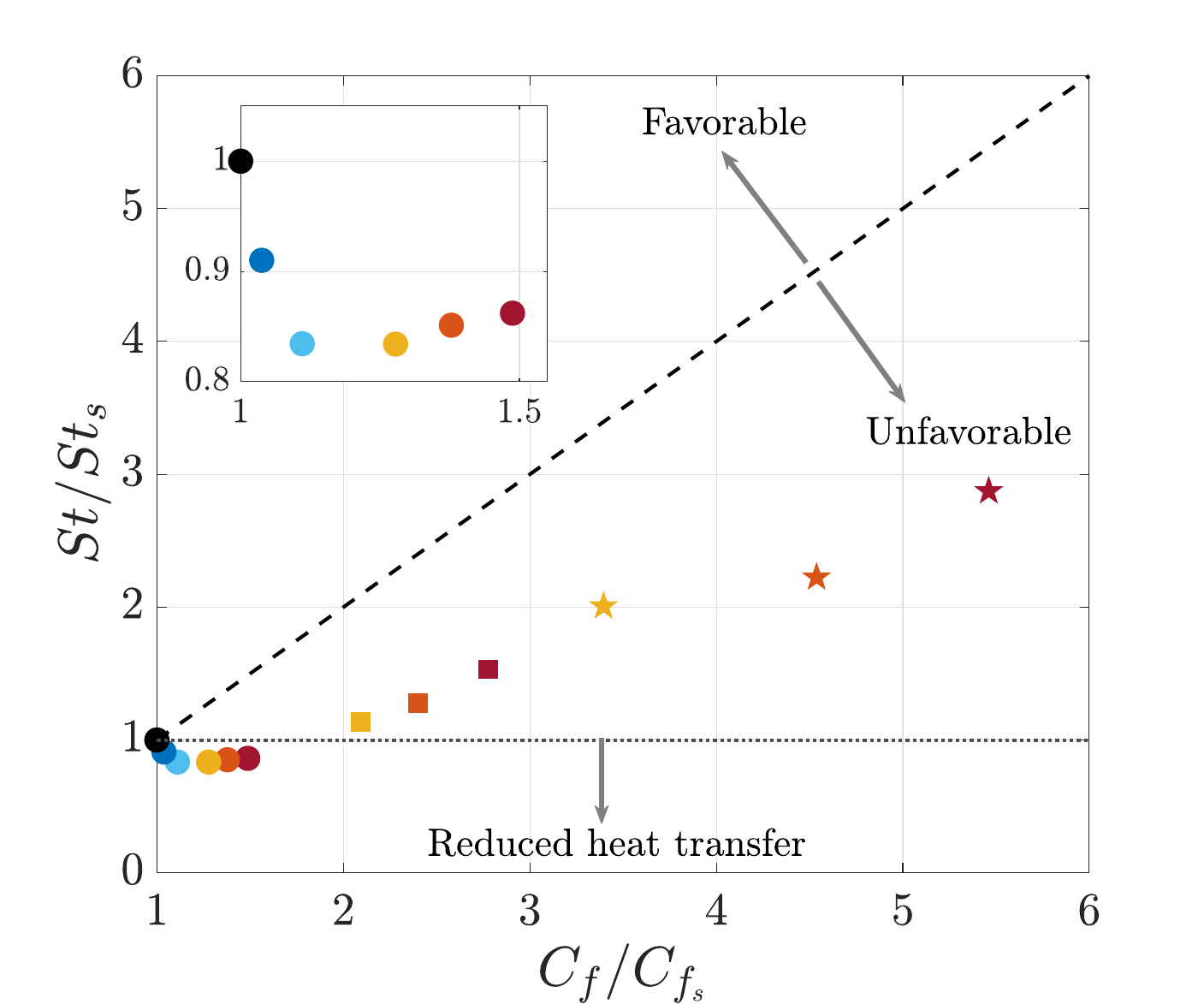}
    \end{subfigure}
    \captionsetup{width=0.98\textwidth, justification=justified}
    \caption{(\emph{a}) Correlation of $St$ against $\sqrt{K_x^+\,K_y^+}$, (\emph{b}) Reynolds analogy plot for the porous walls of \cref{tab:DNS}.}
    \label{fig:St_cf_scalings}
    \end{center}
 \end{figure}

 As described in \cref{sec:intro}, heat-transfer efficiency is technologically important. The porous walls of \cref{tab:DNS} are passive structures which change both the flow and heat transfer characteristics of the system. Typically, heat transfer in systems involving convection is expressed in terms of the Stanton number, defined as
 \begin{gather}\label{eq:St}
        St = \cfrac{q_{\mathrm{tot}}}{\rho\,U_b\,c_p}.
 \end{gather}
 \Cref{fig:St_vs_KxKy} shows the changes in heat transfer performance for the different porous walls by plotting $St$ against $\sqrt{K_x^+\,K_y^+}$. This parameter, obtained by \cite{sharma2017} using linear stability analysis, was shown to well-predict the change in turbulence regimes from smooth-wall-like to K-H-like when applied to the DNS data of \cite{khorasani_2023}, making it a suitable structural parameter for characterising a porous wall.
 The heat transfer changes non-monotonically in \cref{fig:St_vs_KxKy}. It initially decreases as the solid conductive wall is made permeable, but begins to increase as turbulence transitions to the K-H-like regime. Once in this regime, $St$ continues to increase as $\sqrt{K_x^+\,K_y^+}$ becomes larger, with the heat transfer exceeding that of the baseline $ZP$ case from $KP3^{\prime}$ and beyond. A least-squares fit of $St$ to $\sqrt{K_x^+\,K_y^+}$ for the K-H-like cases gives the power-law relation $St = 0.0005\sqrt{K_x^+\,K_y^+}^{0.335}$. This fit is plotted in \cref{fig:St_vs_KxKy}, where its concave quality can be interpreted as diminishing gains in heat transfer being achieved relative to incremental increases in permeability. This can also be deduced by simply considering what physically happens as a wall is made more permeable. At some point, the wall will cease to be a porous structure and no longer affect the turbulent flow as one. In the limit of $100\%$ porosity it will vanish, leaving just a larger channel for the fluid to flow through.

 Moving onto the matter of heat-transfer efficiency, following the approach of \cite{rouhi_endrikat_modesti_sandberg_oda_tanimoto_hutchins_chung_2022} adopted from \cite{Bunker_2017}, the ratio $St/St_{s}$ versus $C_f/C_{f_{s}}$ is examined. Here, $St_{s}$ and $C_{f_{s}}$ are the smooth-wall Stanton and skin-friction coefficient and correspond to case $ZP$ of \cref{tab:DNS}. The skin-friction coefficient is defined as
 \begin{gather}\label{eq:C_f}
        C_f = \cfrac{2\tau_{w}}{\rho\,U_b^2}.
 \end{gather}
 The ratio $St_{s}/C_{f_{s}}$ thus defines the Reynolds analogy factor for the baseline smooth-wall case, which here is $ZP$. For a porous case (or any other target configuration), should $St/St_{s} < C_{f}/C_{f_{s}}$, it represents an unfavorable breakdown in the Reynolds analogy and a lower heat\nobreakdash-transfer efficiency since the amount of energy required to drive the flow is not matched by a proportional increase in heat transfer. In the opposite case of $St/St_{s} > C_{f}/C_{f_{s}}$, the breakdown is favorable and heat-transfer becomes more efficient.
 Plotting $St/St_{s}$ versus $C_{f}/C_{f_{s}}$ in \cref{fig:St_vs_cf} shows that the breakdown in the Reynolds analogy is unfavorable for the porous walls of \cref{tab:DNS}. The initial decrease in heat transfer observed in \cref{fig:St_vs_KxKy} is also reflected in \cref{fig:St_vs_cf}, which coupled with the increase in $C_{f}$, translates into a net unfavorable performance for $SP$, $TP$, $KP3$, $KP2$, and $KP1$. For the remaining walls the heat transfer performance increases and does not attain a maximum limit.
 Recall that in \cref{sec:intro}, it was mentioned that for rough-wall turbulent flows \cite{macdonald_hutchins_chung_2018} observed that $St$ for a constant roughness height of $k/\delta$ began to decrease under fully rough wall conditions, in line with the rough-tube experiments of \cite{Dipprey_1963}. The latter concluded that ``There is a limit for any combination of Reynolds number and Prandtl number beyond which increases in roughness, while increasing $C_{f}$, will no longer increase $St$''. The data of \cite{macdonald_hutchins_chung_2018} and \cite{Dipprey_1963} were included by \cite{rouhi_endrikat_modesti_sandberg_oda_tanimoto_hutchins_chung_2022} in their Reynolds analogy plot (figure 1 of their manuscript), where it shows this limit translating into a constant value of $St/St_{s}\approx2$ beyond $C_{f}/C_{f_{s}}\approx3$. Such a limit is not reached in \cref{fig:St_vs_cf} beyond $C_{f}/C_{f_{s}}=3$ for the porous walls. How far $St/St_{s}$ can be increased before reaching the limit of a vanishing porous wall is something that requires further investigating.

\subsection{Investigation of Reynolds analogy breakdown}

 \begin{figure}
    \begin{center}
    \begin{subfigure}[tbp]{.48\textwidth}
        {\captionsetup{labelfont=it,textfont=normalfont,singlelinecheck=false,justification=raggedright,labelformat=parens}\caption{}
        \label{fig:Pr_t_channel}}
        \includegraphics[width=1\linewidth]{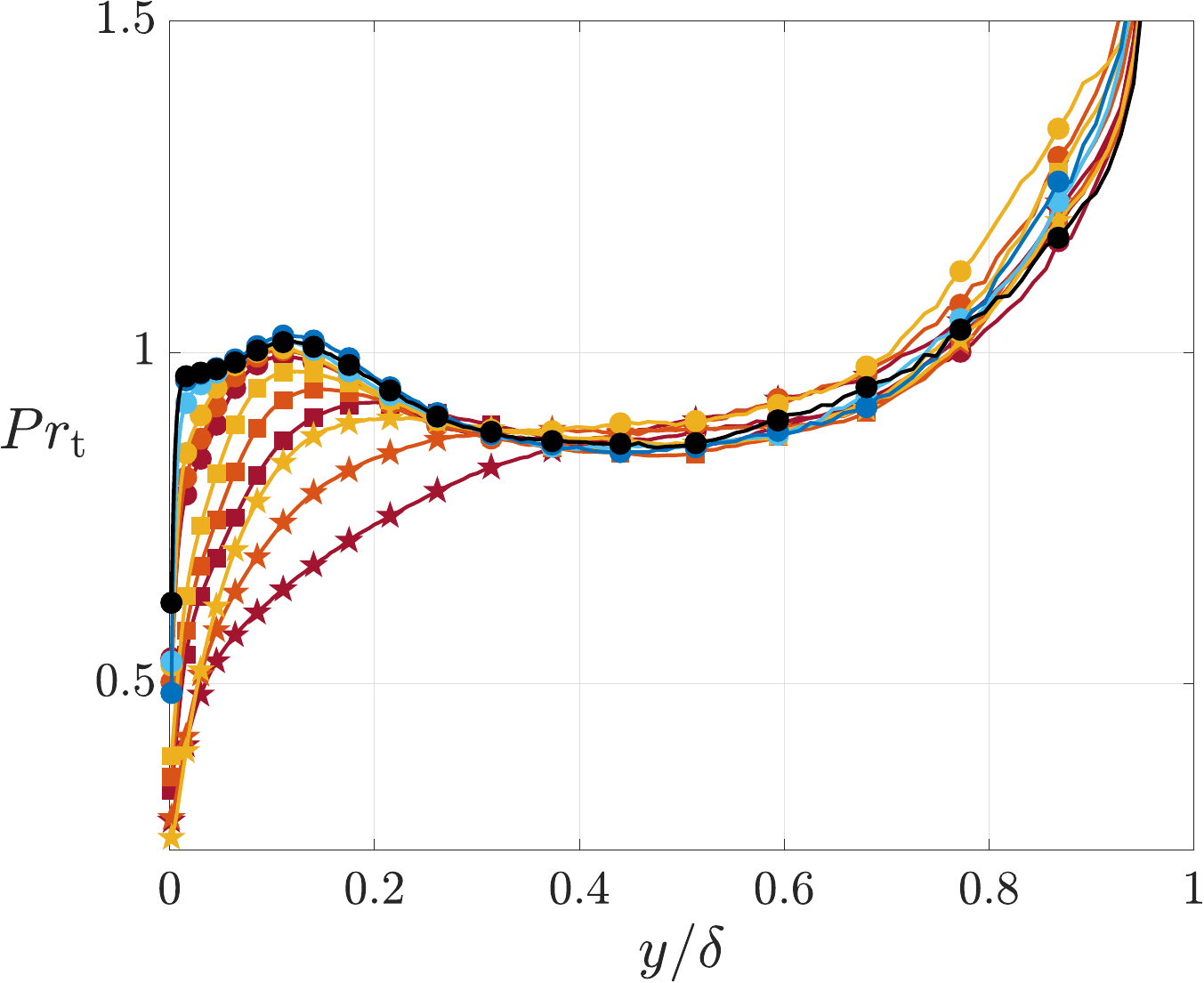}
    \end{subfigure}%
    \hspace*{5pt}
    \begin{subfigure}[tbp]{.5\textwidth}
        {\captionsetup{labelfont=it,textfont=normalfont,singlelinecheck=false,justification=raggedright,labelformat=parens}\caption{}
        \label{fig:Pr_t_surface_region}}
        \includegraphics[width=0.98\linewidth]{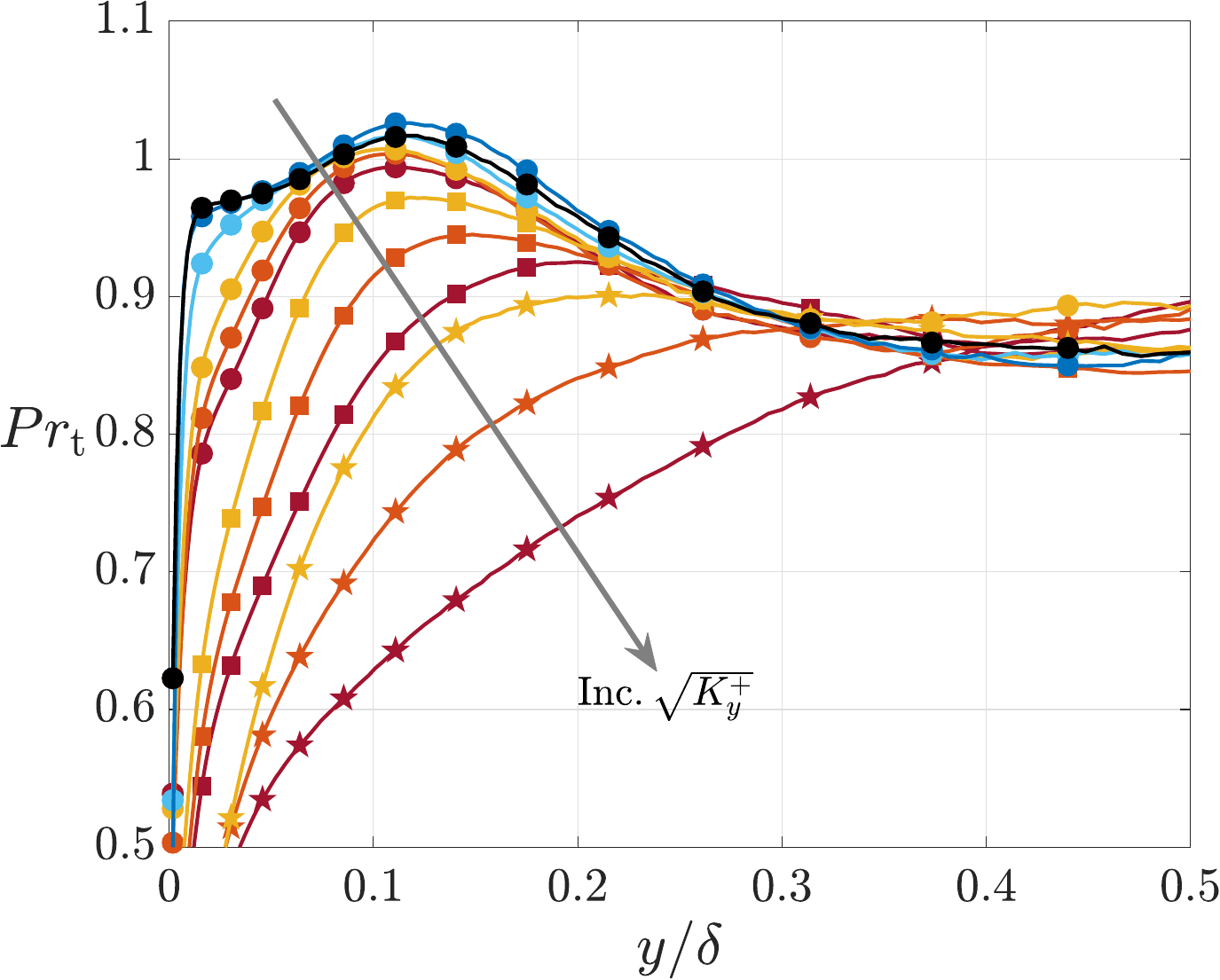}
    \end{subfigure}\\
    \begin{subfigure}[tbp]{.5\textwidth}
        {\captionsetup{labelfont=it,textfont=normalfont,singlelinecheck=false,justification=raggedright,labelformat=parens}\caption{}
        \label{fig:Correlation_coeff_surface_region}}
        \includegraphics[width=0.98\linewidth]{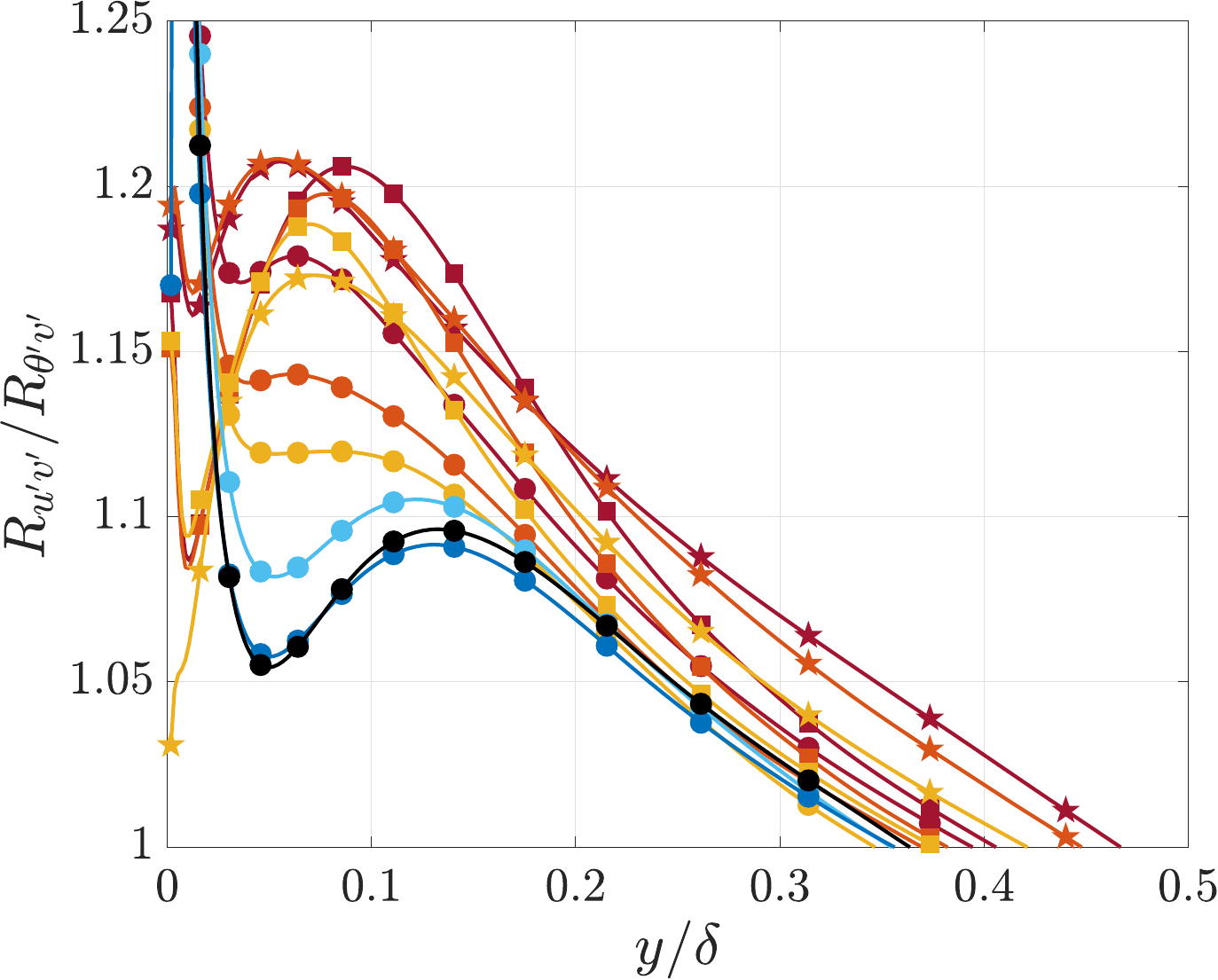}
    \end{subfigure}
    \captionsetup{width=0.98\textwidth, justification=justified}
    \caption{Turbulent Prandtl number (\emph{a}) throughout the entire channel region and (\emph{b}) the lower half of the channel region. (\emph{c}) the ratio of the $u^{\prime}$ and $v^{\prime}$ correlation coefficient to the $\theta^{\prime}$ and $v^{\prime}$ correlation coefficient in the lower half of the channel region.}
    \label{fig:Prandtl_turbulent_Correlation_coeff}
    \end{center}
 \end{figure}

 In the preceding section, the unfavorable breakdown of the Reynolds analogy over the porous walls was established. The potential causes underlying this breakdown will now be investigated. For this purpose, it is helpful to first examine how dissimilar the transport of momentum and heat are throughout the flow. The turbulent Prandtl number, $Pr_T = \nu_T/\alpha_T$, is commonly used as an indicator of the dissimilarity between momentum and heat transport. Here, $\nu_T = {\overline{u^\prime v^\prime}}/(dU/dy)$ and $\alpha_T = {\overline{\theta^\prime v^\prime}}/(d\theta/dy)$ are the momentum and heat eddy diffusivities, respectively.

 \Cref{fig:Prandtl_turbulent_Correlation_coeff} shows the turbulent Prandtl number throughout the channel region. The pattern exhibited by case $ZP$ in \cref{fig:Pr_t_channel} is similar to that reported for smooth-wall channel flows throughout the literature, with a peak value of $Pr_T\approx1$ close to the surface and $Pr_T\approx0.85$ farther away from it (see \citealt{Pirozzoli_Bernardini_Orlandi_2016} and the references contained therein). The porous cases conform to the pattern of $ZP$ away from the surface, but closer to it differences start to emerge, with $Pr_T$ deviating significantly over $0<{y/\delta}<0.2$. The deviations become greater and extend farther into the channel as the wall-normal permeability, $\sqrt{K_y^+}$, increases.
 
 The disparity in $Pr_t$ indicates growing dissimilarity between the turbulent transport of momentum and heat in this region over the porous walls compared to the impermeable/smooth-wall case of $ZP$. However, as was seen in \cref{subsec:analysis_heat_reynolds_analogy}, the increase in heat transfer was always less than the increase in skin-friction, which should translate into $Pr_t>1$ in the vicinity of the porous walls and not $Pr_t<1$. This was indeed the case for the two-dimensional bar-type roughness of \cite{leonardi_orlandi_djenidi_antonia_2015}, where $Pr_T$ was higher than the smooth-wall value at the roughness crests, going as high as $2.5$. The sole exception was the case with a roughness Reynolds number of $k^+\approx45$, where $Pr_T$ went as low as $0.5$. The reason why that particular case did not behave similar to rest was probably due to it not belonging to the fully-rough regime where form-induced pressure drag becomes dominant. The surface of the porous walls of \cref{tab:DNS} are like gratings or perforated plates, with no elements protruding above the surface level. As such, the bulk flow is not subject to strong pressure drag at the surface. This may partially explain why the momentum and heat dissimilarity over porous walls (\cref{fig:Pr_t_surface_region}) differs from that over rough walls, and also why $St/St_{s}$ does not level out like it does for rough-wall turbulence. It does not however explain the incongruity between $Pr_t$ and $St/St_{s}$. Referring to canopy-flow literature, it can be deduced that the cause behind this is the K\nobreakdash-H\nobreakdash-like rollers that emerge over the porous walls. \cite{raupach1996} demonstrated how the turbulent flow above a canopy layer conformed to that of a turbulent mixing layer. In particular, the mixing-layer analogy gave similar low values of $Pr_t\approx0.5$ that had been observed in canopy flows. Thus, the eddy\nobreakdash-diffusivity model of turbulent transport is not valid for such flows, due to the presence of the recurrent cross-stream K\nobreakdash-H\nobreakdash-like rollers that create large-scale counter\nobreakdash-gradient flows. These K\nobreakdash-H\nobreakdash-like rollers also feature over porous walls and therefore explain the contradiction between the calculated $Pr_t$ over the walls and the heat transfer efficiency characterized through $St/St_{s}$ vs. $C_{f}/C_{f_{s}}$. Further clarity is made by examining the correlation coefficient ratio
 \begin{gather}\label{eq:correlation_coeff_ratio}
        \cfrac{R_{u^{\prime}v^{\prime}}}{R_{\theta^{\prime}v^{\prime}}} = \cfrac{{\overline{u^\prime v^\prime}}/{({u^{\prime}}_{\mathrm{rms}}{v^{\prime}}_{\mathrm{rms}})}}{{\overline{\theta^\prime v^\prime}}/{({\theta^{\prime}}_{\mathrm{rms}}{v^{\prime}}_{\mathrm{rms}})}},
 \end{gather}
 in \cref{fig:Correlation_coeff_surface_region}. The ratio grows larger in the vicinity of the surface as the wall is made more permeable, with the position of the local maxima and minima moving closer to the surface. The growing trend of the ratio reflects the growing dissimilarity between the turbulent heat and momentum fluxes while also demonstrating that the latter undergoes greater enhancement compared to the former, in agreement with the results of \cref{subsec:analysis_heat_reynolds_analogy}.
 % Ultimately, caution must be exercised when using turbulence models. If the target flow configuration involves a porous-wall, eddy\nobreakdash-diffusivity\nobreakdash-based turbulence models cannot be relied upon to produce physically realistic results.

%%%%%%%%%%%%%%%%%%%%%%%%%%%%%%%%%%%%%%%%%%%%%%%%%%%% quadrant analysis %%%%%%%%%%%%%%%%%%%%%%%%%%%%%%%%%%%%%%%%%%%%%%%%%%%%%%%%%%%%%%%%%%%%%%%%
\begin{figure}
 \vspace*{-12mm}
 \begin{center}
    $KP1^{\prime\prime}:$\hspace*{4mm}%
    \begin{subfigure}[tbp]{.35\textwidth}
        {\captionsetup{position=bottom, labelfont=it,textfont=it,singlelinecheck=false,justification=centering,labelformat=parens}
        \caption{}\label{fig:jpdf_heat_y_0:KP1-2}}
        \vspace*{-0.8mm}
        \includegraphics[width=1\linewidth]{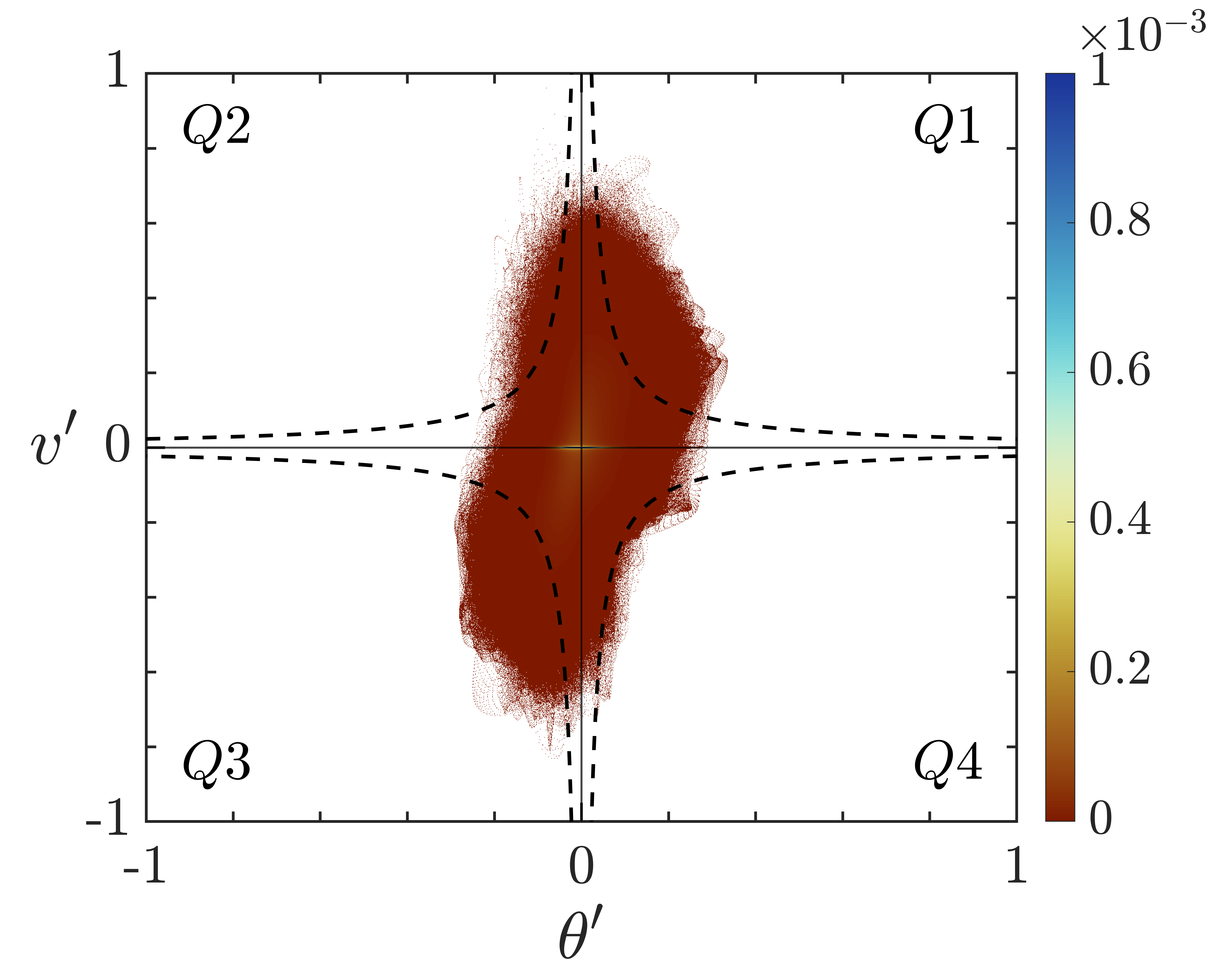}
    \end{subfigure}\hspace*{2mm}%
    \begin{subfigure}[tbp]{.35\textwidth}
        {\captionsetup{position=bottom, labelfont=it,textfont=it,singlelinecheck=false,justification=centering,labelformat=parens}
        \caption{}\label{fig:jpdf_momentum_y_0:KP1-2}}
        \vspace*{-0.8mm}
        \includegraphics[width=1\linewidth]{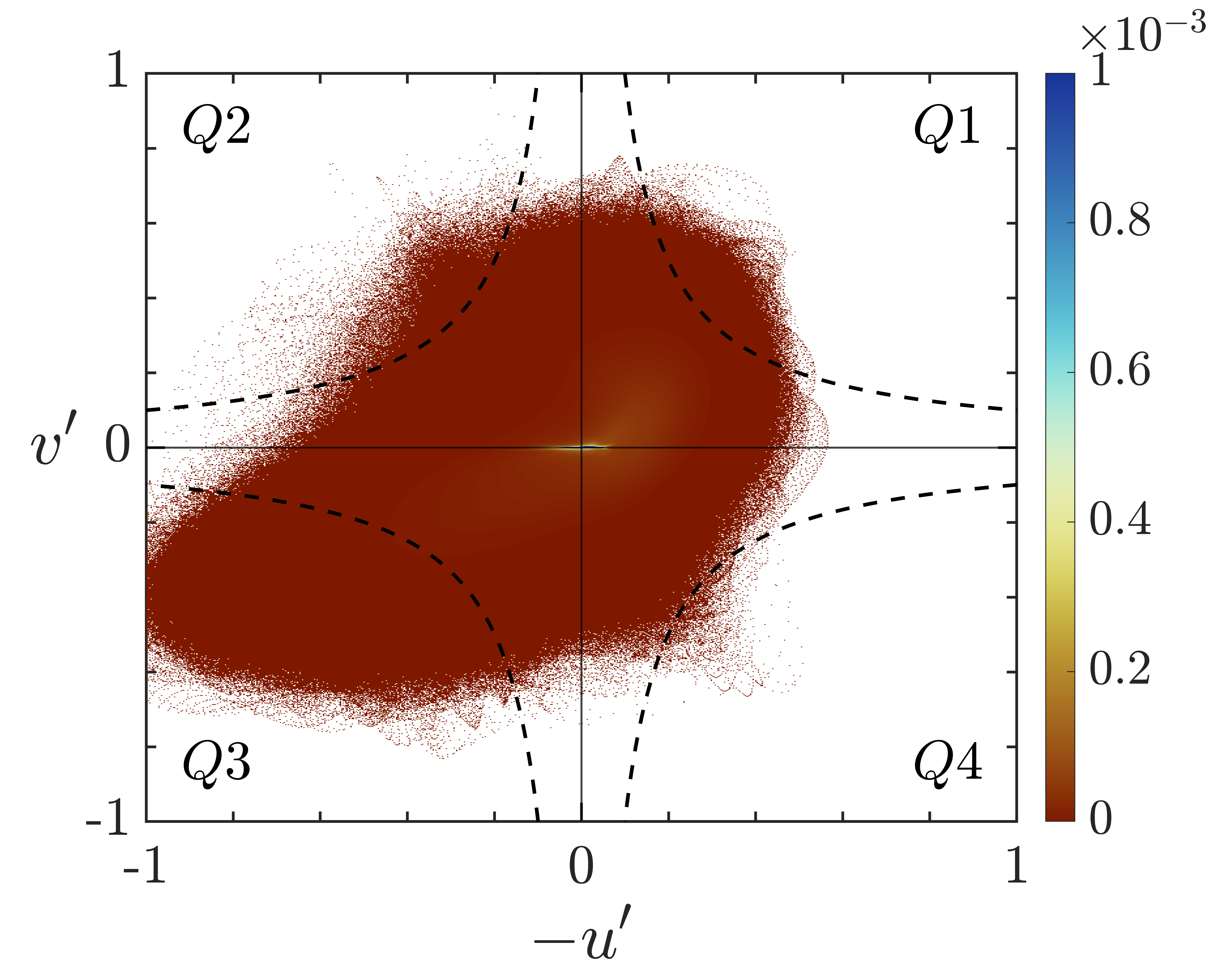}
    \end{subfigure}
    $KP1^{\prime}:$\hspace*{4mm}%
    \begin{subfigure}[tbp]{.35\textwidth}
        {\captionsetup{position=bottom, labelfont=it,textfont=it,singlelinecheck=false,justification=centering,labelformat=parens}
        \caption{}\label{fig:jpdf_heat_y_0:KP1-1}}
        \vspace*{-0.8mm}
        \includegraphics[width=1\linewidth]{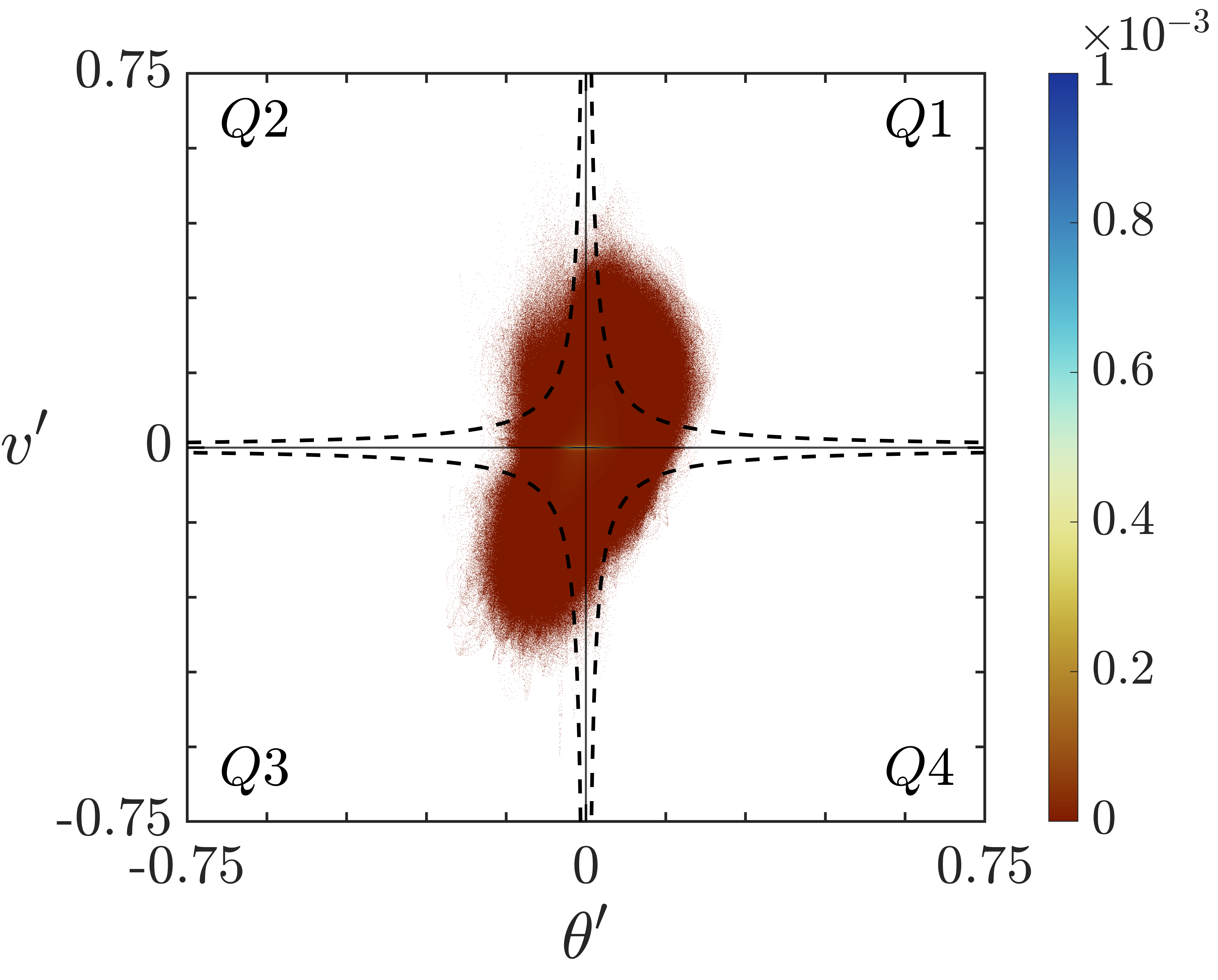}
    \end{subfigure}\hspace*{2mm}%
    \begin{subfigure}[tbp]{.35\textwidth}
        {\captionsetup{position=bottom, labelfont=it,textfont=it,singlelinecheck=false,justification=centering,labelformat=parens}
        \caption{}\label{fig:jpdf_momentum_y_0:KP1-1}}
        \vspace*{-0.8mm}
        \includegraphics[width=1\linewidth]{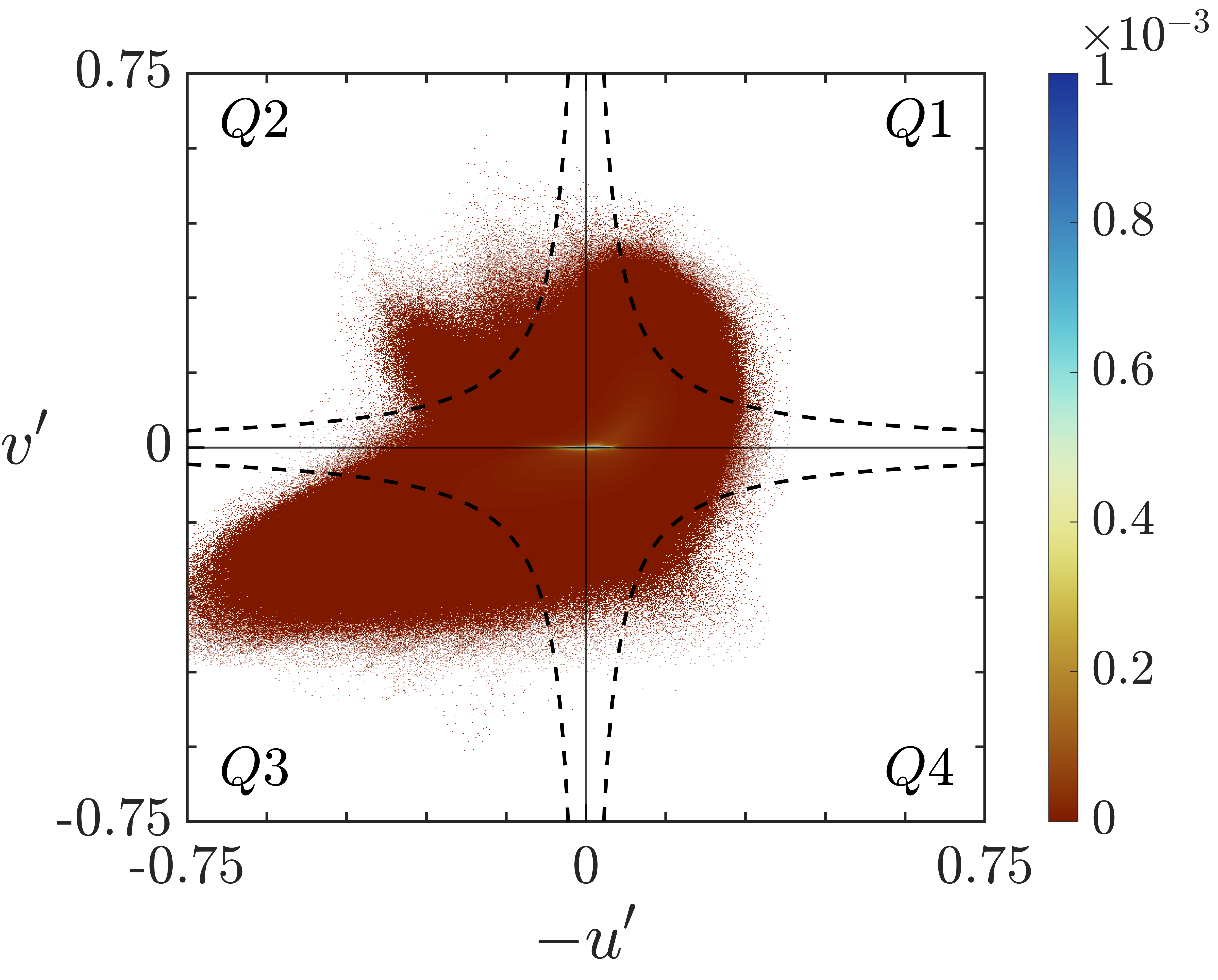}
    \end{subfigure}
    $KP1:$\hspace*{4mm}%
    \begin{subfigure}[tbp]{.35\textwidth}
        {\captionsetup{position=bottom, labelfont=it,textfont=it,singlelinecheck=false,justification=centering,labelformat=parens}
        \caption{}\label{fig:jpdf_heat_y_0:KP1}}
        \vspace*{-0.8mm}
        \includegraphics[width=1\linewidth]{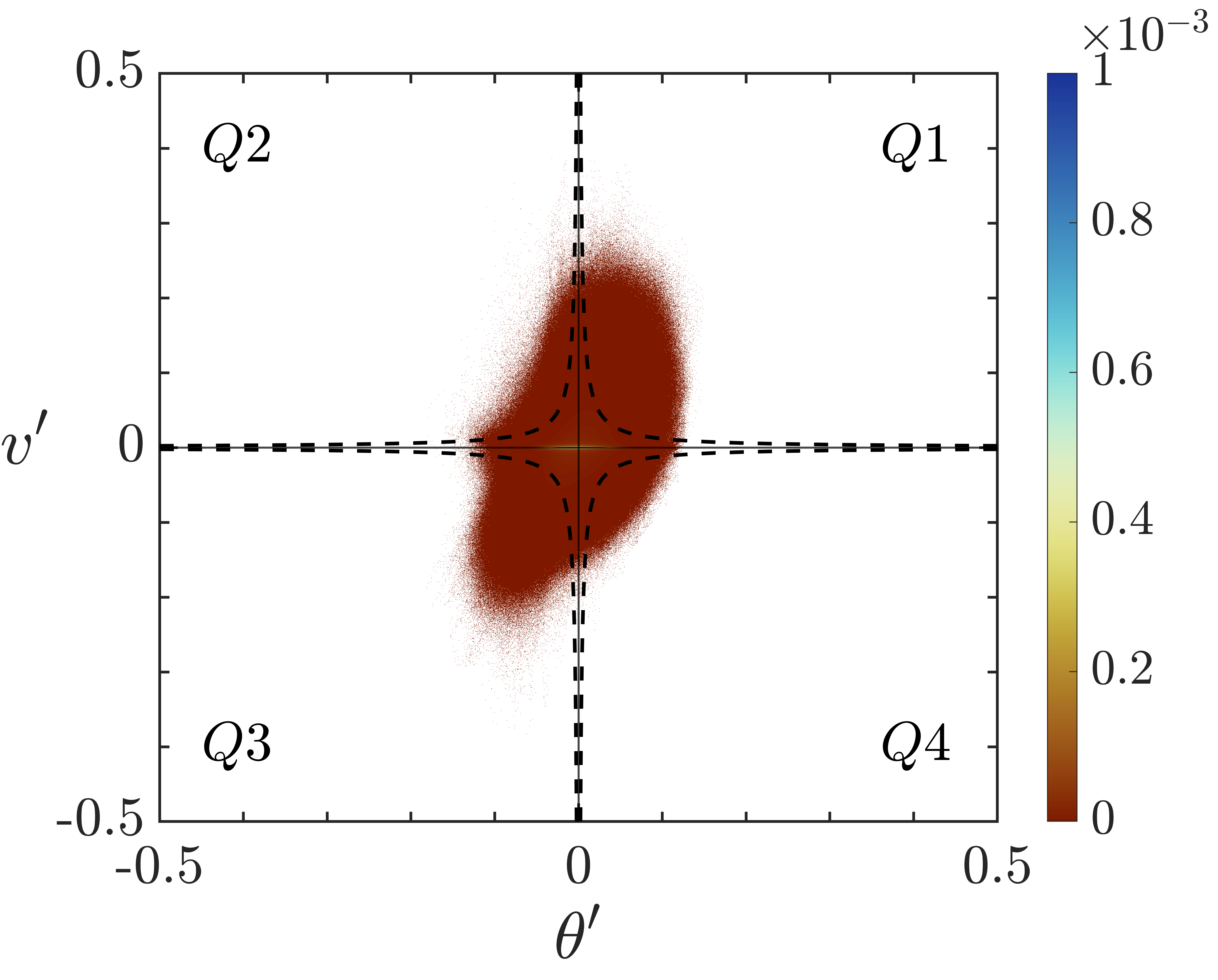}
    \end{subfigure}\hspace*{2mm}%
    \begin{subfigure}[tbp]{.35\textwidth}
        {\captionsetup{position=bottom, labelfont=it,textfont=it,singlelinecheck=false,justification=centering,labelformat=parens}
        \caption{}\label{fig:jpdf_momentum_y_0:KP1}}
        \vspace*{-0.8mm}
        \includegraphics[width=1\linewidth]{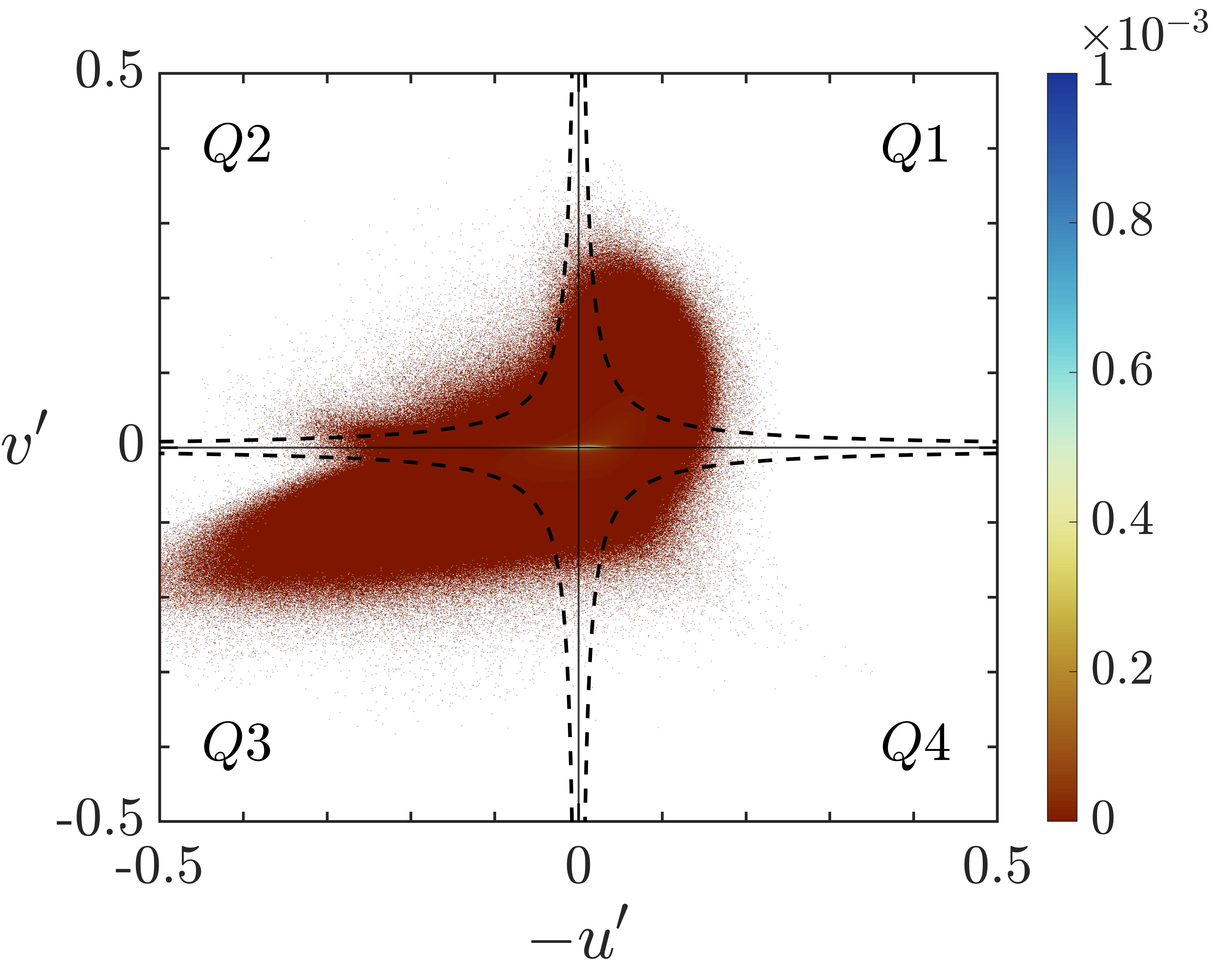}
    \end{subfigure}
    $TP:$\hspace*{4mm}%
    \begin{subfigure}[tbp]{.35\textwidth}
        {\captionsetup{position=bottom, labelfont=it,textfont=it,singlelinecheck=false,justification=centering,labelformat=parens}
        \caption{}\label{fig:jpdf_heat_y_0:TP}}
        \vspace*{-0.8mm}
        \includegraphics[width=1\linewidth]{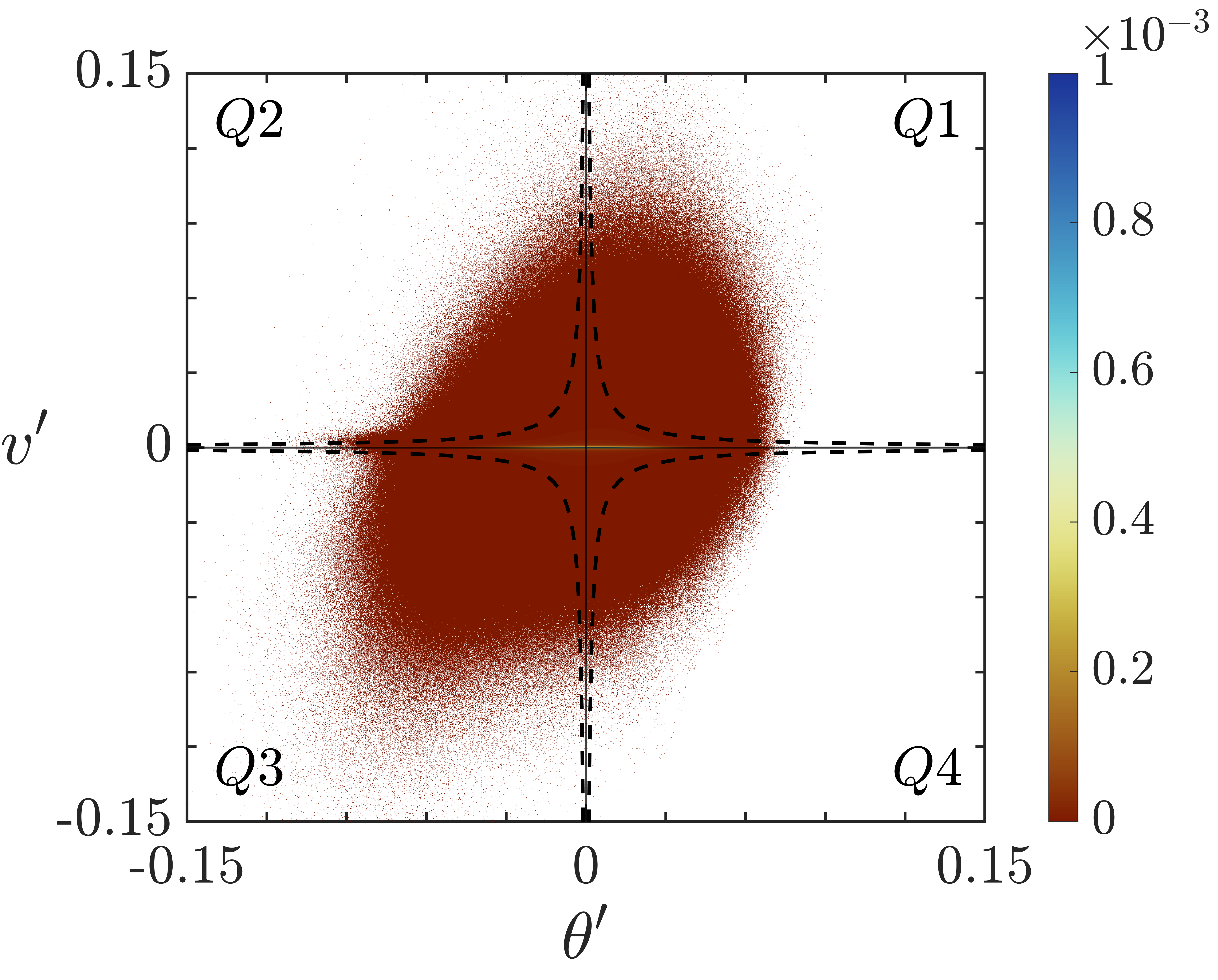}
    \end{subfigure}\hspace*{2mm}%
    \begin{subfigure}[tbp]{.35\textwidth}
        {\captionsetup{position=bottom, labelfont=it,textfont=it,singlelinecheck=false,justification=centering,labelformat=parens}
        \caption{}\label{fig:jpdf_momentum_y_0:TP}}
        \vspace*{-0.8mm}
        \includegraphics[width=1\linewidth]{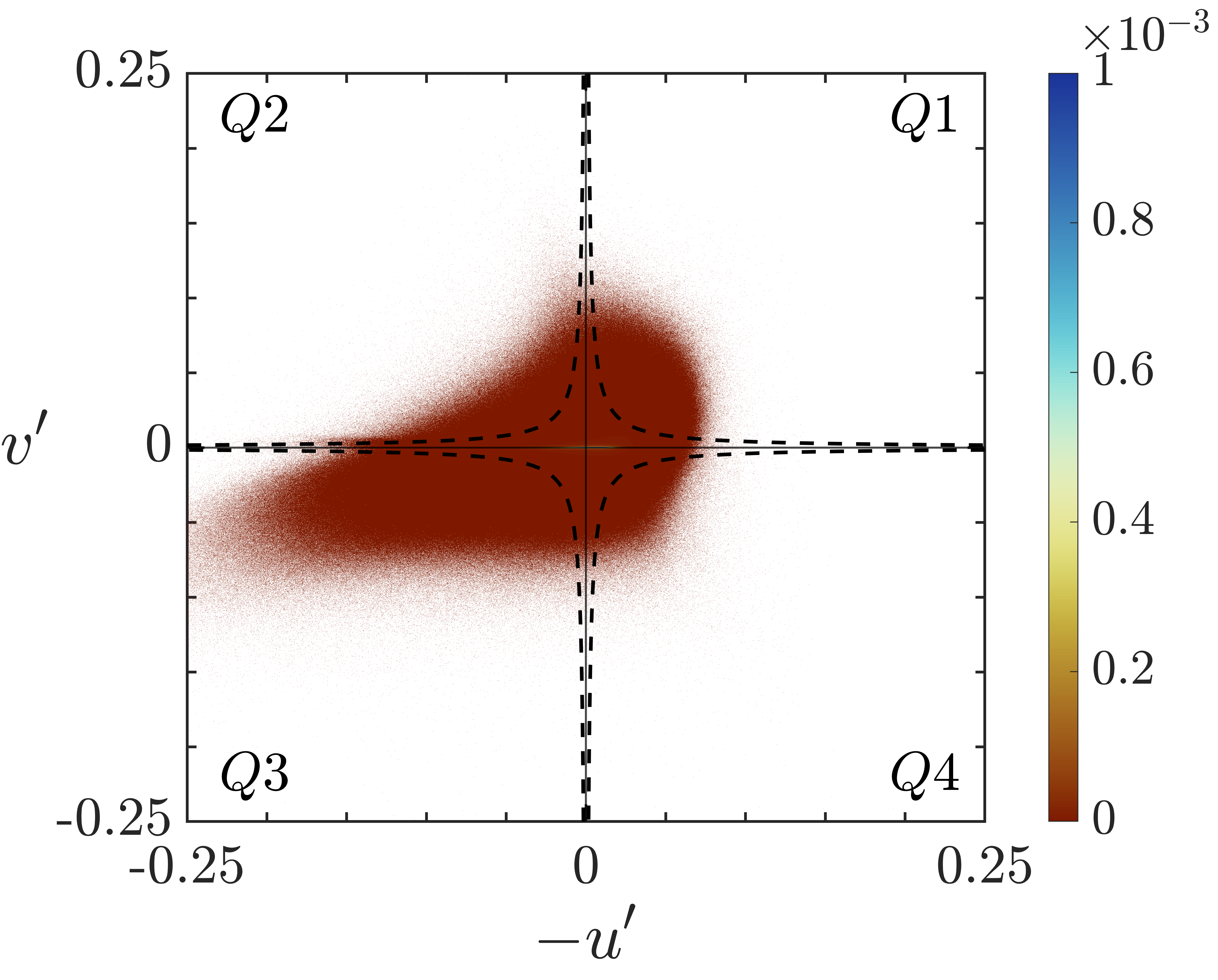}
    \end{subfigure}
    $SP:$\hspace*{5mm}%
    \begin{subfigure}[tbp]{.35\textwidth}
        {\captionsetup{position=bottom, labelfont=it,textfont=it,singlelinecheck=false,justification=centering,labelformat=parens}
        \caption{}\label{fig:jpdf_heat_y_0:SP}}
        \vspace*{-0.8mm}
        \includegraphics[width=1\linewidth]{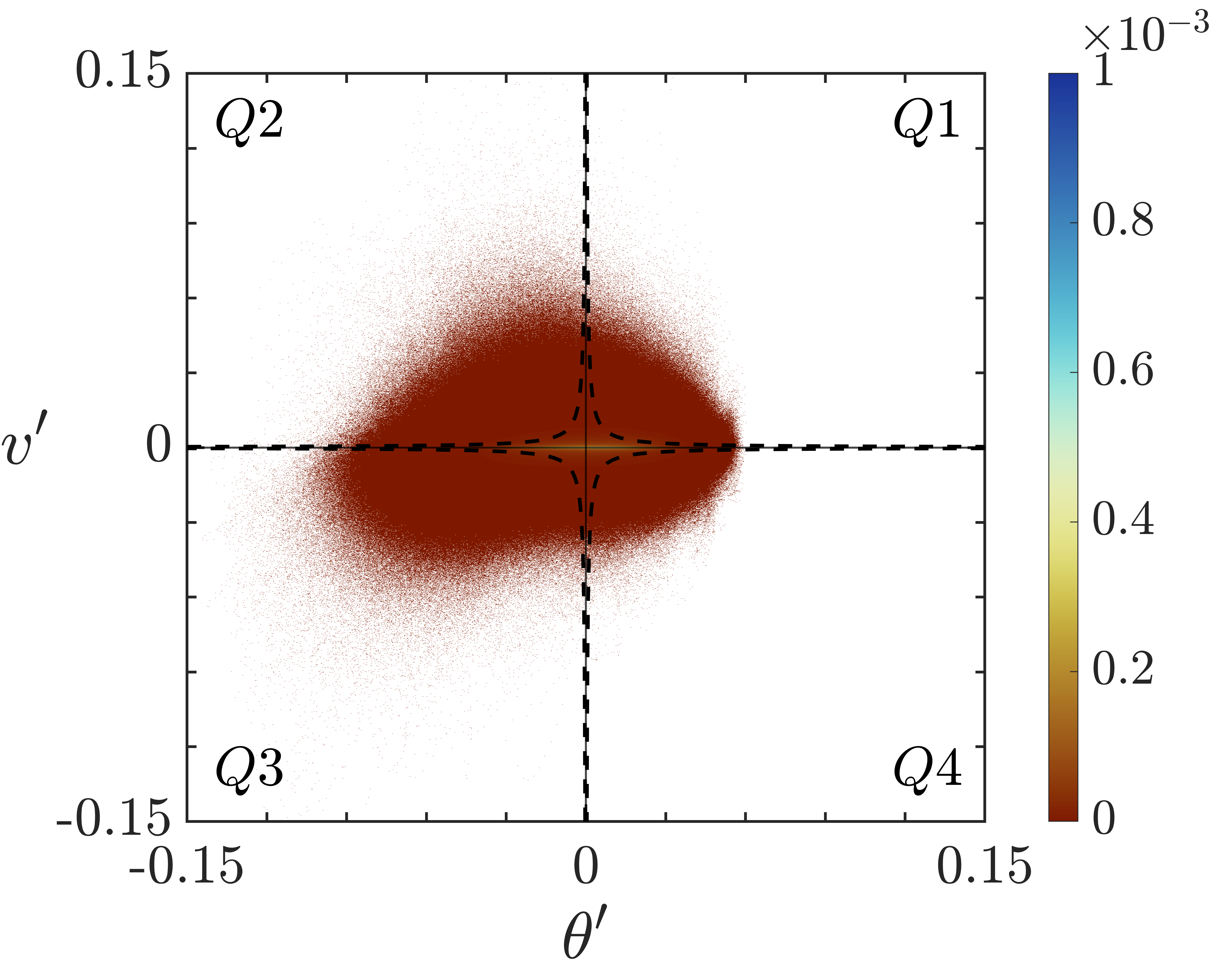}
    \end{subfigure}\hspace*{2mm}%
    \begin{subfigure}[tbp]{.35\textwidth}
        {\captionsetup{position=bottom, labelfont=it,textfont=it,singlelinecheck=false,justification=centering,labelformat=parens}
        \caption{}\label{fig:jpdf_momentum_y_0:SP}}
        \vspace*{-0.8mm}
        \includegraphics[width=1\linewidth]{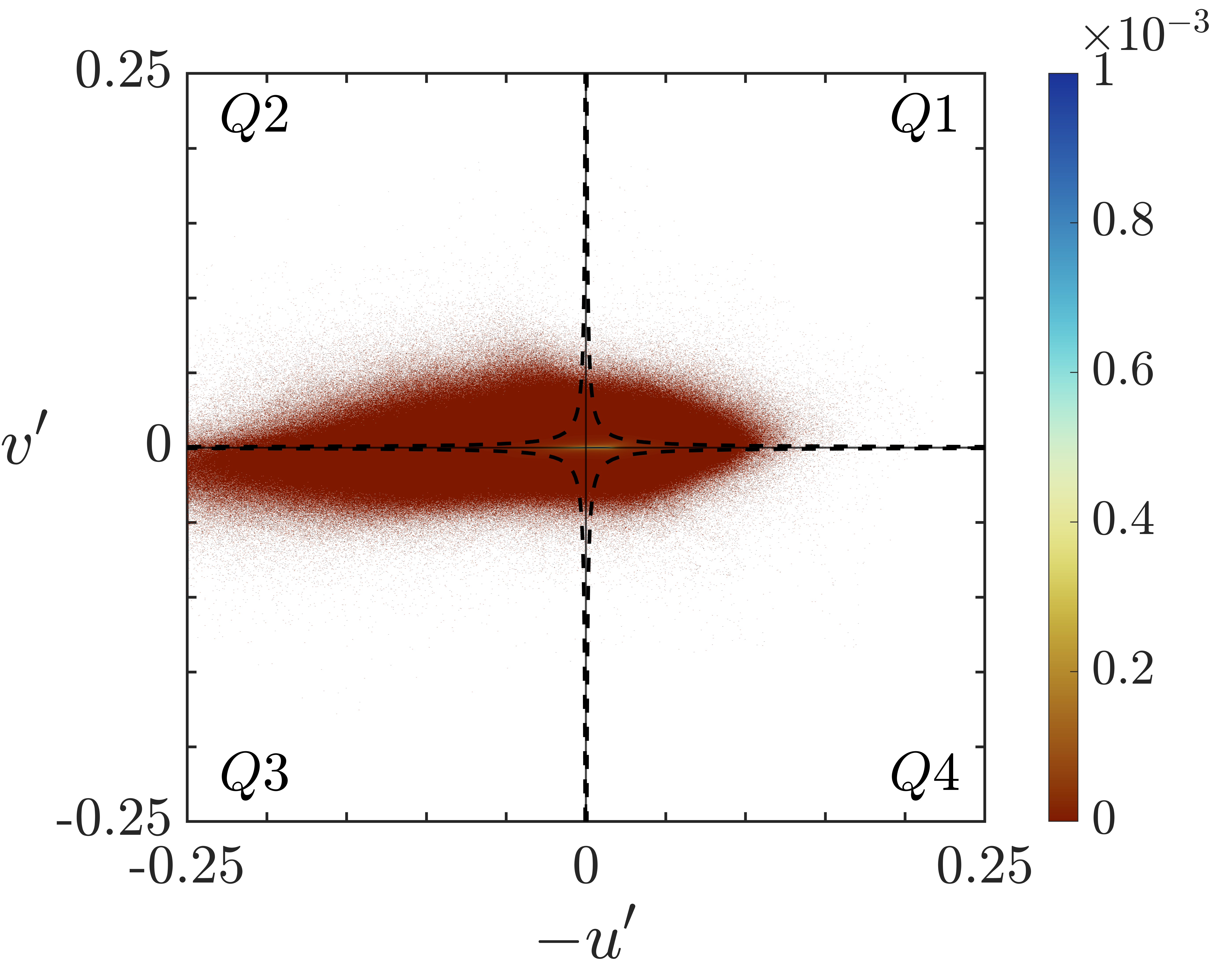}
    \end{subfigure}
    \vspace*{-2mm}
    \captionsetup{width=0.95\textwidth, justification=justified}
    \caption{Quadrant analysis maps for (\emph{a, c, e, g, i}) ${\theta^{\prime}v^{\prime}}$ and (\emph{b, d, f, h, j}) ${u^{\prime}v^{\prime}}$ at $y=0$: (\emph{a, b}) $KP1^{\prime\prime}$, (\emph{c, d}) $KP1^{\prime}$, (\emph{e, f}) $KP1$, (\emph{g, h}) $TP$, (\emph{i, j}) $SP$. Dashed lines are hyperbolas corresponding to $|\theta^{\prime}v^{\prime}| = 8 \times \overline{\theta^{\prime}v^{\prime}}$ and $|u^{\prime}v^{\prime}| = 8 \times -\overline{u^{\prime}v^{\prime}}$. Empty bins are not shown. Note that the range of the axes is not the same in all the plots.}
    \label{fig:jpdf}
 \end{center}
\end{figure}
%%%%%%%%%%%%%%%%%%%%%%%%%%%%%%%%%%%%%%%%%%%%%%%%%%%%%%%%%%%%%%%%%%%%%%%%%%%%%%%%%%%%%%%%%%%%%%%%%%%%%%%%%%%%%%%%%%%%%%%%%%%%%%%%%%%%%%%%%%%%%%%%%%%%%%

 Setting aside $Pr_T$, it is still clear that the turbulence dynamics affect the momentum and heat fields differently in the vicinity of the porous walls. Since the turbulent components of the heat flux term in $St$ and the wall shear stress term in $C_f$ at the surface are $\overline{\theta^{\prime}v^{\prime}}$ and $-\overline{u^{\prime}v^{\prime}}$, flow events should be examined to see how they contribute to these terms. This is done via quadrant analysis of $-\overline{u^{\prime}v^{\prime}}$ and $\overline{\theta^{\prime}v^{\prime}}$ using the joint probabilities of $(-u^{\prime},v^{\prime})$ and $(\theta^{\prime},v^{\prime})$ for select cases in \cref{fig:jpdf}. Since the temperature gradient has the opposite sign of the velocity gradient, events in the first ($Q1$) and third ($Q3$) quadrants of $(\theta^{\prime},v^{\prime})$ contribute positively to the wall-normal heat flux. The negative sign for $u^{\prime}$ in the joint probabilities of $(u^{\prime},v^{\prime})$ is to therefore make the Reynolds shear stress generating events also correspond to $Q1$ and $Q3$. In this manner, comparisons between the quadrant maps of the momentum and heat fluxes are more conveniently made.
 It is evident when going from the smooth\nobreakdash-wall\nobreakdash-like case of $SP$ (\cref{fig:jpdf_momentum_y_0:SP}) to the K\nobreakdash-H\nobreakdash-like case of $KP1$ (\cref{fig:jpdf_momentum_y_0:KP1}), that the overall distribution of events in the quadrant map changes, resulting in a greater number of high-intensity events in $Q1$ and $Q3$. This trend continues in cases $KP1^{\prime}$ (\cref{fig:jpdf_momentum_y_0:KP1-1}) and $KP1^{\prime\prime}$ (\cref{fig:jpdf_momentum_y_0:KP1-2}) where the wall-normal permeability becomes much larger. A greater number of $Q3$ events (large positive $u^\prime$ and large negative $v^\prime$) contribute to the generation of Reynolds shear stress, and the quadrant map has more asymmetry between $Q1$ and $Q3$ than between $Q2$ and $Q4$.
 Examining the quadrant maps for $(\theta^{\prime},v^{\prime})$, they also undergo notable changes in shape when going from $SP$ (\cref{fig:jpdf_heat_y_0:SP}) to $KP1$ (\cref{fig:jpdf_heat_y_0:KP1}), becoming more diagonally symmetric. As a result, while contributions to $\overline{\theta^{\prime}v^{\prime}}$ increase in both $Q1$ and $Q3$, there is no clear preponderance of either event type, with both quadrants largely mirroring each other. This is in contrast to the quadrant maps of $(-u^{\prime},v^{\prime})$, which have notable asymmetry between $Q1$ and $Q3$. The events of $Q1$ and $Q3$ for $(\theta^{\prime},v^{\prime})$ also do not undergo an expansion similar to $Q2$ and $Q4$ for $(-u^{\prime},v^{\prime})$ when going from $KP1$ to $KP1^{\prime}$ (\cref{fig:jpdf_heat_y_0:KP1-1}) and $KP1^{\prime\prime}$ (\cref{fig:jpdf_heat_y_0:KP1-2}).

 The quadrant analysis thus shows that turbulent flow events do not affect heat and momentum flux in a similar way. Events generating $-\overline{u^{\prime}v^{\prime}}$ are more numerous and intense than those generating $\overline{\theta^{\prime}v^{\prime}}$, and this dissimilarity in the turbulent transport of heat is why the Reynolds analogy undergoes an unfavorable breakdown over the porous walls of \cref{tab:DNS}. As to why such a difference in the transport mechanisms emerges, that is a question which demands to be pursued in-depth in a separate study of its own.

\section{Summery and Discussion}\label{sec:conclusions}
Direct numerical simulations of porous-wall turbulence with conjugate heat transfer have been conducted in this study.
One focus of the study has been to assess the heat transfer performance achievable with porous walls compared to other passive structures that have similar effects, particularly rough walls. Unlike rough walls, which experience a limit in heat transfer once fully rough conditions are achieved, porous walls do not experience a similar limitation, at least in-so-far as the walls of this study are considered. The equivalent of a form-drag dominating regime does not develop over a porous wall, which is the main mechanism behind the limitation of heat transfer over rough walls. Pressure has no analogue in the heat equation, leaving only diffusion as the mechanism via which a heat flux can occur over a surface. Under fully rough conditions, characterized by viscous independence, the diffusive mechanism becomes so diminished across the elements of a rough surface that the heat flux becomes limited. Across the surface of a porous wall, this situation does not occur. Unlike a rough wall, the depth of a porous structure permits the penetration of wall-normal velocity, which in turn prevents regions of successive flow separation and reattachment from occurring. This is of course highly dependent on the interfacial topography of a porous wall. The walls examined in this work have no protruding elements at the surface level, minimizing their interfacial roughness. For other surface topographies, form drag may play a more prominent role. Nevertheless, for porous walls an upper limit for heat transfer must also exist, since a wall cannot indefinitely be made more permeable. Before the limit of a vanishing wall is reached, it should cease to affect the overlying turbulent flow as a porous wall does, and any heat-transfer enhancing effects should become lost. This limit remains to be established however.

While heat transfer alone becomes enhanced over the porous walls, this enhancement must be gauged in terms of the overall energy that is expended to maintain the flow conditions necessary to facilitate the heat transfer. For this purpose, the ratio $St/St_{s}$ versus $C_f/C_{f_{s}}$ was examined. This quantifies the fractional increases in heat transfer against the fractional increases in skin friction relative to a smooth-wall turbulent flow. The Reynolds analogy, i.e. similarity in the turbulent transport of momentum and heat, is applicable to smooth-wall flows. Hence, the ratio $St/St_{s}$ versus $C_f/C_{f_{s}}$ measures the breakdown in heat and momentum transfer similarity for a target flow with non-smooth surfaces, and whether this breakdown is favorable or not. As such, it serves as an indicator of heat transfer efficiency. In common with almost all passive surfaces, porous walls also cause an unfavorable breakdown of the Reynolds analogy. As mentioned before however, they surpass the limits of rough surfaces in terms of $St/St_{s}$. The former saturate around $C_f/C_{f_{s}}\approx3$, whereas the porous walls do not. Another important aspect also becomes revealed by examining $St/St_{s}$ versus $C_f/C_{f_{s}}$. In the low-permeability porous-wall flow regime where turbulence remains smooth-wall-like, making a favorably conducting solid wall permeable reduces the overall rate of heat transfer. This is because in this regime, the transport of heat over and within the wall are diffusive dominated. Hence, by introducing voids into the solid wall, the overall conductive capacity of it becomes diminished. Once permeability becomes present to sufficient amounts, the transition to K-H-like regime becomes triggered, and the change in heat transfer becomes positive. The point at which the trend becomes positive however is likely dependent on the thermal properties of the solid material. Therefore, depending on the target application, an investigation is necessary to determine how much permeability is required before positive gains in heat transfer are obtained.

Related to the matter of dissimilarity in the turbulent transport of heat and momentum, the commonly employed turbulent Prandtl number, $Pr_t$, showed behavior that contradicted the results of the Reynolds analogy analysis. This however is attributable to the fact that porous-wall flows are very similar to those of canopies. For canopy flows, it has been established that the eddy\nobreakdash-diffusivity concept is not applicable, due to such flows being similar to mixing-layers. As such, they are characterized by strong counter-gradient motions caused by the cross-stream K-H-like rollers. Such structures also occur over a porous wall and therefore lead to the same characteristics. Therefore, caution is advised in using eddy-diffusivity turbulence models that take prescribed values of $Pr_t$ as inputs. For non-canonical flows involving non-smooth surfaces such models cannot be relied upon.

As to why the heat transfer does scale proportionally to the flow drag, quadrant analysis of turbulent shear stress and heat flux reveals that high intensity events feature much more prominently in the generation of Reynolds shear stress compared to the turbulent heat flux. This disparity in flow events becomes greater as the permeability of wall grows larger. It remains an open question as to why flow turbulence promotes such strongly dissimilar behavior in the heat and momentum fields. The work carried out here will hopefully generate interest into this aspect of non-canonical turbulent heat transfer and lead to more research directed at uncovering the causes behind this dissimilar development.

\FloatBarrier

\backsection[Acknowledgements]{The numerical simulations in this work were conducted using the computational resources of the PDC Center for High Performance Computing, KTH Royal Institute of Technology, and the National Supercomputer Centre (NSC), Linköping University. Access to these computing centers was provided by the National Academic Infrastructure for Supercomputing in Sweden (NAISS) through project number 2023/1-19.}

\backsection[Funding]{S. Bagheri acknowledges the funding provided by the Swedish Foundation for Strategic Research (SSF) through grant SSF\nobreakdash-FFL15\nobreakdash-0001. G. Berthouwer acknowledges the funding provided by the Swedish Research Council (VR) through grant 2021-03967.}

\backsection[Declaration of interests]{The authors report no conflicts of interest.}

% \backsection[Author ORCIDs]{\\
% S. M. Habibi Khorasani, \hyperlink{https://orcid.org/0000-0001-6520-3261}{https://orcid.org/0000-0001-6520-3261}\\
% G. Berthouwer, \hyperlink{https://orcid.org/0000-0002-9819-2906}{https://orcid.org/0000-0002-9819-2906}\\
% S. Bagheri, \hyperlink{https://orcid.org/0000-0002-8209-1449}{https://orcid.org/0000-0002-8209-1449}}

\pagebreak

\appendix

\bibliographystyle{jfm}
\bibliography{jfm}

\end{document}